\documentclass[a4paper,11pt]{article}
\pdfoutput=1
\usepackage[utf8]{inputenc}
\usepackage{jheppub}
\usepackage{amsmath}
\usepackage{amssymb}
\usepackage{color}
\usepackage{graphicx}
\usepackage{xspace,slashed}
\usepackage{stackengine}
\usepackage{booktabs}
\usepackage{mathtools}
\usepackage{etoolbox}
\usepackage{tikz}
\usepackage[compat=1.1.0]{tikz-feynman}
\usetikzlibrary{backgrounds}
\tikzfeynmanset{/tikzfeynman/warn luatex=false}
\usepackage{hyperref}

\makeatletter
\g@addto@macro\bfseries{\boldmath}
\makeatother

\newcommand{\be}{\begin{equation}}
\newcommand{\ee}{\end{equation}}
\newcommand{\MSbar}{\overline{\text{MS}}}
\newcommand{\ep}{\epsilon}
\newcommand{\zb}{\bar{z}}
\newcommand{\pb}{\bar{p}}
\newcommand{\R}{\text{R}}
\newcommand{\RR}{\text{RR}}
\newcommand{\RV}{\text{RV}}
\newcommand{\RRR}{\text{RRR}}
\newcommand{\RRV}{\text{RRV}}
\newcommand{\RVV}{\text{RVV}}
\newcommand{\RVsq}{\text{RV}^2}
\newcommand{\I}{\mathrm{i}}
\newcommand{\E}{\mathrm{e}}
\newcommand{\rmd}{\mathrm{d}}
\newcommand{\progname}[1]{\textsc{#1}}
\newcommand{\mat}[1]{\mathbf{#1}}
\DeclareMathOperator*{\BaikCut}{Cut}
\DeclareMathOperator{\Tr}{Tr}
\renewcommand{\Re}{\operatorname{Re}}

\newcommand{\BareBeamUnren}[2][]{B_{#2}^{\text{bare}\ifblank{#1}{}{,#1}}}
\newcommand{\BareBeam}[2][]{\bar{B}_{#2}^{\text{bare}\ifblank{#1}{}{,#1}}}
\newcommand{\Beam}[2][]{B_{#2}^{#1}}
\newcommand{\MC}[2][]{\mathcal{I}_{#2}^{#1}}
\newcommand{\PDFs}[2][]{f_{#2}^{#1}}
\newcommand{\dPhiB}[1][]{\rmd\Phi_B^{\ifblank{#1}{}{(#1)}}}
\newcommand{\pfD}[2]{D^{\text{#1}}_{#2}}
\newcommand{\pfDt}[2]{\tilde{D}^{\text{#1}}_{#2}}

\allowdisplaybreaks

\newcommand{\CERN}{Theoretical Physics Department, CERN,
  1211 Geneva 23, Switzerland}
\newcommand{\KIT}{Institute for Theoretical Particle Physics, KIT,
  76128 Karlsruhe, Germany}
\newcommand{\UZH}{Physik Institut, Universit\"at Z\"urich,
  Winterthurerstrasse 190, 8057 Z\"urich, Switzerland}
\newcommand{\TUM}{Physics Department, Technical University of Munich,
  James-Franck-Stra\ss{}e~1, 85748 Garching, Germany}
\newcommand{\Bosch}{Corporate Sector Research and Advanced Engineering,
  Robert Bosch GmbH, Robert-Bosch-Campus~1, 71272 Renningen, Germany}

\preprint{
  {\raggedleft
  TTP22-067 \\
  P3H-22-108 \\
  TUM-HEP-1429/22 \\
  CERN-TH-2022-178 \\
  ZU-TH 51/22 \\
  }
}

\title{
Beam functions for $N$-jettiness at N$^3$LO in perturbative QCD
}
\author[a,b]{Daniel Baranowski,}
\author[a,c]{Arnd Behring,}
\author[a]{Kirill Melnikov,}
\author[d]{Lorenzo Tancredi,}
\author[d,e]{Christopher Wever}
\affiliation[a]{\KIT}
\affiliation[b]{\UZH}
\affiliation[c]{\CERN}
\affiliation[d]{\TUM}
\affiliation[e]{\Bosch}

\abstract{
  We present a calculation of all matching coefficients for $N$-jettiness beam
  functions at next-to-next-to-next-to-leading order (N$^3$LO) in perturbative
  quantum chromodynamics (QCD). Our computation is performed starting from the
  respective collinear splitting kernels, which we integrate using the axial
  gauge. We use reverse unitarity to map the relevant phase-space integrals to
  loop integrals, which allows us to employ multi-loop techniques including
  integration-by-parts identities and differential equations. We find a
  canonical basis and use an algorithm to establish non-trivial partial
  fraction relations among the resulting master integrals, which allows us to
  reduce their number substantially. By use of regularity conditions, we
  express all necessary boundary constants in terms of an independent set,
  which we compute by direct integration of the corresponding integrals in the
  soft limit. In this way, we provide an entirely independent calculation of
  the matching coefficients which were previously computed in
  Ref.~\cite{Ebert:2020unb}.
}

\begin{document}

\maketitle

\setcounter{footnote}{0}
\renewcommand*{\thefootnote}{\arabic{footnote}}%
\clearpage

\section{Introduction}
\label{sec:intro}

The high energy and luminosity of the Large Hadron Collider, as well as the
excellent performance of the ATLAS and CMS detectors allow one to conduct
high-precision studies of a large number of processes with the goal to
stress-test the Standard Model and to search for possible deviations from its
predictions. Perturbative QCD enables an accurate description of hadron
collisions through a systematically improvable computation of partonic cross
sections and kinematic distributions, and plays a central role in this
endeavour.

However, making theoretical predictions for partonic QCD processes is
complicated. One reason for that are the infra-red and collinear divergences
which have to be carefully extracted and cancelled between elastic and
inelastic contributions to the final result in any fixed-order perturbative
computation. While such divergences naturally appear as poles in the
dimensional regularisation parameter $\ep$ when elastic contributions to a
given process are computed, the situation is more complex for inelastic ones,
where additional partons appear in the final state. In such cases, the
singularities arise when these partons become either soft or collinear to other
partons, but they only turn into poles in $\ep$ once an integration over the
energies and angles of these additional partons is performed. Since such
integrations are not suitable for computations that aim at describing arbitrary
infra-red safe distributions of final state particles, special methods have to
be designed to allow the extraction of soft and collinear singularities without
the integration over the kinematic variables of the resolved partons.

Two distinct methods to do this have been proposed and developed since the
early days of perturbative computations; one is called subtraction and the
other one is called slicing. The idea of the subtraction method is to subtract
and add back an approximate expression of the product of the relevant matrix
elements squared and the multi-particle phase space. The difference between the
exact and approximate expressions should be integrable in four dimensions. The
integral of the subtraction term over the unresolved phase space of final state
particles should be performed in $d=4-2 \ep$ dimensions either numerically or
analytically. Several subtraction methods for generic hadron collider processes
have been worked out at next-to-leading order (NLO) \cite{Catani:1996vz,%
Frixione:1995ms} and at next-to-next-to-leading order (NNLO) in perturbative
QCD \cite{Somogyi:2005xz,Gehrmann-DeRidder:2005btv,Czakon:2010td,%
Caola:2017dug,Herzog:2018ily,Magnea:2018hab} but their extension to
next-next-to-next-to-leading order (N$^3$LO) in QCD is not available.

The slicing method seeks to split the phase space of final state particles into
the (most) singular and non-singular (or less singular) parts. The most
singular contributions typically arise when all particles in the final state,
beyond those present in the Born process, become soft or collinear. The less
singular parts correspond to processes which contain resolved partons in
addition to the Born ones; such processes can be dealt with by computing
lower-order perturbative corrections to higher-multiplicity processes. Hence,
if one is interested in computing N$^3$LO QCD corrections to the partonic
process $pp \to X$ using the slicing method, one needs to know NNLO QCD
corrections to the process $pp \to X+\text{jet}$ and N$^3$LO QCD contribution
to $pp \to X$ which comes from the fully-unresolved region of phase space.

The exact definition of resolved and unresolved phase-space regions requires
the introduction of a slicing variable. The choice of the slicing variable can
be arbitrary but recently two such variables have been used for several NNLO
and N$^3$LO QCD computations. For processes where colour-singlet (e.g., $H$,
$Z$, $W$, $ZZ$, $WW$ etc.) or heavy colour-charged (e.g., $t \bar{t}$ etc.)
particles are produced in hadron collisions at the Born level, one can use
their total transverse momentum $q_\perp$ \cite{Catani:2007vq} to distinguish
elastic and inelastic contributions. Another option is to use the $N$-jettiness
variable \cite{Stewart:2009yx,Stewart:2010tn}, first introduced in the context
of Soft Collinear Effective Theory (SCET)~\cite{Bauer:2001ct,Bauer:2001yt,%
Bauer:2002nz,Beneke:2002ph,Beneke:2002ni}, to distinguish processes with a
different number of jets. The advantage of using $N$-jettiness for slicing
stems from the fact that it can be employed for processes with jets at leading
order, whereas dealing with such processes remains a challenge for
$q_\perp$-slicing.

The construction of a slicing scheme that uses the $N$-jettiness variable
benefits from the existence of a factorisation theorem that describes the
behaviour of relevant cross sections at small values of the $N$-jettiness
variable. To present this theorem, we consider the case of colour-singlet
production in proton-proton collision. Since the Born process in this case
contains no final-state jets, the generic $N$-jettiness variable becomes
zero-jettiness. It is defined as
\begin{align}
  \mathcal{T}_0
    &= \sum \limits_{j=1}^{N} \min_{i \in \{1,2\}}
       \left[ \frac{2 p_i \cdot k_j}{Q_i} \right]
  \label{eq1.1}
  \,.
\end{align}
In Eq.~\eqref{eq1.1} the sum runs over all final state QCD partons. For each
final state parton $j$, the smallest scalar product of its momentum $k_j$ and
the momenta of the incoming partons $p_{1,2}$ contributes to the value of
$\mathcal{T}_0$. The $Q_i$ are the so-called hardness variables; they can be
chosen in different ways and are of no relevance for the computations described
in this paper.

Schematically, the cross section of the process $pp \to V+X$, where $V$ is a
colour-singlet system, at small values of zero-jettiness can be written
as~\cite{Stewart:2009yx,Stewart:2010tn,Stewart:2010qs}
\begin{align}
  \lim_{\mathcal{T}_0 \to 0} \rmd\sigma_{pp \to V+X}^{\text{N$^3$LO}}
    &\approx B \otimes B \otimes S_0 \otimes H
      \otimes \rmd\sigma_{pp \to V}^{\text{LO}}
  \label{eq1.2}
  \,.
\end{align}
In Eq.~\eqref{eq1.2}, the summation over different partonic species is implied
and $\otimes$ stands for the convolutions. Furthermore, $H$ is a
process-specific hard function which, essentially, accounts for loop
corrections to the Born process, $S_0$ is the zero-jettiness soft function and
$B$ is the beam function that accounts for the effects of the collinear QCD
radiation off the incoming partons.

We note that in contrast to the soft and hard functions, the beam function in
Eq.~\eqref{eq1.2} is universal in that it does not depend on the process and on
the number of jets in the final state. Eq.~\eqref{eq1.2} suggests that, in
order to calculate the cross section for small values of $\mathcal{T}_0$
through, say, N$^3$LO in perturbative QCD, one has to compute the beam
function, the soft function and the hard function to third perturbative order
and then combine them to obtain the unresolved contribution to $pp \to V+X$
cross section.

We note that, in principle, beam functions are non-perturbative objects.
However, at leading power in $\Lambda_{\text{QCD}}/\mathcal{T}$, their
non-perturbative parts are related to parton distribution functions (PDFs)
$\PDFs{j}$ through the following formula
\begin{align}
  \Beam{i}
    &= \sum \limits_{\text{partons}~j} \MC{ij} \otimes \PDFs{j}
  \,, \quad \text{where} \quad i,j \in \{g, u, \bar{u}, d, \bar{d}, \dots\}
  \label{eq1.3}
  \,.
\end{align}
The quantities $\MC{ij}$ are the so-called matching coefficients; they can be
computed in perturbative QCD and used to describe fixed-order cross sections at
small values of $N$-jettiness.

The calculation of the matching coefficients $\MC{ij}$ has a long history. The
NLO and NNLO results for the matching coefficients were obtained in
Refs.~\cite{Gaunt:2014cfa,Gaunt:2014xga,Boughezal:2017tdd}. The N$^3$LO QCD
computations were initiated in Refs.~\cite{Melnikov:2019pdm,Melnikov:2018jxb}
and first physics results for $\MC{qq}$ in the generalised large-$N_c$
approximation were presented in Ref.~\cite{Behring:2019quf}. In
Ref.~\cite{Ebert:2020unb} the matching coefficients for all partonic channels
were computed through N$^3$LO in perturbative QCD using the method described in
Ref.~\cite{Ebert:2020lxs}.

The goal of this paper is to complete the calculation described in
Ref.~\cite{Behring:2019quf}. We do this by going beyond the generalised
large-$N_c$ approximation and by computing beam-function matching coefficients
for all partonic channels. As explained in the next section, we perform this
calculation by utilising the connection between partonic beam functions and
integrals of the collinear splitting kernels pointed out in
Ref.~\cite{Ritzmann:2014mka} (see also Ref.~\cite{Catani:2022sgr} for a recent
discussion). We note that this method of computing the matching coefficients is
very different from the method used in Refs.~\cite{Ebert:2020lxs,%
Ebert:2020unb}. Thus, our calculation provides a fully independent check of the
results reported in Ref.~\cite{Ebert:2020unb} and demonstrates the practical
utility of working in a ghost-free physical gauge.

The rest of the paper is organised as follows. In Section~\ref{sec:calc}, we
discuss the calculation of the matching coefficients. We begin with the
description of the computational setup. Then, we discuss the derivation of the
differential equations for master integrals and their solutions, the
computation of the boundary conditions and the assembly of the partonic beam
functions. In Section~\ref{sec:matching-coeff}, we describe the renormalisation
of the beam function and the extraction of the matching coefficients. We
present the results of the computation in Section~\ref{sec:res} and conclude in
Section~\ref{sec:conc}. Additional discussion of particular aspects of the
calculation can be found in the appendices.

\section{Calculation}
\label{sec:calc}

Beam functions were originally defined in SCET as matrix elements of particular
operators calculated with respect to external hadronic states
\cite{Stewart:2009yx,Stewart:2010qs}. The matching relation in
Eq.~\eqref{eq1.3} arises from an operator product expansion (OPE) in the limit
$\Lambda_{\text{QCD}}/\mathcal{T} \ll 1$. Since the OPE is independent
of the external states, the matching coefficients $\MC{ij}$ remain the same if
we replace the hadronic external states by the partonic ones. The matching
relation between partonic beam functions $\Beam{ij}$ and partonic PDFs
$\PDFs{ij}$ reads
\begin{align}
  \Beam{ij}
    &= \sum_{k \in \{g, u, \bar{u}, d, \bar{d}, \dots\}}
       \MC{ik} \otimes \PDFs{kj}
  \label{eq:partonic-matching-relation}
  \,.
\end{align}
In comparison to hadronic quantities, the partonic ones carry an additional
index $j$ which specifies the flavour of the external parton. Thus, one
possibility to compute the matching coefficients is to directly calculate the
matrix elements of SCET operators on both sides of the matching relation with
external partonic states.

However, it is more practical to forgo the SCET-based definition and to
calculate the matching coefficients directly by integrating collinear splitting
functions over momenta of the collinear partons subject to certain phase-space
constraints \cite{Ritzmann:2014mka}. The advantage of this procedure is that it
allows one to work with familiar objects such as Feynman diagrams and
scattering amplitudes instead of matrix elements of complicated SCET operators.

Let us sketch the computation of the matching coefficient $\MC{ij}$. According
to Eqs.~\eqref{eq1.2} and \eqref{eq1.3}, the matching coefficient is related to
the behaviour of the cross section at small values of zero-jettiness.
Zero-jettiness becomes small if final state partons are soft or collinear to
incoming partons. By forcing the final state partons to be in the collinear
limit to one of the incoming ones, we effectively project the differential
cross section onto beam functions. As explained above, replacing the external
hadronic states by partonic states introduces partonic beam functions. The
partonic beam function describes a process where a parton $j$ in the initial
state emits collinear partons, loses some of its original momentum, goes
slightly off-shell, changes its identity to a parton $i$ and continues into the
hard process. The final state kinematics of this collinear-splitting process is
subject to a constraint on the zero-jettiness variable. According to
Eq.~\eqref{eq:partonic-matching-relation}, the matching coefficients are then
obtained by removing contributions, that have to be associated with partonic
PDFs, from the partonic beam functions. In the rest of this section, we
describe how to calculate the fully unrenormalised, bare partonic beam
functions $\BareBeamUnren{ij}$ from phase-space integrals over splitting
functions. We will explain how to use these results to derive the matching
coefficients in Section~\ref{sec:matching-coeff}.

In total, there are five independent flavour combinations to consider for $i$
and $j$ and, therefore, also five different matching coefficients, $(i j) \in
\{q_i q_j, q g, g q, g g, q_i \bar{q}_j\}$. In addition, depending on the order
of QCD perturbation theory, there are contributions with different numbers of
final-state partons. To compute the N$^3$LO contribution to the matching
coefficient, we need to consider final states with up to three additional
partons, depending on how many virtual loops appear in a particular amplitude.
We will refer to these different processes as triple-real (RRR), double-real
virtual (RRV) and real double-virtual (RVV) contributions.

\begin{figure}
  \centering
  \begin{tikzpicture}
    \begin{feynman}
      \vertex (p) {$j(p)$};
      \vertex (e1) at ($(p)+(-25:1.5)$);
      \vertex (e2) at ($(e1)+(-25:1.2)$);
      \vertex (e3) at ($(e2)+(-25:1.2)$);
      \vertex (hard) at ($(e3)+(-25:1.2)$);
      \vertex (V) at ($(hard)+(1,0)$) {$V$};
      \vertex (k3) at ($(V)+(0,1.2)$) {$f_3(k_3)$};
      \vertex (k2) at ($(k3)+(0,1.2)$) {$f_2(k_2)$};
      \vertex (k1) at ($(k2)+(0,1.2)$) {$f_1(k_1)$};
      \vertex (pb) at ($(hard)+(-155:2)$) {$\bar{f}(\pb)$};

      \diagram*{
        (p) -- [fermion] (e1)
            -- [fermion] (e2)
            -- [fermion] (e3)
            -- [fermion,edge label={$i^*(p_i)$}] (hard)
            -- [fermion] (pb),
        (k1) -- [gluon,bend right] (e1),
        (k2) -- [gluon,bend right] (e2),
        (k3) -- [gluon,bend right] (e3),
        (hard) -- [boson] (V)
      };
    \end{feynman}
  \end{tikzpicture}
  \caption{Sketch of a triple-real contribution to the process
    $j(p) + \bar{f}(\pb) \to V + f_1(k_1) + f_2(k_2) + f_3(k_3)$, for $j = q$,
    $\bar{f} = \bar{q}$ and $f_{1,2,3} = g$, to illustrate the notation for the
    momenta and flavours.}
  \label{fig:notation}
\end{figure}
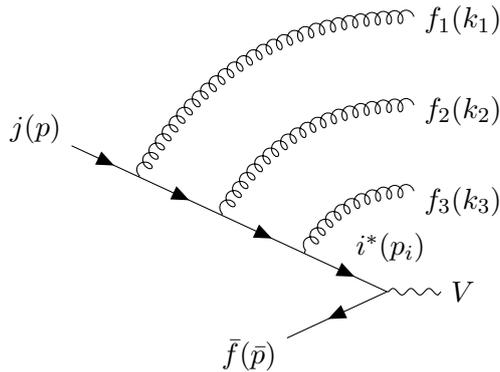
For the sake of concreteness, we consider a triple-real contribution to a
partonic beam function. To compute it, we study the partonic process
\begin{align}
  j(p) + \bar{f}(\pb) &\to V + f_1(k_1) + f_2(k_2) + f_3(k_3)
\end{align}
for a colour-singlet state $V$, and investigate its squared matrix element in
the collinear limit $k_1 || k_2 || k_3 || p$. We illustrate our notations in
Figure~\ref{fig:notation}. It is well-known \cite{Catani:1999ss} that in this
limit the squared matrix element factorises into a product of the splitting
function that describes a transition of a parton $j$ to a parton $i$ along with
three collinear partons $f_{1,2,3}$, and the hard matrix element of the process
$i + \bar{f}(\pb) \to V$. We write
\begin{align}
  \lim_{k_1 || k_2 ||k_3 || p}
    |\mathcal{M}(j(p), \bar{f}(\pb); f_1, f_2, f_3)|^2
    &\sim \frac{P_{ij}(p, \pb, \{k_1,k_2,k_3\})}{s^2_{123}}
      |\mathcal{M}(i(z p), \bar{f} (\pb))|^2
  \label{eq2.2}
  \,,
\end{align}
where $s_{123} = (p-k_1-k_2-k_3)^2 $ is the off-shell propagator of the parton
$i$ which enters the hard process and $P_{ij}$ is the corresponding splitting
function. In principle, the factorisation formula shown in Eq.~\eqref{eq2.2} is
not exact because of spin correlations but since, eventually, we integrate over
momenta of collinear partons, the spin correlations average out.

To define the variable $z$ in Eq.~\eqref{eq2.2}, we use the Sudakov
decomposition of the momentum $p_i = p - k_1 - k_2 - k_3$ and write
\begin{align}
  p_i &= z p + y \pb + k_{\perp}
  \,,
\end{align}
where $p \cdot k_\perp = \pb \cdot k_\perp = 0$. Thanks to momentum
conservation, we find
\begin{align}
  k_{123} &= k_1 + k_2 + k_3 = (1-z) p - y \pb - k_{\perp}
  \,.
\end{align}
Multiplying this equation with $\pb$ and using $\pb^2 = 0$, we obtain
\begin{align}
  1-z &= \frac{\pb \cdot k_{123}}{\pb \cdot p}
  \,.
\end{align}

In the collinear limit where $k_{1,2,3} || p$, the zero-jettiness defined in
Eq.~\eqref{eq1.1} simplifies and becomes\footnote{We note that this formula
applies to an arbitrary $N$-jettiness variable $\mathcal{T}$, not only to the
zero-jettiness, in the limit where unresolved final state partons are collinear
to the incoming parton with momentum $p$.}
\begin{align}
  \mathcal{T}_0 &\approx \frac{2 p \cdot k_{123}}{Q_p}
  \label{eq2.4}
  \,.
\end{align}
To compute the matching coefficient at fixed zero-jettiness, we introduce a new
variable, the so-called transverse virtuality $t$, defined as
\begin{align}
  t &= 2 z p \cdot k_{123}
  \,.
\end{align}
Once the matching coefficients at fixed $t$ are available, it is
straightforward to change variables from $t$ to $\mathcal{T}_0$ using
Eq.~\eqref{eq2.4}, if needed.

To compute the partonic beam functions, we need to integrate over energies and
angles of the emitted partons, keeping $z$ and $t$ fixed. To account for these
constraints, we follow the common practice and introduce two delta functions
into the phase-space integral, writing the triple-real contribution to the
partonic beam function as follows
\begin{align}
  \BareBeamUnren{ij}
    &\sim \int \dPhiB[3,0] \, \frac{P_{ij}(k_1,k_2,k_3)}{s^2_{123}}
  \label{eq2.5}
  \,.
\end{align}
In Eq.~\eqref{eq2.5} $\dPhiB[3,0]$ is defined as
\begin{align}
  \dPhiB[3,0]
    &= \prod \limits_{m=1}^{3} [\rmd k_m]
       \delta\left(2 p \cdot k_{123} - \frac{t}{z}\right)
       \delta\left(\frac{2 \pb \cdot k_{123}}{s} - (1-z)\right)
  \label{eq2.6}
  \,,
\end{align}
with $s = 2 p \cdot \pb$ and
\begin{align}
  [\rmd k_m] &= \frac{\rmd^d k_m}{(2 \pi)^{d-1}} \delta^+(k_m^2),
\end{align}
is the phase-space element for the parton $f_m$ with the momentum $k_m$. We
note that contributions of lower-multiplicity final states can also be computed
using Eq.~\eqref{eq2.5} except that the corresponding splitting functions
should include loop contributions, and the number of final-state partons should
be reduced accordingly.

Since NLO and NNLO QCD splitting functions are known
\cite{Catani:1999ss,Bern:2004cz,Badger:2004uk}, one can integrate them directly
to compute the partonic beam functions. In principle, many ingredients required
for an N$^3$LO QCD computation are also known \cite{DelDuca:2019ggv,%
DelDuca:2020vst,Catani:2003vu,Badger:2015cxa,Sborlini:2014mpa,Czakon:2022fqi}
but the results at this order are so complex that using them for our purposes
does not appear to be beneficial. We thus decided to compute the collinear
projections of the relevant matrix elements and their contributions to the
various N$^3$LO QCD splitting functions on our own, to have full control over
their possible simplifications.

To do that, we follow Ref.~\cite{Catani:1999ss} where it is explained how to
extract the splitting functions by applying collinear projections to parton
scattering amplitudes. To perform the calculation in a process-independent way,
we need to use the \emph{axial gauge} for real and virtual gluons since, if
such a gauge is used, collinear singularities only appear in diagrams where
virtual and real gluons are emitted and absorbed by the \emph{same} external
parton $j$. Hence, when computing the sum over polarisations for a gluon with
the momentum $k$, we use the following formula
\begin{align}
  \sum_{\lambda} {\epsilon_\lambda}^\mu(k) {\epsilon^*_\lambda}^{\nu}(k)
    &= -g^{\mu \nu} + \frac{k^\mu \pb^\nu + k^\nu \pb^\mu}{k \cdot \pb}
  \,,
\end{align}
where have chosen $\pb$ as the auxiliary light-like vector required to define
the axial gauge. We note that this formula also defines the numerator of the
gluon propagator which we use to compute the virtual corrections and also,
since we use the axial gauge, no ghost particles need to be included.

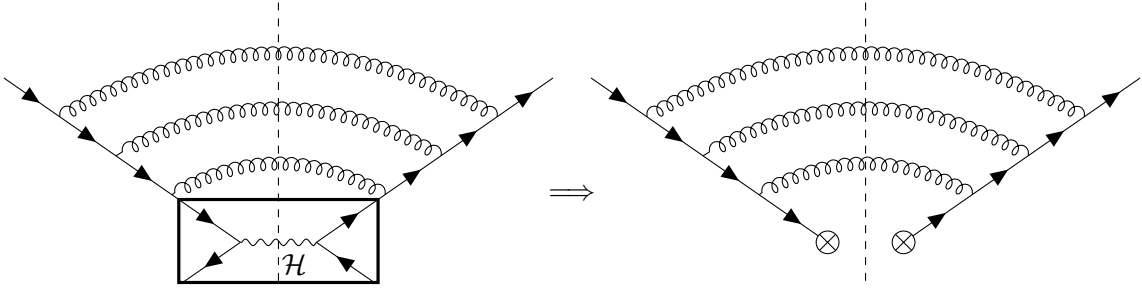
\begin{figure}
  \centering
  \begin{tikzpicture}
    \begin{feynman}
      \vertex (p);
      \vertex (e1) at ($(p)+(-35:0.9)$);
      \vertex (e2) at ($(e1)+(-35:0.9)$);
      \vertex (e3) at ($(e2)+(-35:0.9)$);
      \vertex (hard) at ($(e3)+(-35:1.1)$);
      \vertex (pb) at ($(hard)+(-145:0.9)$);
      \vertex (hardc) at ($(hard)+(1,0)$);
      \vertex (e3c) at ($(hardc)+(35:1.1)$);
      \vertex (e2c) at ($(e3c)+(35:0.9)$);
      \vertex (e1c) at ($(e2c)+(35:0.9)$);
      \vertex (pc) at ($(e1c)+(35:0.9)$);
      \vertex (pbc) at ($(hardc)+(-35:0.9)$);

      \diagram*{
        (p) -- [fermion] (e1)
            -- [fermion] (e2)
            -- [fermion] (e3)
            -- [fermion] (hard)
            -- [fermion] (pb),
        (pbc) -- [fermion] (hardc)
              -- [fermion] (e3c)
              -- [fermion] (e2c)
              -- [fermion] (e1c)
              -- [fermion] (pc),
        (e1c) -- [gluon,bend right] (e1),
        (e2c) -- [gluon,bend right] (e2),
        (e3c) -- [gluon,bend right] (e3),
        (hard) -- [boson] (hardc)
      };
    \end{feynman}
    \draw[dashed] ($(pc)!0.5!(p)+(0,1)$) -- ($(pc)!0.5!(p)+(0,-2.8)$);
    \coordinate (Hbox1) at ($(e3)!0.1!(hard)$);
    \coordinate (Hbox2) at ($(e3c)!0.1!(hardc)$);
    \coordinate (Hbox3) at ($(Hbox2)+(0,-1.1)$);
    \coordinate (Hbox4) at ($(Hbox1)+(0,-1.1)$);
    \draw[very thick] (Hbox1) -- (Hbox2) -- (Hbox3) -- (Hbox4) -- cycle;
    \node at ($(Hbox3)+(-1.1,0.25)$) {$\mathcal{H}$};
    \begin{feynman}
      \vertex (Pp) at ($(pc)+(0.5,0)$);
      \vertex (Pe1) at ($(Pp)+(-35:0.9)$);
      \vertex (Pe2) at ($(Pe1)+(-35:0.9)$);
      \vertex (Pe3) at ($(Pe2)+(-35:0.9)$);
      \vertex (Phard) at ($(Pe3)+(-35:1.1)$);
      \vertex (Phardc) at ($(Phard)+(1,0)$);
      \vertex (Pe3c) at ($(Phardc)+(35:1.1)$);
      \vertex (Pe2c) at ($(Pe3c)+(35:0.9)$);
      \vertex (Pe1c) at ($(Pe2c)+(35:0.9)$);
      \vertex (Ppc) at ($(Pe1c)+(35:0.9)$);

      \diagram*{
        (Pp) -- [fermion] (Pe1)
             -- [fermion] (Pe2)
             -- [fermion] (Pe3)
             -- [fermion] (Phard),
        (Phardc) -- [fermion] (Pe3c)
                 -- [fermion] (Pe2c)
                 -- [fermion] (Pe1c)
                 -- [fermion] (Ppc),
        (Pe1c) -- [gluon,bend right] (Pe1),
        (Pe2c) -- [gluon,bend right] (Pe2),
        (Pe3c) -- [gluon,bend right] (Pe3),
      };
      \node[crossed dot,fill=white] at (Phard) {};
      \node[crossed dot,fill=white] at (Phardc) {};
    \end{feynman}
    \draw[dashed] ($(Ppc)!0.5!(Pp)+(0,1)$) -- ($(Ppc)!0.5!(Pp)+(0,-2.8)$);
    \node at ($(e3c)!0.5!(Pe3)$) {$\Longrightarrow$};
  \end{tikzpicture}
  \caption{Sketch of a squared diagram from the RRR contribution of
    $q\bar{q} \to V + ggg$ (left-hand side) and the same contribution to the
    partonic beam function where the hard process ($\mathcal{H}$) has been
    replaced by a projector denoted by the crossed dots (right-hand side). The
    dashed vertical line represents the final state cut. The Feynman rules used
    for the projector are explained in the main text.}
  \label{fig:projector}
\end{figure}

The next step in constructing the splitting functions consists in choosing the
``hard process'' in a smart way. As explained in Ref.~\cite{Catani:1999ss},
one has to distinguish between cases where a quark or a gluon enters the hard
scattering. In the quark case, $i = q$, we obtain the splitting functions by
simply replacing the hard scattering kernel defined in
Figure~\ref{fig:projector} with $\hat{\pb} /(4 \, p_i \cdot \pb) = \hat{\pb} /
(4 z \, p \cdot \pb)$ where $\hat{\pb} = \pb^\mu \gamma_\mu$. Hence, the matrix
element becomes
\begin{align}
  \Tr \left[\dots \hat{p}_i \, \mathcal{H} \, \hat {p}_i \dots\right]
    &\to \frac{\Tr\left[\dots \hat{p}_i \, \hat{\pb} \, \hat{p}_i
         \dots\right]}{4 z \, p \cdot \pb}
  \,,
\end{align}
where ellipses stand for other contributions stemming from the Feynman diagrams
that describe the $j \to i^*+f_1+f_2+f_3$ transition. The trace originates from
the summation over polarisation states of initial or final state quarks. Thanks
to the universal nature of the splitting functions and to the properties of the
collinear limit, this procedure projects the relevant amplitudes onto singular
contributions which arise when the emitted partons become collinear to the
incoming ones \cite{Catani:1999ss}. Finally, we note that in order to compute
partonic beam functions which describe processes where gluons enter the hard
process, i.e. $\Beam{gq}$ and $\Beam{gg}$, we simply use $-g^{\mu \nu}/(d-2)$
as a proxy for the hard matrix element.

This procedure allows us to compute the contributions to the splitting
functions from individual Feynman diagrams which describe a process where
partons are emitted by an incoming parton $j$ with momentum $p$, and an
off-shell parton $i$ with momentum $z p$ enters the hard process. Once these
contributions to the splitting functions are computed, they need to be
integrated over the unresolved phase space of all final state partons while
keeping the transverse virtuality $t$ and the $z$-parameter fixed, c.f.
Eqs.~(\ref{eq2.5},\ref{eq2.6}). We perform this integration by employing the
method of reverse unitarity \cite{Anastasiou:2002yz}. The main idea behind this
method is to map all delta functions that appear in a given integrand, for
example the on-shell conditions for final state partons $\delta(k^2)$ or the
kinematic constraints such as $\delta(2 p \cdot k_{123} - t/z)$, onto
propagator-like structures using the identity
\begin{align}
  \delta(X)
    &= \frac{1}{2\pi \I} \left[\frac{1}{X-\I 0} - \frac{1}{X+\I 0}\right]
  \,.
\end{align}

This mapping is beneficial because it turns constrained phase-space integrals
into ``loop-like'' integrals, enabling the use of standard multi-loop
technologies, such as the integration-by-parts method (IBPs)
\cite{Tkachov:1981wb,Chetyrkin:1981qh} and differential equations technique
\cite{Kotikov:1990kg,Bern:1993kr,Remiddi:1997ny,Gehrmann:1999as}, to compute
them.

The rest of the calculation proceeds in a relatively standard way. We use
\progname{QGRAF}~\cite{Nogueira:1991ex} to generate diagrams that describe the
various partonic processes, e.g., $j(p) \to i^*(p_i)+ f_1(k_1) + f_2(k_2) +
f_3(k_3) $. We note that since the parton $i$ is off-shell,
we have to account for self-energy corrections on this leg. We use
\progname{FORM}~\cite{Vermaseren:2000nd,Kuipers:2012rf,Kuipers:2013pba,%
Ruijl:2017dtg} to deal with the Dirac and Lorentz algebra and
\progname{color.h}~\cite{vanRitbergen:1998pn} for the colour algebra. For each
partonic process and final state multiplicity we define suitable integral
families which close under IBPs. We use
\progname{Reduze2}~\cite{vonManteuffel:2012np} and
\progname{Kira}~\cite{Maierhofer:2017gsa,Maierhofer:2018gpa,Maierhofer:2019goc,%
Klappert:2020nbg} to solve the system of IBP identities and express all
integrals in terms of a (relatively) small number of master integrals.

Since the Laporta algorithm requires the propagators of each integral family to
be linearly independent, we have to apply a partial fraction decomposition to
integrals where linearly dependent propagators appear. In our calculation,
linear dependencies between propagators arise because of the phase-space delta
functions and the axial-gauge propagators. For example, the delta functions in
the RRV contributions imply the following constraint
\begin{align}
  2 (k_1 + k_2) \cdot \pb &= s (1-z)
  \,,
\end{align}
and it gives rise to partial fraction identities such as
\begin{align}
  \frac{\delta (2 k_{12} \cdot \pb/s - (1-z))}{%
    (k_1 \cdot \pb) (k_2 \cdot \pb)}
    &= \frac{2 \, \delta (2 k_{12} \cdot \pb/s - (1-z)) }{s(1-z)} \left[
         \frac{1}{k_1 \cdot \pb}
         +\frac{1}{k_2 \cdot \pb}
       \right]
  \,.
\end{align}

Moreover, the propagator of a gluon with momentum $k$ in the axial gauge
features a term proportional to $(k \cdot \pb)^{-1}$. As every integral has to
contain the cut propagator corresponding to $\delta(2 k_{1 \dots n_R} \cdot
\pb/s - (1-z))$, where $k_{1 \dots n_R} = \sum \limits_{i=1}^{n_R} k_i$ is the
total momentum of final state partons, and since at N$^3$LO there are at most
three linearly independent scalar products $k_i \cdot \pb$, $i=1,2,3$, diagrams
with sufficiently many gluon propagators will contain linearly dependent
propagators. We remove such linear dependencies by systematically applying
partial fraction relations using polynomial reduction over Gr\"obner bases. The
algorithm that we use has been described in
Refs.~\cite{Pak:2011xt,Hoff:2015kub} (see also Refs.~\cite{Meyer:2017joq,%
Abreu:2019odu,Heller:2021qkz} for related work). We discuss an example of the
application of this algorithm in Appendix~\ref{sec:parfrac}.

The above discussion applies to the mapping of integrals that appear in the
transition amplitudes onto the integral families. However, it turns out to be
important to identify linearly-dependent master integrals which belong to
\emph{different} families. This step reduces the number of integrals that need
to be computed and it contributes towards significant simplifications of
systems of differential equations that the master integrals satisfy. An example
of such a relation between three RRR master integrals is
\begin{align}
  \MoveEqLeft{\overbrace{\int \frac{\dPhiB[3,0]}{(k_1-p)^2 (k_{13}-p)^2
    (k_3 \cdot \pb) (k_{13} \cdot \pb)}}^{=I^\text{T29}_{3391}}}
  \notag \\
    &= \underbrace{\int \frac{\dPhiB[3,0]}{(k_1-p)^2 (k_{12}-p)^2
         (k_1 \cdot \pb) (k_2 \cdot \pb)}}_{=I^\text{T1}_{3263}}
       -\underbrace{\int \frac{\dPhiB[3,0]}{(k_1-p)^2 (k_{13}-p)^2
         (k_1 \cdot \pb) (k_{13} \cdot \pb)}}_{=I^\text{T28}_{3291}}
  \,,
  \label{eq:pfrel}
\end{align}
where $I_s^f$ denotes an integral from family $f$ and sector $s$ and
$\dPhiB[3,0]$ is the phase-space measure, c.f. Eq.~\eqref{eq2.6}. This relation
becomes obvious after relabeling $k_2 \leftrightarrow k_3$ in the first
integral on the right hand side of Eq.~\eqref{eq:pfrel}; it arises because the
propagators $1/(k_1 \cdot \pb)$, $1/(k_3 \cdot \pb)$ and $1/(k_{13} \cdot \pb)$
are linearly dependent.

To construct partial fraction relations between master integrals we generate a
list of seed integrals from \emph{all sectors} of \emph{all} integral families.
We then calculate a Gr\"obner basis for the overcomplete set of all propagators
obtained by joining all integral families. After polynomial reduction of the
seed integrals with respect to this Gr\"obner basis, we map the integrals to
integral families and apply IBP reduction to all integrals. This yields a large
system of linear equations which involve integrals that are identified as
master integrals by the Laporta algorithm. Finally, we solve the system taking
into account the Laporta ordering of the integrals. The non-trivial solutions
of this system give us linear relations between master integrals. By applying
these relations, we observe a significant reduction in the number of master
integrals that are required to express the transition amplitude.

Nevertheless, the question whether \emph{all} linear relations are found in
this way remains open. As we explained earlier, this is an important question
because eliminating linear dependencies is crucial for achieving the simplest
possible form of the differential equations for the master integrals.
Unfortunately, some of these linear dependencies only become apparent after
relabeling and shifting real and virtual momenta. Since the Gr\"obner basis is
calculated for one particular choice of the integration momenta, it is
conceivable that some partial dependencies are not captured by the procedure
described above. Extending it in a way that avoids loopholes related to choices
of integration momenta and symmetry transformations of the integrals is an
interesting question worthy of further investigation.

We note here that, as a matter of fact, we do find further relations between
master integrals but only a posteriori, by looking for linear combinations of
master integrals that fulfil purely \emph{homogeneous} differential equations,
see Section~\ref{sec:calc-candeq}. More details are given in
Ref.~\cite{Melnikov:2019pdm}.

To simplify the computation of master integrals, we remark that their dependence
on the variables $t$ and $s$ is homogeneous; this allows us to treat the
integrals as functions of the single variable $z$. To see this, we note that,
for example, a triple-real integral can be schematically written as follows
\begin{align}
  I(s,t,z)
    &= \int \prod \limits_{m=1}^{3} [\rmd k_m]
       \delta\left(2 p\cdot k_{123} - \frac{t}{z}\right)
       \delta\left(\frac{2 \pb \cdot k_{123}}{s} - (1-z)\right)
       F(p,\pb,\{k_i\})
  \label{eq2.13}
  \,,
\end{align}
where $F$ is given by a product of propagators constructed from inverse powers
of
\begin{align}
  &(p-l_A)^2 \quad \text{and} \quad \pb \cdot l_A
  \,, & \text{where~}
  l_A \in \{k_1,k_2,k_3,k_{12},k_{13},k_{23},k_{123}\}
  \label{eq2.14}
  \,,
\end{align}
and
\begin{align}
  &l_A^2
  \,, & \text{with~} l_A \in \{k_{12},k_{13},k_{23},k_{123}\}
  \label{eq2.15}
  \,.
\end{align}
It is now easy to see that Eq.~\eqref{eq2.13} simplifies if we introduce new
(tilded) momenta according to
\begin{align}
  k_i &= \sqrt{\frac{t}{z}} \tilde k_i
  \,, &
  p &= \sqrt{\frac{t}{z}} \, \tilde p
  \,, &
  \pb &= s \sqrt{\frac{z}{t}} \tilde{\pb}
  \,.
\end{align}
Using the tilded momenta in Eq.~\eqref{eq2.13}, we obtain
\begin{align}
  I(s,t,z)
    &= s^{n_s} \left(\frac{t}{z}\right)^{n_t-3 \ep}
       \int \rmd \tilde{\Phi}_B
       F(\tilde{p},\tilde{\pb},\{\tilde{k}_i\})
  \notag \\
    &= s^{n_s} \left(\frac{t}{z}\right)^{n_t-3\ep} \, I(1,1,z)
  \label{eq2.21}
  \,,
\end{align}
where $\rmd \tilde{\Phi}_B$ can be obtained from Eq.~\eqref{eq2.6} by
substituting $t \to 1$, $s \to 1$, $p \to \tilde{p}$ and $\pb \to \tilde{\pb}$
there. Also, $n_{s,t}$ are integral-dependent integers and the last step in
Eq.~\eqref{eq2.21} follows from the fact that $ \tilde{p} \cdot \tilde{\pb} =
1/2$. By a slight abuse of notation, we will occasionally use $p$ and $\pb$
instead of $\tilde{p}$ and $\tilde{\pb}$; it should be clear from the context
which normalisation is used in a particular part of the calculation. Note that
an analogous discussion also holds for RRV integrals.

To derive the differential equations for the master integrals with respect to
the variable $z$, we follow the standard procedure. We differentiate the
integrands with respect to $z$ and use IBP identities and partial fraction
relations to express the derivatives through the master integrals. For solving
the differential equations, we need to supply the boundary conditions; to do
that, we compute the required master integrals in the $z \to 1$ limit where all
emitted partons become soft. In the next sections we first describe how we
choose the basis of master integrals to simplify the solution of the
differential equations and then elaborate on the details of the computation of
the relevant boundary conditions.

\subsection{Differential equations for master integrals}
\label{sec:calc-candeq}

We use differential
equations~\cite{Kotikov:1990kg,Bern:1993kr,Remiddi:1997ny,Gehrmann:1999as} to
compute the master integrals. As the integrals depend non-trivially on a single
variable $z$, we have to deal with the following system of differential
equations
\begin{align}
  \partial_z \vec{I}(z,\ep)
    &= \mat{A}(z,\ep) \vec{I}(z,\ep)
  \label{eq:deq-laporta}
  \,.
\end{align}
In Eq.~\eqref{eq:deq-laporta} $\vec{I}(z,\ep)$ is the vector of master
integrals and $\mat{A}(z,\ep)$ is a matrix of (rational) functions of $z$ and
of the dimensional regularisation parameter $\ep$.

It is well known that different choices of master integrals can significantly
impact the form of the matrix $\mat{A}(z,\ep)$ and the complexity of the system
of differential equations. When it exists, a canonical basis of master
integrals~\cite{Henn:2013pwa} is a particularly convenient choice.

For integrals in a canonical basis
the differential equations assume the especially simple form
\begin{align}
  \partial_z \vec{I}_c (z,\ep) = \ep \mat{A}_c(z) \vec{I}_c(z,\ep)
  \,, \quad
  \text{where~} \quad
  \mat{A}_c(z) = \sum_i f_i(z) \mat{A}_{c,i}
  \,.
  \label{eq:deq-can}
\end{align}
A further crucial property of a canonical basis is that the matrix
$\mat{A}_c(z) $ must be in the so-called $\text{dlog}$ form. This means that in
Eq.~\eqref{eq:deq-can} $f_i(z) = \rmd\log{g_i(z)}/\rmd z$ for some algebraic
functions $g_i(z)$, and the entries of the $z$-independent matrices
$\mat{A}_{c,i}$ are just numbers.

Since the $\ep$-dependence of the right-hand side in Eq.~\eqref{eq:deq-can} is
completely factorised, the solutions of the homogeneous equation at each order
in $\ep$ are just constants, and one can construct an iterative solution of
this equation in a straightforward manner. Inserting a Laurent-series ansatz in
$\ep$ for the integrals $\vec{I}_c(z,\ep) = \sum \limits_{k=0}^{} \ep^k
\vec{I}_c^{(k)}(z)$, we obtain the solutions
\begin{align}
  \vec{I}_c^{(k)}(z)
    &= \vec{B}^{(k)}
       +\sum_i \mat{A}_{c,i} \int\limits^{z} \rmd z' \,
         f_i(z') \vec{I}_c^{(k-1)}(z')
  \label{eq:canonical-mi-epexp}
  \,.
\end{align}
Here, $\vec{B}^{(k)}$ are the vectors of integration constants which need to be
determined by computing the integrals for a particular value of $z$. The
$z$-dependence of the solution is calculated through iterated integration over
$z$. The kernels of such iterated integrals are specified by the logarithmic
differential forms in $\mat{A}_c(z)$. A famous result~\cite{Chen:1977oja} in
the theory of iterated integrals guarantees that, as long as the logarithmic
forms are independent, the resulting iterated integrals will also be linearly
independent of each other. Besides the simplicity of the differential
equations, another benefit of having a canonical basis is that the cancellation
of spurious poles in $\ep$ in the amplitude is made explicit so that we do not
have to compute the master integrals to higher orders in $\ep$ than strictly
necessary.

Finding a canonical basis for master integrals is a non-trivial task. Starting
from a generic choice of master integrals, one performs a basis transformation
\begin{align}
  \vec{I}_c(z,\ep) &= \mat{T}(z,\ep) \vec{I}(z,\ep)
  \,.
\end{align}
The differential equation for $\vec{I}_c(z,\ep)$ becomes
\begin{align}
  \partial_z \vec{I}_c (z,\ep)
    &= \tilde{\mat{A}}(z,\ep) \vec{I}_c(z,\ep)
  \,,
\end{align}
where
\begin{align}
  \tilde{\mat{A}}(z,\ep)
    &= \mat{T}(z,\ep) \mat{A}(z,\ep) \mat{T}^{-1}(z,\ep)
       +(\partial_z \mat{T}(z,\ep)) \mat{T}^{-1}(z,\ep)
  \label{eq:deq-basis-transform}
  \,.
\end{align}
If one is able to find a matrix $\mat{T}(z,\ep)$ such that
$\tilde{\mat{A}}(z,\ep) = \ep \mat{A}_c(z)$, one arrives at a canonical basis.

Various semi-automated methods have been developed to find the transformation
to a canonical basis. Iterative approaches to constructing the matrix $\mat{T}$
starting from a general matrix $\mat{A}(z,\ep)$, have been discussed in
Refs.~\cite{Gehrmann:2014bfa,Lee:2014ioa,Argeri:2014qva,Gituliar:2017vzm,%
Meyer:2016slj,Meyer:2017joq,Prausa:2017ltv,Lee:2020zfb}. An alternative
procedure based on constructing candidates for the canonical basis through the
analysis of the so-called leading singularities was suggested in
Refs.~\cite{Arkani-Hamed:2010pyv,Henn:2013pwa,Henn:2020lye}. Other approaches
to finding canonical bases were discussed in Refs.~\cite{Hoschele:2014qsa,%
Dlapa:2020cwj,Chen:2020uyk,Chen:2022lzr}.

While the advantage of having canonical bases is clear and significant
progress in developing algorithms to determine them has been achieved, their
application to non-trivial problems, such as the calculation of the N$^3$LO QCD
contributions to beam functions, remains a difficult task. In fact, upon
inspecting the differential equations for our problem, we observe that their
solutions involve rational functions of $z$ and three different square roots
\begin{align}
  \sqrt{z}\sqrt{4-z} \,, &&
  \sqrt{z}\sqrt{4+z} \,, &&
  \text{and} &&
  \sqrt{4+z^2} \,.
  \label{eq:square-roots}
\end{align}
The presence of algebraic functions renders the application of the majority of
automated algorithms for finding a canonical basis either impossible or highly
non-trivial. Hence, we have decided to adopt a pragmatic approach to finding
the canonical basis for our system of differential equations. It is
based on the following steps:

\begin{enumerate}
  \item As a starting point, we choose a basis of integrals whose differential
        equations do not contain denominators which mix the kinematic variable
        $z$ with the dimensional regularisation parameter $\ep$. Such
        denominators unnecessarily complicate the calculation and can be avoided
        by choosing the master integrals appropriately (see
        Refs.~\cite{Melnikov:2016qoc,Usovitsch:2020jrk,Smirnov:2020quc} for
        related discussions). Here, we encountered such denominators only for a
        handful of integrals so that we searched for appropriate replacements on
        a case-by-case basis. It was sufficient to replace integrals for which
        this occurs either by other integrals from the same sector\footnote{As
        usual, the term \emph{sector} denotes the set of propagators that is
        present in the denominator of an integral. We call a sector $S'$ a
        \emph{subsector} of another sector $S$ if the set of denominators of
        $S'$ is a subset of those present in $S$ and a \emph{supersector} if
        the set of denominators in $S'$ is a superset of those present in $S$.}
        or, in a small number of cases, by integrals from a supersector which
        does not contain any master integrals.\footnote{A heuristic for choosing
        candidates for the replacement integrals was whether or not the
        reductions of the candidate integrals to the original master integrals
        contain the offending factors in the denominator.}
  \item To make use of the automated packages for the determination of a
        canonical basis, it was beneficial to rationalise square roots. We
        note that according to Eq.~\eqref{eq:square-roots} all
        square roots involve second-degree polynomials in the variable $z$.
        Hence, each of them can be easily rationalised by a particular variable
        transformation. On the other hand, we could not find a transformation
        that rationalises all three square roots at the same time.

        To take full advantage of the possibility to rationalise individual
        roots, we split up the differential equations for the master integrals
        into subsystems which close and contain at most one square root. It
        turns out that this can be achieved for the majority of the relevant
        integrals. For the sake of completeness, we present the variable
        transformations which rationalise three square roots in
        Eq.~\eqref{eq:square-roots}
        \begin{align}
          \sqrt{z}\sqrt{4-z}: && z \to z = \frac{(1+x)^2}{x}
          \label{eq:sqrt-rat-x}
          \,, \\
          \sqrt{z}\sqrt{4+z}: && z \to z = \frac{(1-y)^2}{y}
          \label{eq:sqrt-rat-y}
          \,, \\
          \sqrt{4+z^2}:       && z \to z = \frac{w^2-1}{w}
          \label{eq:sqrt-rat-w}
          \,.
        \end{align}
        The last transformation is a generalisation of the Landau transformation
        introduced in the first two.

        While the process of rationalisation can make previously simple linear
        letters more complicated, once the equations are rationalised, we can
        use \progname{Fuchsia} \cite{Gituliar:2017vzm} and
        \progname{CANONICA}~\cite{Meyer:2016slj,Meyer:2017joq} to
        algorithmically construct a canonical basis.
  \item For a relatively small number of equations where multiple square roots
        appear simultaneously, we constructed a canonical basis without
        rationalising the square roots.
        To achieve this, we employed a heuristic strategy based on the analysis
        of some of the leading singularities of the corresponding integrals,
        combined with the procedure described in Ref.~\cite{Gehrmann:2014bfa}
        that allows the simplification of possible remaining non-canonical
        entries in the differential equations.
        We performed the analysis of the leading singularities for both
        triple-real and double-real virtual integrals using the Baikov
        representation~\cite{Baikov:1996rk,Baikov:1996iu}. For completeness, we
        present an example of such a construction in
        Appendix~\ref{sec:leading-sing}.
  \item Once canonical systems of differential equations are constructed, we
        switch back to the original variable $z$ even if the canonical basis
        was calculated using $x,y$ or $w$ variables. This allows us to integrate
        all master integrals in a uniform way, as we explain below.
\end{enumerate}

Once the equations are written in canonical form, we can easily solve them in
terms of iterated integrals. As already hinted to above, a general result from
the theory of iterated integrals ensures that, as long as the $\text{dlog}$
forms that we integrate over are independent, also the corresponding iterated
integrals are linearly independent of each other. We stress for clarity that
this is only true once all products of different iterated integrals evaluated
at the same argument have been removed using the standard shuffle product
relations, which hold for any (properly regulated) iterated integral. This
implies that our integration procedure is literally as simple as in the
well-known case of multiple polylogarithms~\cite{Goncharov:1998kja,%
Remiddi:1999ew,Vollinga:2004sn,Gehrmann:2000zt}, being entirely reduced to the
iterative addition of a new ``index'' to the iterated integrals for each
differential form. We define iterated integrals as follows
\begin{align}
  L_{a,\vec{b}}(x)
    &= \int\limits_0^x f_a(t) \, L_{\vec{b}}(t) \, \rmd t
  \,, &
  L_{a}(x)
    &= \int\limits_0^{x} f_a(t) \, \rmd t
  \,, &
  L_{\underbrace{0,...,0}_n}(x)
    &= \frac{1}{n!} \log^n{x}
  \label{eq:iterint}
  \,,
\end{align}
where the functions $f_a(t)$ are differentials of logarithms, as discussed
right after Eq.~\eqref{eq:deq-can}. Note that this definition makes sense
since, in our problem there is only one function $f(t)$ that diverges at zero,
c.f. Eq.~\eqref{eq:letters} below.

Of course, as already mentioned earlier, to completely determine the integrals
from their system of differential equations, we need to know their values
at a certain point in order to fix all the relevant boundary conditions. For all
the integrals considered, it is convenient to choose their soft ($z \to 1$)
limit as a reference point. To simplify the evaluation of the iterated
integrals close to the soft limit, we rewrite the differential equations using
the auxiliary variable $\zb = 1-z$. In fact, when using $\zb$ we find that we
need to consider iterated integrals defined over the following alphabet
\begin{align}
  f_i(\zb) \in \biggl\{&
    \frac{1}{\zb-1},
    \frac{1}{\zb},
    \frac{1}{\zb+1},
    \frac{1}{\zb-2},
    \frac{1}{\zb-3},
    \frac{1}{\zb+3},
    \frac{1}{\zb-5},
    \frac{2\zb-3}{\zb^2-3\zb+3},
    \frac{2\zb-2}{\zb^2-2\zb+5},
  \notag \\ &
    \frac{1}{\sqrt{1-\zb}\sqrt{5-\zb}},
    \frac{1}{\sqrt{1-\zb}\sqrt{3+\zb}},
    \frac{1}{\sqrt{\zb^2-2\zb+5}},
    \frac{1}{(\zb-1) \sqrt{\zb^2-2\zb+5}}
  \biggr\}\,. \label{eq:letters}
\end{align}
We note here that iterated integrals of rational functions and square roots of
polynomials up to degree two have been studied extensively in the
literature~\cite{Aglietti:2004tq,Weinzierl:2004bn,Ablinger:2011te,%
Bonciani:2010ms,Ablinger:2014bra,Ablinger:2021fnc}.

Since we defined the iterated integrals by integrating from $0$ to $\zb$, c.f.
Eq.~\eqref{eq:iterint}, their $\zb \to 0$ limits either vanish or behave as
$\log^n(\zb)$ for some $n \in \mathbb{N}$. Moreover, the fact that the only
letter in Eq.~\eqref{eq:letters} that diverges in the soft $\zb \to 0$ limit is
$1/\zb$, guarantees that all logarithmic $\log^n(\zb)$ singularities can be
systematically extracted by un-shuffling all the occurrences of this letter,
similar to what is done for standard multiple polylogarithms.

In this way, we obtain the canonical master integrals as expansions in $\ep$
through $\mathcal{O}(\ep^5)$, i.e. including $I_c^{(5)}(\zb)$, as functions of
$\zb$. We note that, although the rational letters with quadratic denominators
can be further factorised at the expense of introducing complex numbers, we
keep the quadratic forms to achieve a more compact representation. To
manipulate iterated integrals we make extensive use of the packages
\progname{HarmonicSums}~\cite{Ablinger:2010kw,Ablinger:2013hcp,%
Vermaseren:1998uu,Remiddi:1999ew,Blumlein:2009ta,Ablinger:2011te,%
Ablinger:2013cf,Ablinger:2014rba,Ablinger:2016ll,Ablinger:2017rad,%
Ablinger:2019mkx}, \progname{HPL}~\cite{Maitre:2005uu,Maitre:2007kp} and
\progname{PolyLogTools}~\cite{Duhr:2019tlz}, as well as
\progname{GiNaC}~\cite{Bauer:2000cp,Vollinga:2004sn} for the numerical
evaluation of multiple polylogarithms.

Since the soft ($\zb \to 0)$ singularities of the beam functions are
regularised dimensionally, we have to solve the differential equations for the
master integrals in this limit in closed form in $\ep$, but as generalised
power series expansions in $\zb$. Due to spurious poles in $1/\zb$ in the
amplitudes, in some cases we need the expansions through $\zb^2$ terms. Since
we work with a canonical basis, obtaining a solution in the limit $\zb \to 0$
is particularly simple. The leading behaviour of the canonical integrals in
this limit can be obtained by solving
\begin{align}
  \partial_{\zb} \vec{I}_c^{(0)}(\zb,\ep)
    &= \ep \frac{\mat{A}_{c,0}}{\zb} \vec{I}_c^{(0)}(\zb,\ep)
  \label{eq:deq-soft-limit-lo}
  \,,
\end{align}
where $\mat{A}_{c,0}$ is the coefficient matrix of the letter $f_0(\zb) =
1/\zb$. The solution of Eq.~\eqref{eq:deq-soft-limit-lo} is expressed through
the matrix exponential
\begin{align}
  \vec{I}_c^{(0)}
    &= \mat{\Phi}(\zb,\ep) \vec{B}_{\text{soft}}(\ep)
  \,, & \text{where~}
  \mat{\Phi}
   &= \exp\left(\ep \log(\zb) \mat{A}_{c,0}\right)
    = \zb^{\ep \mat{A}_{c,0}}
  \label{eq:fundsys}
  \,.
\end{align}
The matrix $\mat{\Phi}(\zb,\ep)$ contains the fundamental system of the
solutions to Eq.~\eqref{eq:deq-soft-limit-lo}. The subleading terms are then
obtained by expanding the coefficient matrix $\mat{A}_c(\zb)$ around $\zb=0$ as
$\mat{A}_c(\zb) = \sum_{n=0}^\infty \zb^{n-1} \mat{A}_c^{(n)}$ and then solving
the differential equations
\begin{align}
 \partial_z \vec{I}_c^{(n)}(\zb,\ep)
    &= \ep \sum_{k=0}^{n} \zb^{k-1} \mat{A}_c^{(k)} \vec{I}_c^{(n-k)}(\zb,\ep)
  \notag \\
    &= \ep \Biggl[
         \underbrace{\sum_{k=1}^{n} \zb^{k-1} \mat{A}_c^{(k)}
           \vec{I}_c^{(n-k)}(\zb,\ep)}_{\equiv \vec{R}^{(n)}(\zb,\ep)}
         +\frac{\mat{A}_{c,0}}{\zb} \vec{I}_c^{(n)}(\zb,\ep)
       \Biggr]
  \label{eq:deq-soft-limit-n}
  \,.
\end{align}
It is obvious that, for $n=0$, Eq.~\eqref{eq:deq-soft-limit-n}
simplifies to Eq.~\eqref{eq:deq-soft-limit-lo}.

We note that we require solutions $\vec{I}_c^{(n)}(\zb,\ep)$ of
Eq.~\eqref{eq:deq-soft-limit-n} that behave asymptotically as $\zb^n$ in the
limit $\zb \to 0$. This implies that for $n>0$ we only need to consider the
inhomogeneous solutions of the above equation. We obtain them using the method
of variation of constants. We note that the fundamental system of the
homogeneous equation that is needed to construct the inhomogeneous solution is
always $\mat{\Phi}(\zb,\ep)$ defined in Eq.~\eqref{eq:fundsys}. We find
\begin{align}
  I_c^{(n)}(\zb,\ep)
    &= \mat{\Phi}(\zb,\ep) \int \limits_{0}^{\zb} \rmd \zb' \,
       \mat{\Phi}^{-1}(\zb',\ep) \vec{R}^{(n)}(\zb',\ep)
  \,,
\end{align}
where $\vec{R}^{(n)}(\zb,\ep)$ is defined in Eq.~\eqref{eq:deq-soft-limit-n}.
Having computed the boundary constants, we find that the soft limit of the
master integrals can be written as
\begin{align}
  \vec{I}_{c,\text{soft}}(\zb,\ep)
    &= \sum_{n=0}^\infty \left[
         \vec{B}_{\text{soft},3,n}(\ep) \zb^{n-3\ep}
         +\vec{B}_{\text{soft},2,n}(\ep) \zb^{n-2\ep}
       \right]
  \,.
\end{align}
In the triple-real case we observe that $\vec{B}_{\text{soft},2,n}(\ep) =
\vec{0}$ for all $n$. Finally, we merge the solutions in the soft limit with
those expanded order by order in $\ep$ via
\begin{align}
  \vec{I}_c(\zb,\ep)
    &= \left[\sum_i \ep^k \vec{I}_c^{(k)}(\zb)
         -\vec{I}_{c,\text{soft}}(\zb,\ep)\right]_{\ep-\text{exp}}
       +\vec{I}_{c,\text{soft}}(\zb,\ep)
  \,.
\end{align}
This way, the soft singularity stays dimensionally regulated, which is
important for computing convolution integrals over $z$.

\subsection{Boundary constants}
\label{sec:calc-boundaries}

The calculation of the boundary conditions is one of the most demanding parts
of this computation and we have used different methods to derive them. We have
discussed the computation of some of these constants in
Refs.~\cite{Melnikov:2019pdm,Melnikov:2018jxb} which the interested reader
should consult. Here we address the problem from a slightly different
perspective.

We start with the discussion of the triple-real integrals which appear in the
calculation. We will show that their soft, $z \to 1$, limits can be related to
integrals which appear in the calculation of the Higgs cross section in the
threshold limit. A generic triple-real integral $I$ can be written as
\begin{align}
  I &= \int \prod_{i=1}^{3} \, [\rmd k_i] \,
       \delta\left(2 p \cdot k_{123} - \frac{t}{z}\right)
       \delta\left(\frac{2 \pb \cdot k_{123}}{s} - (1-z)\right) \,
       F(p,\pb,\{k_i\})
  \label{eq3.1}
  \,,
\end{align}
where the function $F(p,\pb,\{k_i\})$ describes the collection of propagators
that are displayed in Eqs.~\eqref{eq2.14} and \eqref{eq2.15}.

The boundary constant $C$ for the integral $I$ is defined as follows
\begin{align}
  \lim_{z \to 1} I
   &= C s^{n_1} t^{n_2-3\ep} (1-z)^{n_3 - 3\ep} \,
      \left\{1 + \mathcal{O}(1-z)\right\}
  \label{eq3.2}
  \,,
\end{align}
where $n_{1,..,3}$ are integers that depend on the exact definition of the
integral $I$. As explained in Ref.~\cite{Melnikov:2019pdm}, the constant $C$
can be computed by replacing all propagators that appear in Eq.~\eqref{eq3.1}
with their eikonal counterparts. For example,
\begin{align}
  \frac{1}{(p-k_{123})^2} &\to \frac{1}{-2 p \cdot k_{123}}
  \,, &
  \frac{1}{(p-k_{12})^2} &\to \frac{1}{-2 p \cdot k_{12}}
  \,, &
  \text{etc.}
\end{align}

Once these simplifications are performed, the integral becomes a homogeneous
function of $(1-z)$ in addition to being a homogeneous function of $s$ and
$t/z$. Hence, we can write
\begin{align}
  I_{\text{eik}}
    &= \int \prod_{i=1}^{3} \; \left[\rmd k_i\right]
       \delta\left(2 p \cdot k_{123} - \frac{t}{z}\right)
       \delta\left(\frac{2 \pb \cdot k_{123}}{s} - (1-z)\right) \,
       F(p,\pb,\{k_i\})|_{\text{eik}}
  \notag \\
    &= C \, s^{n_1} \left(\frac{t}{z}\right)^{n_2-3 \ep} (1-z)^{n_3 - 3\ep}
  \label{eq3.4}
  \,.
\end{align}
We emphasise that the constant $C$ in Eq.~\eqref{eq3.4} is the same as in
Eq.~\eqref{eq3.2} and that the dependence of $I_{\text{eik}}$ on $z$, as
displayed in the second line of Eq.~\eqref{eq3.4}, is exact.

We now explain how to make use of this fact to simplify the computation of the
integration constant. To this end, we take $t = s z^2$ in Eq.~\eqref{eq3.4}
and integrate that equation over $z$. We find
\begin{align}
  \bar{I}
     = \int_{0}^{1} \, \rmd z \, I_{\text{eik}}
    &= \int \prod_{i=1}^{3} \, \left[\rmd k_i\right]
       \delta\left(2 P \cdot k_{123} - s\right) \,
       F(p,\pb,\{k_i\})|_{\text{eik}}
  \notag \\
    &= C \, s^{n_1+n_2-3\ep} \frac{\Gamma(n_2+1-3\ep) \Gamma(n_3+1-3\ep)}{%
         \Gamma(n_2+n_3+2-6\ep)}
  \label{eq3.5}
  \,,
\end{align}
where $P = p + \pb$.

Eq.~\eqref{eq3.5} is useful because the integral there can be related to the
triple-real soft integrals calculated for the Higgs boson production in gluon
fusion in Refs.~\cite{Anastasiou:2013srw,Li:2014bfa,Zhu:2014fma,%
Duhr:2022cob}. Hence, many of the boundary constants required for the
calculation of the N$^3$LO $N$-jettiness beam functions can be compared
with the boundary constants computed in
Ref.~\cite{Anastasiou:2013srw,Li:2014bfa,Zhu:2014fma,Duhr:2022cob}. Moreover,
we can use the integration-by-parts identities for triple-real soft Higgs
integrals to relate the various boundary constants required in our case.

In fact, we have checked that our computation of the boundary constants for
triple-real integrals covers all of the ten integrals described in
Ref.~\cite{Anastasiou:2013srw} as well as the 13 inclusive RRR integrals
calculated recently in Ref.~\cite{Duhr:2022cob}. We checked that our results
agree with those references up to the $\ep$-order required for our calculation.
We note that whereas Refs.~\cite{Anastasiou:2013srw,Duhr:2022cob} heavily
relied on using the Mellin-Barnes representation to compute the triple-real
soft integrals, in Ref.~\cite{Melnikov:2019pdm} we calculated most of them by a
direct integration over Sudakov parameters. Hence, in addition to serving its
original purpose, our computation of the boundary constants provides an
independent confirmation of the triple-real soft integrals for Higgs boson
production calculated in Ref.~\cite{Anastasiou:2013srw}.

\bigskip
The absolute majority of the RRV boundary constants can be computed by a direct
integration over Feynman and Sudakov parameters following the discussion in
Ref.~\cite{Melnikov:2018jxb}. We used a convenient Feynman parameter
representation of the one-loop integrals and calculated the
relevant\footnote{We have found empirically that only the branches
$(1-z)^{m_1-2\ep}$ and $(1-z)^{m_2 - 3\ep}$ with $m_1,m_2 \in \mathbb{Z}$
appear in RRV integrals. If the homogeneous part of the differential equation
for a RRV integral allows a different branch, its coefficient can be
immediately set to zero.} $(1-z)$-branches by making suitable approximations.
Most of the required boundary conditions are discussed in
Ref.~\cite{Melnikov:2018jxb}; however, we discovered that a few additional
boundary constants are needed.

All but one of these additional constants can be computed following the
discussion in Ref.~\cite{Melnikov:2018jxb}. The RRV integral for which this
approach fails reads
\begin{align}
  I ={}& \int \prod_{i=1}^{2} \, \left[\rmd k_i\right]
         \delta\left(2 p \cdot k_{12} - \frac{t}{z}\right)
         \delta\left(\frac{2 \pb \cdot k_{12}}{s} - (1-z)\right)
         \frac{1}{(k_1-p)^2} \frac{1}{k_2 \cdot \pb}
  \notag \\ &
         \times \int \frac{\rmd^d k_3}{(2\pi)^d} \,
         \frac{1}{k_3^2 k_{13}^2 k_{123}^2 \, (k_{123}-p)^2 \,
           (k_3 \cdot \pb)}
  \label{eq3.6}
  \,,
\end{align}
where $k_{13} = k_1+k_3$ and $k_{123} = k_1+k_2+k_3$. Similar to the
triple-real case, this integral is a homogeneous function of $t/z$ and $s$.
Hence, in what follows we will set $t/z \to 1$, $s \to 1$ and $p \cdot \pb \to
1/2$ when discussing the RRV integral in Eq.~\eqref{eq3.6}.

To determine the boundary constant, we need to know the coefficient of the
$(1-z)^{-2-3\ep}$ branch of this integral, i.e.
\begin{align}
  \lim_{z \to 1} I &\approx C (1-z)^{-2-3\ep} + \dots
\end{align}
One can show that, in the $z \to 1$ limit, this branch can be calculated from a
simplified integral that is obtained by using the soft approximation for
\emph{both} real and virtual-loop momenta (the \emph{soft region} in the
terminology of the method of expansion by regions \cite{Beneke:1997zp}). This
amounts to the replacement $1/(k_{123} -p)^2 \to 1/(-2 p \cdot k_{123})$ in
Eq.~\eqref{eq3.6}. Denoting the approximate integral by $I_s$, and making use
of the fact that it is a homogeneous function of $(1-z)$ we can write
\begin{align}
  I_s &= C (1-z)^{-2-3\ep}
  \,.
\end{align}

To proceed further, we write the soft approximation for the integral shown
in Eq.~\eqref{eq3.6} in the following way
\begin{align}
  I_s &= \int \rmd^d q \,
         \delta\left(2 p \cdot q - 1\right)
         \delta\left(2 \pb \cdot q - (1-z)\right) \,
         F_s(p \cdot q, \pb \cdot q, q^2)
  \label{eq2.25}
  \,,
\end{align}
where
\begin{align}
  F_s(p \cdot q, \pb \cdot q, q^2)
    ={}& \int \prod_{i=1}^{2} \,\left[\rmd k_i\right]
         \delta^{(d)}\left(q - k_1 - k_2\right)
         \frac{1}{(-2 k_1 \cdot p) } \, \frac{1}{k_2 \cdot \pb}
  \notag \\ &
         \times \int \frac{\rmd^d k_3}{(2\pi)^d} \,
         \frac{1}{k_3^2 k_{13}^2 k_{123}^2 \, (-2 p \cdot k_{123}) \,
           (\pb \cdot k_3)}
  \label{eq2.26}
  \,.
\end{align}

To integrate over $q$ in Eq.~\eqref{eq2.25}, we introduce the Sudakov
decomposition
\begin{align}
  q_\mu &= \alpha_q \, p_\mu + \beta_q \, \pb_\mu + q_{\perp,\mu}
  \,,
\end{align}
and write
\begin{align}
   \rmd^d q
     &= \frac{1}{2} \rmd\alpha_q \, \rmd\beta_q \, \rmd^{d-2} q_\perp
      = \frac{1}{4} \, \rmd q^2 \, \rmd\alpha_q \, \rmd\beta_q \,
        \rmd\Omega^{(d-2)} \, (\alpha_q \beta_q - q^2)^{-\ep}
  \,,
\end{align}
where in the last step we traded the integration over $q_\perp^2$ for the
integration over $q^2$. Using the fact that the function $F_s$ in
Eq.~\eqref{eq2.25} is independent of the directions of the vector $q_\perp$, we
integrate over directions of the vector $q_\perp$ and the variables $\alpha_q$
and $\beta_q$, and obtain
\begin{align}
  I_s &= \frac{\Omega^{(d-2)}}{4} \,
         \int_{0}^{1-z} \rmd q^2 \, \left((1-z) - q^2\right)^{-\ep}
         \left. F_s(p \cdot q, \pb \cdot q, q^2)
         \right|_{p \cdot q=1/2, \; \pb \cdot q = (1-z)/2}
  \label{eq2.30}
  \,.
\end{align}

We now discuss how this integral can be computed. We note that, in spite of its
appearance, the function $F_s$ in Eq.~\eqref{eq2.26} is a non-trivial function
of a \emph{single variable} $x = q^2/(4 (p \cdot q) \, (\pb \cdot q))$; the
remaining dependences on the other two variables, say $p \cdot q$ and $\pb
\cdot q$, follow from dimensional analysis and are homogeneous. Hence, we can
write
\begin{align}
  F_s(p \cdot q, \pb \cdot q, q^2)
    &= (2 p \cdot q)^{\omega_1} (2 \pb \cdot q)^{\omega_2} \,
       \tilde{F}_s\left(\frac{q^2}{4 (p \cdot q) \, (\pb \cdot q)}\right)
  \,,
\end{align}
where $\omega_{1,2}$ are two, potentially $\ep$-dependent, constants.

Using this representation in Eq.~\eqref{eq2.30} and observing that the factor
$(2 \pb \cdot q)^{\omega_2}$ provides the only source of the $(1-z)$-dependence
in the function $F_s$, we find
\begin{align}
  I_s &= \frac{\Omega^{(d-2)}}{4} (1-z)^{-2-3\ep} \,
         \int_{0}^{1} \rmd x \, \left(1-x\right)^{-\ep}
         \tilde{F}_s(x)
  \label{eq2.32}
  \,.
\end{align}
Hence, to compute the boundary constant $C$ we need to determine the function
$\tilde{F}_s(x)$ and integrate it over $x$ from zero to one with the weight
shown in Eq.~\eqref{eq2.32}.

It is quite challenging to compute the function $\tilde{F}_s(x)$ by directly
integrating Eq.~\eqref{eq2.26}. A more elegant way to do this is to use the
definition of the function $\tilde{F}_s$ to construct a differential equation
that this function satisfies; we do this by using integration-by-parts
identities. Of course, closing the system of differential equations requires
the introduction of many more integrals in addition to the one in
Eq.~\eqref{eq2.26}, but these integrals are simpler. We find that 22 integrals,
including $\tilde{F}_s$, are needed to close the system of differential
equations with respect to the variable $x$.

To solve these differential equations we require boundary constants; we
determine them by considering the $x \to 1$ limit. Physically, this limit
corresponds to vanishing transverse momentum $q_\perp$. We will now discuss
a few examples of integrals that need to be calculated to determine the
boundary conditions.

The simplest integral reads
\begin{align}
  I_1 &= \int [\rmd k_1] [\rmd k_2] (2\pi)^{d-1} \delta^{(d)}(q - k_1 - k_2)
         \int \frac{\rmd^d k_3}{(2\pi)^d} \frac{1}{k_3^2 (k_3 + q)^2}
  \,.
\end{align}
In contrast to other cases considered below, this integral can be computed
exactly. We begin by integrating over the loop momentum $k_3$ and find
\begin{align}
  \int \frac{\rmd^d k_3}{(2\pi)^d} \frac{1}{k_3^2 (k_3 + q)^2}
    &= \frac{\I}{(4\pi)^{d/2}}
       \frac{\Gamma^2(1-\ep) \Gamma(\ep)}{\Gamma(2-2\ep)}
       \left(-q^2\right)^{-\ep}
  \,.
\end{align}

The remaining integrations over $k_{1,2}$ are also elementary. Performing them,
we obtain
\begin{align}
  I_1 &= \frac{\I \Gamma^2(1+\ep)}{(4 \pi)^{d}} \E^{\I \pi \ep} \,
         \frac{\Gamma^3(1-\ep)}{\ep \, \Gamma^2(2-2\ep) \Gamma(1+\ep)} \,
         x^{-2\ep} \left(4 (p \cdot q) \, (\pb \cdot q) \right)^{-2\ep}
  \,.
\end{align}

\bigskip
A slightly more complicated integral reads
\begin{align}
  I_2 &= \int [\rmd k_1] [\rmd k_2] (2\pi)^{d-1} \delta^{(d)}(q - k_1 - k_2)
         \int \frac{\rmd^d k_3}{(2\pi)^d} \,
         \frac{1}{k_{13}^2 \, (\pb \cdot k_3) \, (p \cdot k_{123})}
  \,.
\end{align}
To compute it, we calculate the loop integral and find
\begin{align}
  \int \frac{\rmd^d k_3}{(2\pi)^d} \,
    \frac{1}{k_{13}^2 \, (\pb \cdot k_3) \, (p \cdot k_{123})}
    &= -\frac{\I}{(4 \pi)^{d/2}} \frac{4 \, \E^{\I \pi \ep} \,
         \Gamma^2(1+\ep) \Gamma(1-\ep)}{\ep^2}
       \left(2 p \cdot k_2\right)^{-\ep} \left(2 k_1 \cdot \pb\right)^{-\ep}
  \,.
\end{align}

To integrate over $k_{1,2}$, we consider the rest frame of the vector $q$ and
determine the energies of the two partons with momenta $k_{1,2}$ by removing
the delta function $\delta^{(d)}(q - k_1 - k_2)$. We obtain
\begin{align}
  I_2 &= -\frac{\I \Gamma^2(1+\ep)}{(4 \pi)^{d}} \E^{\I \pi \ep} \,
         \frac{4 \Gamma^2(1-\ep) }{\ep^2 \Gamma(2 - 2\ep)}
         \left((q \cdot p) \, (q \cdot \pb)\right)^{2\ep}
         (4 x)^{-\ep} I_{\Omega}
  \label{eq2.36}
  \,,
\end{align}
where
\begin{align}
  I_{\Omega}
    &= \frac{1}{\Omega^{(d-1)}} \int \frac{\rmd \Omega^{(d-1)}_{\vec{n}}}{%
         (1 + \vec{a} \cdot \vec{n})^{\ep} (1 - \vec{b} \cdot \vec{n})^{\ep}}
  \label{eq2.37}
  \,.
\end{align}
In Eq.~\eqref{eq2.37}, $\vec{n}$, $\vec{a}$ and $\vec{b}$ are
$(d-1)$-dimensional\footnote{As indicated by the arrow.} unit vectors which
describe the directions of vectors $k_1$, $p$ and $\pb$ in the rest frame of
the vector $q$. Furthermore, when writing Eq.~\eqref{eq2.37}, we have used the
fact that in this frame $\vec{k}_1 $ and $\vec{k}_2$ are back-to-back.

It follows from Eq.~\eqref{eq2.36} that, in order to determine the boundary
condition for the integral $I_2$, we need to analyse the behaviour of $I_\Omega$
in the $x \to 1$ limit. This can be done by noticing that, since
$\vec a^2 = \vec b^2 = 1$, $I_{ \Omega}$ is a function of the scalar product
$\vec{a} \cdot \vec{b}$. To relate this scalar product to the variable $x$, we
compute the scalar product $p \cdot \pb$ in the rest frame of $q$. We find
\begin{align}
  1 - \vec{a} \cdot \vec{b}
    &= \frac{p \cdot \pb}{p^0 \pb^0}
     = \frac{(p \cdot \pb) q^2}{(p \cdot q) (\pb \cdot q)}
     = 2 x
  \,.
\end{align}
Hence, we conclude that the limit $x \to 1$ corresponds to vectors $\vec{a}$
and $\vec{b}$ being back-to-back. Therefore, computation of the boundary
conditions for the integral $I_2$ requires extracting relevant branches from
$I_{\Omega}$ in the limit $\vec{a} \to -\vec{b}$.

The relevant branches can be analysed by introducing a Feynman parameter to
combine the two denominators which appear in the integrand in
Eq.~\eqref{eq2.37}. We write
\begin{align}
  \frac{1}{(1 + \vec{a} \cdot \vec{n})^{\ep} (1 - \vec{b} \cdot \vec{n})^{\ep}}
    &= \frac{\Gamma(2\ep)}{\Gamma^2(\ep)}
       \int_{0}^{1} \rmd y \, \frac{y^{\ep-1} (1-y)^{\ep-1}}{%
         (1 - \vec{\eta} \cdot \vec{n})^{2\ep}}
  \,,
\end{align}
where $\vec{\eta} = \vec{b} (1-y) - \vec{a} y$. We use this representation in
Eq.~\eqref{eq2.37} and integrate over the directions of $\vec{n}$ choosing the
$z$-axis along the vector $\vec{\eta}$. We find
\begin{align}
  I_{\Omega}
    &= \frac{\Gamma(2\ep)}{\Gamma^2(\ep)} \int_{0}^{1} \rmd y \,
       \frac{y^{\ep-1} (1-y)^{\ep - 1}}{(1 + \eta)^{2\ep}} \, \;
       {}_{2}F_{1}\left(2\ep , 1-\ep, 2-2\ep, \frac{2 \eta}{1+\eta}\right)
  \label{eq2.40}
  \,,
\end{align}
where $\eta = |\vec{\eta}| = \sqrt{1 - 2 y (1-y)(1+\vec{a} \cdot \vec{b})} =
\sqrt{1 - 4(1-x) y (1-y)}$. Hence, $\eta = 1$ at $x = 1$.

Although Eq.~\eqref{eq2.40} has no explicit $x \to 1$ or $\eta \to 1$ branches,
implicit branches are hidden in the hypergeometric function. To expose them and
to extract the relevant $x \to 1$ branches, it is sufficient to rewrite the
hypergeometric function in Eq.~\eqref{eq2.40} as follows
\begin{align}
  {}_{2}F_{1}&\left(2\ep, 1-\ep, 2-2\ep, \frac{2 \eta}{1+\eta}\right)
    = \frac{\Gamma(1-3\ep) \Gamma(2-2\ep)}{\Gamma(2-4\ep) \Gamma(1-\ep)}
         {}_{2}F_{1}\left(2\ep, 1-\ep, 3\ep, \frac{1-\eta}{1+\eta}\right)
  \notag \\ &
         +\left(\frac{1-\eta}{1+\eta}\right)^{1-3 \ep}
         \frac{\Gamma(2-2 \ep) \Gamma(-1+3 \ep)}{\Gamma(1-\ep) \Gamma(2 \ep)}
         {}_{2}F_{1}\left(2-4 \ep, 1-\ep, 2-3 \ep, \frac{1-\eta}{1+\eta}\right)
  \label{eq2.41}
  \,.
\end{align}
Both hypergeometric functions in Eq.~\eqref{eq2.41} can be expanded in Taylor
series around $\eta = 1$ which corresponds to $x = 1$. Furthermore, the two
terms in Eq.~\eqref{eq2.41} provide two distinct branches that arise in the $x
\to 1$ limit, i.e. $\mathcal{O}((1-x)^0)$ and $\mathcal{O}((1-x)^{1-3\ep})$.

It follows from the system of differential equations, that coefficients of
\emph{both} of these branches can be used to fix some of the boundary
constants. Hence, we require the two coefficients $C_{1,2}$ defined through the
following equation
\begin{align}
  \lim_{x \to 1} I_2 &\sim C_1 + C_2 (1-x)^{1-3\ep} + \dots
  \,,
\end{align}

These two constants can be readily computed using Eqs.~\eqref{eq2.40} and
\eqref{eq2.41}. Indeed, to compute $C_1$ we can simply set $x=1$ in these
equations. Since we work at fixed $\ep$, $(1-x)^{1-3\ep} = 0$ at $x =1$ and
we obtain
\begin{align}
  I_{\Omega}|_{x=1}
    &= \frac{2^{-2\ep} \Gamma(1-3\ep) \Gamma(2-2\ep)}{%
       \Gamma(2-4\ep) \Gamma(1-\ep)}
  \label{eq2.45}
  \,,
\end{align}
from where $C_1$ is easily determined.

For $C_2$, we need to consider the second term in Eq.~\eqref{eq2.41}. To
extract the $(1-x)^{1-3\ep}$ branch, we write
\begin{align}
  \left(\frac{1-\eta}{1+\eta}\right)^{1-3\ep}
    &= \left(\frac{1-\eta^2}{(1+\eta)^2}\right)^{1-3\ep}
     \approx (1-x)^{1-3\ep} \, \left(y (1-y)\right)^{1-3\ep}
  \label{eq2.46}
  \,.
\end{align}
Hence, by taking the second term in Eq.~\eqref{eq2.41}, setting the
hypergeometric function that appears there to one, and using the
simplifications indicated in Eq.~\eqref{eq2.46}, we obtain
\begin{align}
  I_{\Omega}|_{\mathcal{O}((1-x)^{1-3\ep})}
    &= \frac{2^{-2\ep} \Gamma(2\ep)}{\Gamma^2(\ep)}
       \frac{\Gamma(2-2 \ep) \Gamma (-1+3 \ep)}{\Gamma(1-\ep) \Gamma(2 \ep)}
       (1-x)^{1-3\ep} \int_{0}^{1} \rmd y \, y^{-2\ep} \, (1-y)^{-2\ep}
  \notag \\
    &= \frac{2^{-2\ep} \Gamma(2-2 \ep) \Gamma(-1+3 \ep)}{%
         \Gamma(1-\ep) \Gamma^2(\ep)} \,
       \frac{\Gamma^2(1-2\ep)}{\Gamma(2-4\ep)} \, (1-x)^{1-3\ep}
  \label{eq2.47}
  \,,
\end{align}
The required boundary conditions are then obtained by combining
Eqs.~\eqref{eq2.45} and \eqref{eq2.47} with Eq.~\eqref{eq2.36}.

Finally, we note that a similar analysis of boundary conditions can be
performed for all integrals needed for the computation of the original integral
$\tilde{F}_s$ shown in Eq.~\eqref{eq2.26}. However, since $\tilde{F}_s$
\emph{itself} needs to be determined from the system of differential equations,
it could have required a calculation of its own boundary constant. Luckily,
this is not the case. Indeed, the analysis of the homogeneous part of the
differential equations for $\tilde{F}_s$ shows that
\begin{align}
  \tilde{F}_s &= C_s \frac{x^{-1-\ep}}{1-x} + \dots
  \label{eq2.61}
  \,,
\end{align}
where the ellipses stand for the integral of the inhomogeneous contributions to
the differential equation for $\tilde{F}_s$. The striking feature of the
homogeneous contribution is that it predicts an $\ep$-unregulated singularity
in the $x \to 1$ limit and one can use an integral representation for
$\tilde{F}_s$ to show that such a singularity does not occur. In fact, in the
$x \to 1$ limit, $\tilde{F}_s$ is described by two branches $(1-x)^{-1-3\ep}$
and $(1-x)^{-1-\ep}$ whose coefficients can be computed but are not needed to
construct the solution of the differential equation for $\tilde{F}_s$.

The result for $\tilde{F}_s(x)$ is written in terms of harmonic polylogarithms.
To compute the required boundary condition for the original integral $I$, we
need to substitute $\tilde{F}_s(x)$ into Eq.~\eqref{eq2.32} and integrate over
$x$. In principle, this procedure is straightforward but it is made complicated
by the fact that, after expansion in $\ep$, $\tilde{F}_s(x)$ develops
non-integrable singularities at $x = 0$. Hence, before the integration can be
completed, one has to re-sum singular $x \to 0$ terms in the expression for
$\tilde{F}_s(x)$. We achieve this using the $x \ne 0$ solution as well as the
differential equation which predicts the structure of branches of
$\tilde{F}_s(x)$ at small values of $x$.

After integrating over $x$, we obtain the following result for the soft limit
of the integral
\begin{align}
  I_s
    ={}& \I \left(\frac{\E^{\ep \gamma_E}}{(4 \pi)^\ep} 4 \pi^2\right)^{-3}
         (1-z)^{-2-3\ep} \Biggl[
           -\frac{25}{64 \ep^4}
           -\frac{23}{32 \ep^3}
           +\frac{\frac{11}{4}+\frac{533 \pi^2}{768}}{\ep^2}
  \notag \\ &
           +\frac{1}{\ep} \biggl(
             -\frac{85}{8}
             +\frac{139 \pi^2}{128}
             +\frac{853 \zeta_3}{64}
           \biggr)
           +\biggl(
             \frac{331}{8}
             -4 \pi^2
             +\frac{361 \zeta_3}{32}
             +\frac{997 \pi^4}{30720}
           \biggr)
  \notag \\ &
           +\ep \biggl(
             -\frac{1297}{8}
             +\frac{479 \pi^2}{32}
             -\frac{157 \zeta_3}{4}
             -\frac{7157 \pi^4}{15360}
             -\frac{15025 \pi^2 \zeta_3}{768}
             +\frac{13015 \zeta_5}{64}
           \biggr)
  \label{eq:result-Is} \\ &
           +\ep^2 \biggl(
             \frac{5107}{8}
             -\frac{1817 \pi^2}{32}
             +\frac{1115 \zeta_3}{8}
             +\frac{3401 \pi^4}{1920}
             +\frac{5739 \zeta_5}{160}
             -\frac{2069 \pi^2 \zeta_3}{128}
  \notag \\ &
             +\frac{416141 \pi^6}{1548288}
             -\frac{14289 \zeta_3^2}{128}
           \biggr)
           +\mathcal{O}\left(\ep^3\right)
         \Biggr]
  \notag
  \,.
\end{align}
According to the above discussion, this result provides the required boundary
condition for the integral $I$ defined in Eq.~\eqref{eq3.6}.

Overall, the RRV master integrals require explicit calculation for eight
boundary constants in addition to those discussed in
Ref.~\cite{Melnikov:2018jxb}. The method for relating boundary constants of RRR
integrals to integrals for Higgs production in gluon fusion at threshold,
discussed at the beginning of this section, can also be applied to the soft
region of RRV integrals. This allows us to map these integrals onto those
for Higgs production computed in Ref.~\cite{Anastasiou:2015yha}. We have not
systematically employed this connection to calculate the RRV boundary
constants, but we did use it to cross-check our results for some of the
boundary constants, including the one for $I_s$ in Eq.~\eqref{eq:result-Is}
which can be mapped on to the most complicated integral $\mathcal{M}_{13}^S$
from Ref.~\cite{Anastasiou:2015yha}. More recently, results up to weight 8 for
the soft RRV integrals have been published in Ref.~\cite{Duhr:2022cob}.

Once the boundary constants are computed and incorporated into the solutions of
the differential equations following the steps outlined in the previous
section, we obtain the final results for the master integrals that are used to
compute the bare beam function. We discuss how this is done in the next
section. Before proceeding with this discussion, we note that the results for
master integrals can be checked numerically as described in
Refs.~\cite{Melnikov:2019pdm,Melnikov:2018jxb}. We performed this check for
many master integrals to ensure their correctness.

\subsection{Assembly of the bare beam functions}
\label{sec:calc-assembly}

Having discussed the computation of the amplitudes and the integrals required
for the calculation of the beam function, we explain how the different
contributions are put together. The expansion of the fully unrenormalised, bare
partonic beam functions in the strong coupling constant reads
\begin{align}
  \BareBeamUnren{ij}(t,z)
    &= \sum_{k=0}^\infty \left(\frac{\alpha_s^{(0)}}{4 \pi}\right)^k
       \BareBeamUnren[(k)]{ij}(t,z)
\label{eq2.73}
       \,.
\end{align}
In Eq.~\eqref{eq2.73} $\alpha_s^{(0)}$ denotes the unrenormalised strong
coupling constant and the expansion coefficients through N$^3$LO read
\begin{align}
  \BareBeamUnren[(0)]{ij}(t,z)
    ={}& \delta_{ij} \delta(t) \delta(1-z)
  \,, \\
  \BareBeamUnren[(1)]{ij}(t,z)
    ={}& \int \dPhiB[1,0] \, |\mathcal{A}^{\R}_{ij}|^2
  \,, \\
  \BareBeamUnren[(2)]{ij}(t,z)
    ={}& \int \dPhiB[2,0] \sum_f |\mathcal{A}^{\RR}_{ij,f}|^2
         +\int \dPhiB[1,1] \, 2 \Re\left[\mathcal{A}^{\RV}_{ij}
            (\mathcal{A}^{\R}_{ij})^*\right]
  \,, \\
  \BareBeamUnren[(3)]{ij}(t,z)
    ={}& \int \dPhiB[3,0] \sum_f |\mathcal{A}^{\RRR}_{ij,f}|^2
         +\int \dPhiB[2,1] \sum_f 2 \Re\left[\mathcal{A}^{\RRV}_{ij,f}
           (\mathcal{A}^{\RR}_{ij,f})^*\right]
  \notag \\ &
         +\int \dPhiB[1,2] \, 2 \Re\left[\mathcal{A}^{\RVV}_{ij}
           (\mathcal{A}^{\R}_{ij})^*\right]
         +\int \dPhiB[1,2] \, |\mathcal{A}^{\RV}_{ij}|^2
  \,.
\end{align}
The sums over $f$ run over different partonic final states. For example, the
term $\sum \limits_{f}^{} |\mathcal{A}^{\RRR}_{q_i q_j,f}|^2$ includes the
processes $q_j \to q_i^* + ggg$ and $q_j \to q_i^* + g \bar{q}_k q_k$. The
squared amplitudes for the splitting functions can be generated from diagrams
for the partonic process $j \to i^* + f$ as explained in the beginning of this
section. Finally, the integration measure is defined as
\begin{align}
  \dPhiB[n_R,n_V]
    &= \prod_{i=1}^{n_R} [\rmd k_i]
       \prod_{j=1}^{n_V} \frac{\rmd^d l_j}{(2\pi)^d}
       \delta\left(2 p \cdot k_{1\dots n_R} - \frac{t}{z}\right)
       \delta\left(\frac{2 \pb \cdot k_{1\dots n_R}}{s} - (1-z)\right)
  \,,
\end{align}
where $k_{1\dots n_R} = \sum \limits_{i=1}^{n_R} k_i$, $n_{R}$ is the number of
final state partons and $n_V$ is the number of virtual loops that appear in the
particular contribution.

While we calculate the $\RRR$ and $\RRV$ contributions in the way we described
in the previous sections, we use a different approach to compute the $\RVV$
contribution. As we previously described in Ref.~\cite{Behring:2019quf}, we can
bypass a two-loop calculation in a physical gauge by making use of the fact that
the splitting functions are gauge invariant and that for the single-collinear
limit at two-loop order they have been calculated from limits of $2 \to 2$
scattering matrix elements in Refs.~\cite{Bern:2004cz,Badger:2004uk,%
Duhr:2014nda}. We use the expressions for the two-loop single-collinear
splitting functions $P_{a^* \to a_1 a_2}(z)$ from Ref.~\cite{Duhr:2014nda}
together with the soft current from Ref.~\cite{Duhr:2013msa} and cross them to
the case of initial state splitting $a_1 \to a^* a_2$ via
\begin{align}
  P_{a_1 \to a^* a_2}(z)
    &= (-1)^{2 s_a+2 s_{a_1}} \frac{n_a}{n_{a_1}} \,
       z \, P_{a^* \to a_1 a_2}\left(\frac{1}{z}\right)
  \label{eq2.75a}
  \,.
\end{align}
In Eq.~\eqref{eq2.75a} $a^*$, $a_1$ and $a_2$ are the flavours of the partons
involved in the splitting and $s_a$ and $s_{a_1}$ are the respective spin
quantum numbers. The factors $n_a$ and $n_{a_1}$ are the corresponding spin and
colour averaging factors, i.e. $n_q = 2 N_c$ and $n_g = (d-2) (N_c^2-1)$.
Finally, we integrate them over the constrained single-emission phase space
\begin{align}
  \int \dPhiB[1,2]
    2 \Re\left[\mathcal{A}^{\RVV}_{ij} (\mathcal{A}^{\R}_{ij})^*\right]
    ={}& \int [\rmd^d k] \, \delta\left(2 k \cdot p - \frac{t}{z}\right)
         \delta\left(\frac{2 k \cdot \bar{p}}{s} - (1-z)\right)
  \notag \\ &
         \times (64 \pi^2) \frac{2}{\tilde{s}_{12}}
         \left(-\frac{\tilde{s}_{12}}{\mu^2}\right)^{-2\ep}
         \frac{2 \Re[P^{(2)}_{j \to i^* f}(z)]}{z}
  \,,
\end{align}
where $\tilde{s}_{12} = (p-k)^2 = -t/z$. The phase-space integration is trivial
because of the delta functions.

For the $\RVsq$ contribution, which contains the square of one-loop amplitudes,
we generate the relevant amplitudes while keeping track of which propagators
belong to $\mathcal{A}^{\RV}_{ij}$ and which to $(\mathcal{A}^{\RV}_{ij})^*$.
It turns out that we can re-use the master integrals calculated in
Ref.~\cite{Baranowski:2020xlp} provided that we do not use symmetries of
integrals which mix propagators with their complex conjugate counterparts.

\begin{table}
  \centering
  \caption{
    We show, for individual partonic processes, the number of combinations of
    relevant diagrams, the number of scalar integrals before IBP reduction, as
    well as the number of master integrals before and after eliminating partial
    fraction (PF) relations between master integrals.}
  \begin{tabular}{llllll}
    \toprule
    Contrib. & Process & Diagram comb. & Scalar int. & MIs before PF & MIs after PF \\
    \midrule
    $\RRR$  & $q \to q^* + ggg$                   & $16 \times 16 = 256$ & $126255$ & $139$ & $91$  \\
            & $q \to q^* + gq\bar{q}$             & $10 \times 10 = 100$ & $33700$  & $241$ & $200$ \\
            & $\bar{q} \to q^* + g\bar{q}\bar{q}$ & $10 \times 10 = 100$ & $3649$   & $207$ & $175$ \\
            & $g \to q^* + \bar{q}gg$             & $16 \times 16 = 256$ & $212882$ & $329$ & $214$ \\
            & $g \to q^* + \bar{q}q\bar{q}$       & $10 \times 10 = 100$ & $25707$  & $136$ & $123$ \\
            & $q \to g^* + qgg$                   & $16 \times 16 = 256$ & $146630$ & $335$ & $222$ \\
            & $q \to g^* + qq\bar{q}$             & $10 \times 10 = 100$ & $18151$  & $73$  & $65$  \\
            & $g \to g^* + ggg$                   & $25 \times 25 = 625$ & $394415$ & $399$ & $219$ \\
            & $g \to g^* + gq\bar{q}$             & $16 \times 16 = 256$ & $49468$  & $169$ & $153$ \\
            & overall                             &                      & $404086$ & $431$ & $278$ \\
    \midrule
    $\RRV$  & $q \to q^* + gg$                    & $30 \times 3 = 90$   & $97801$  & $271$ & $115$ \\
            & $q \to q^* + q\bar{q}$              & $18 \times 2 = 36$   & $24029$  & $178$ & $119$ \\
            & $\bar{q} \to q^* + \bar{q}\bar{q}$  & $18 \times 2 = 36$   & $5056$   & $184$ & $110$ \\
            & $g \to q^* + \bar{q}g$              & $30 \times 3 = 90$   & $131856$ & $367$ & $166$ \\
            & $q \to g^* + qg$                    & $33 \times 3 = 99$   & $112301$ & $363$ & $152$ \\
            & $g \to g^* + gg$                    & $68 \times 4 = 272$  & $282894$ & $365$ & $149$ \\
            & $g \to g^* + q\bar{q}$              & $33 \times 3 = 99$   & $23796$  & $72$  & $63$  \\
            & overall                             &                      & $290843$ & $420$ & $195$ \\
    \bottomrule
  \end{tabular}
  \label{tab:statistics}
\end{table}
For the $\RRR$ and $\RRV$ contributions we generate the amplitudes for the
splitting functions using the setup described above. The number of combinations
of relevant diagrams, as well as the number of scalar integrals before IBP
reduction, and the number of master integrals before and after applying partial
fraction relations are collected in Table~\ref{tab:statistics}. Once the master
integrals are known, we first rewrite the fully unrenormalised, bare partonic
beam function in terms of canonical master integrals
\begin{align}
  \BareBeamUnren{ij}
    &= \sum_n c_n I_n
     = \sum_{n,m} c_n (T^{-1})_{nm} I_{c,m}
  \,,
\end{align}
which makes cancellations of spurious poles in $\ep$ explicit. Next, we insert
the solutions for the canonical master integrals in terms of iterated integrals
of the variable $\zb$. As we explained earlier, the $\zb \to 0$ as well as
the $t \to 0$ singularities are regulated dimensionally. To construct their
expansions in $\ep$, we have to use the distributional identities
\begin{align}
  \zb^{-1+b\ep}
    &= \frac{\delta(\zb)}{b \ep}
       +\sum_{k=0}^\infty \frac{b^k \ep^k}{k!} \mathcal{D}_k(\zb)
  \label{eq:zb-distributions}
  \,, \\
  \frac{1}{\mu^2} \left(\frac{t}{\mu^2}\right)^{-1+b\ep}
    &= \frac{\delta(t)}{b \ep}
       +\sum_{k=0}^\infty \frac{b^k \ep^k}{k!}
         \mathcal{L}_k\left(\frac{t}{\mu^2}\right)
  \label{eq:t-distributions}
  \,,
\end{align}
where we use the notation
\begin{align}
  \mathcal{D}_k(\zb)
    &= \left[\frac{\log^k(\zb)}{\zb}\right]_+
  \,, &
  \mathcal{L}_k\left(\frac{t}{\mu^2}\right)
    &= \frac{1}{\mu^2} \left[\frac{\log^k(t/\mu^2)}{t/\mu^2}\right]_+
  \label{eq:plus-distributions}
  \,.
\end{align}
In Eq.~\eqref{eq:plus-distributions} $[\dots]_+$ denotes standard
plus-distributions which regulate limits when their arguments become
infinite.\footnote{The exact definitions are
\begin{align*}
  \int \limits_{0}^{1} \rmd \zb \, \mathcal{D}_k(\zb) f(\zb)
    &= \int \limits_{0}^{1} \frac{\rmd \zb}{\zb} \log^k \zb \,
       \left(f(\zb) - f(0)\right)
  \,, &
  \int \limits_{0}^{\infty} \rmd t \,
    \mathcal{L}_k\left(\frac{t}{\mu^2}\right) f(t)
    &= \int \limits_{0}^{\infty} \frac{\rmd t}{t} \,
       \log^k\left(\frac{t}{\mu^2}\right) \left(f(t) - f(0)\right)
  \,.
\end{align*}}

The highest possible pole of the amplitude is $\ep^{-6}$ so that, in principle,
the canonical master integrals have to be known up to weight six, i.e.
including $\vec{I}_c^{(6)}(z)$. However, if we express the amplitude in terms
of canonical master integrals, the highest pole that appears there is
$\ep^{-4}$. The additional poles can only arise from the terms proportional to
delta functions in Eqs.~\eqref{eq:zb-distributions} and
\eqref{eq:t-distributions}. Therefore, the weight six terms
$\vec{I}_c^{(6)}(z)$ of the canonical master integrals are only needed for $\zb
= 0$, i.e. for terms proportional to $\delta(\zb)$. As such, they can be
obtained from the soft limit of the amplitudes and master integrals.

\section{Matching coefficients}
\label{sec:matching-coeff}

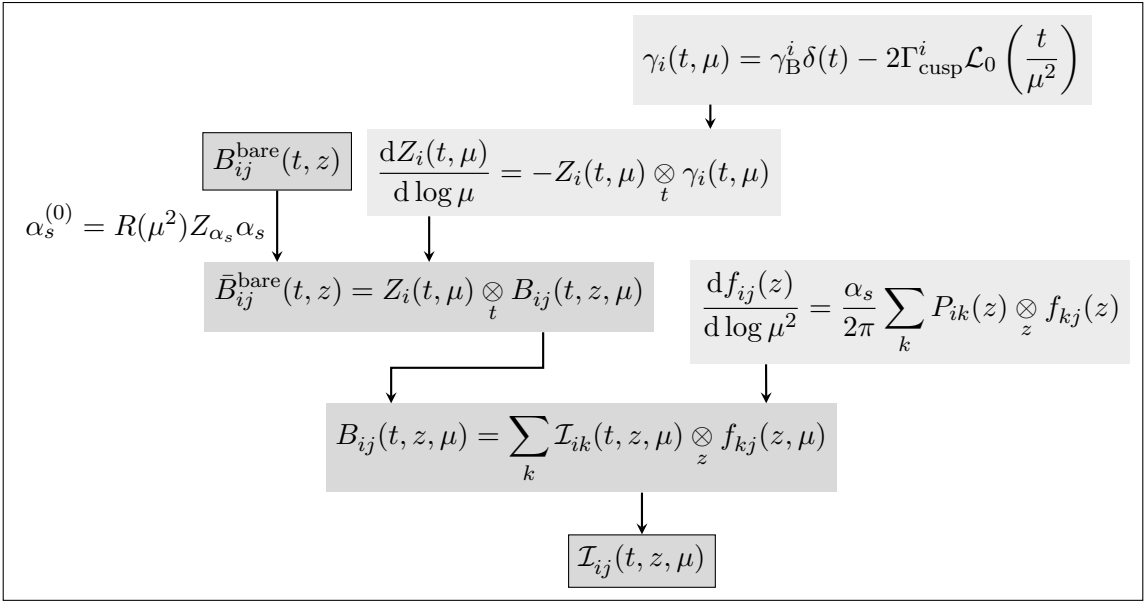
\begin{figure}
  \centering
  \begin{tikzpicture}[%
      arr/.style = {->,>=stealth,thick},
      box/.style = {fill=gray!30},
      box2/.style = {fill=gray!15},
      framed
    ]
    \node[box,draw=black] (amp) at (0,0)
               {$\displaystyle \BareBeamUnren{ij}(t,z)$};
    \node[box] (Bijbare to Bij) at ($(amp)+(2,-1.8)$)
               {$\displaystyle \BareBeam{ij}(t,z)
               = Z_i(t,\mu) \underset{t}{\otimes} \Beam{ij}(t,z,\mu)$};
    \node[box] (Bij to Iij) at ($(Bijbare to Bij)+(2,-2)$)
               {$\displaystyle \Beam{ij}(t,z,\mu)
               = \sum_{k} \MC{ik}(t,z,\mu) \underset{z}{\otimes}
                 \PDFs{kj}(z,\mu)$};
    \node[box,draw=black] (Iij) at ($(Bij to Iij)+(0.8,-1.5)$)
               {$\displaystyle \MC{ij}(t,z,\mu)$};

    \coordinate (amp out) at (amp.south);
    \coordinate (amp in) at ($(Bijbare to Bij.north)+(-2,0)$);
    \draw[arr] (amp out) -- (amp in) node[pos=0.45,left]
               {$\alpha_s^{(0)} = R(\mu^2) Z_{\alpha_s} \alpha_s$};

    \coordinate (Bij out) at ($(Bijbare to Bij.south)+(1.5,0)$);
    \coordinate (Bij in) at ($(Bij to Iij.north)+(-2.5,0)$);
    \draw[arr] (Bij out) |-
               ($(Bij out)!0.5!(Bij in)$) -|
               (Bij in);

    \coordinate (Iij out) at ($(Bij to Iij.south)+(0.8,0)$);
    \coordinate (Iij in) at (Iij.north);
    \draw[arr] (Iij out) -- (Iij in);

    \node[box2] (Zdef) at ($(Bijbare to Bij)+(1.9,1.65)$)
               {$\displaystyle \frac{\rmd Z_i(t,\mu)}{\rmd \log\mu}
                = -Z_i(t,\mu) \underset{t}{\otimes} \gamma_i(t,\mu)$};
    \node[box2] (gammadef) at ($(Zdef)+(3.8,1.5)$)
               {$\displaystyle \gamma_i(t,\mu)
                = \gamma_{\text{B}}^i \delta(t) - 2 \Gamma_{\text{cusp}}^i
                  \mathcal{L}_0\left(\frac{t}{\mu^2}\right)$};
    \node[box2] (pdfdef) at ($(Bij to Iij)+(4.333,1.8)$)
               {$\displaystyle \frac{\rmd \PDFs{ij}(z)}{\rmd \log\mu^2}
                = \frac{\alpha_s}{2\pi} \sum_{k} P_{ik}(z)
                  \underset{z}{\otimes} \PDFs{kj}(z)$};

    \coordinate (Z out) at ($(Zdef.south)+(-1.9,0)$);
    \coordinate (Z in) at (Bijbare to Bij.north);
    \draw[arr] (Z out) -- (Z in);

    \coordinate (gamma out) at ($(gammadef.south)+(-2,0)$);
    \coordinate (gamma in) at ($(Zdef.north)+(1.8,0)$);
    \draw[arr] (gamma out) -- (gamma in);

    \coordinate (pdf out) at ($(pdfdef.south)+(-1.9,0)$);
    \coordinate (pdf in) at ($(Bij to Iij.north)+(2.435,0)$);
    \draw[arr] (pdf out) -- (pdf in);
  \end{tikzpicture}
  \caption{Steps required for the extraction of the matching coefficients
           $\MC{ij}(t,z,\mu)$ from the fully unrenormalised, bare partonic beam
           functions $\BareBeamUnren{ij}(t,z,\mu)$, whose calculation is
           described in the previous section. For a detailed description of the
           steps see the main text.}
  \label{fig:renormalisation}
\end{figure}

Having computed the fully unrenormalised, bare partonic beam functions, we
discuss the extraction of the matching coefficients. As explained in
Ref.~\cite{Stewart:2009yx}, this is done by absorbing certain collinear
$1/\ep$-poles of the partonic beam functions into the corresponding parton
distribution functions and by removing the remaining $1/\ep$-poles through an
appropriate renormalisation. A detailed discussion of this procedure was
provided in our earlier paper~\cite{Behring:2019quf}, but we repeat it here for
completeness. A schematic overview of the required steps is shown in
Figure~\ref{fig:renormalisation}.

As the first step, we replace the bare QCD coupling constant with the
renormalised one. We use
\begin{align}
  \alpha_s^{(0)} = R(\mu^2) Z_{\alpha_s} \alpha_s(\mu^2)
  \,,
\end{align}
where $R(\mu^2) = \left(\mu^2 \mathrm{e}^{\gamma_E}/(4 \pi)\right)^\ep$
and
\begin{align}
  Z_{\alpha_s}
    &= 1 - \frac{\alpha_s}{4\pi} \frac{\beta_0}{\ep}
       +\left(\frac{\alpha_s}{4\pi}\right)^2
         \left(\frac{\beta_0^2}{\ep^2} - \frac{\beta_1}{2\ep}\right)
       +\mathcal{O}(\alpha_s^3)
  \,.
\end{align}
Once this is done, we obtain the bare partonic beam functions $\BareBeam{ij}$
from the original \emph{fully unrenormalised} bare partonic beam functions
$\BareBeamUnren{ij}$.

Expanding $\BareBeam{ij}$ in the renormalised strong coupling yields
\begin{align}
  \BareBeam{ij}
    &= \sum_{k=0}^\infty \left(\frac{\alpha_s}{4 \pi}\right)^k
       \BareBeam[(k)]{ij}
  \,,
\end{align}
where the expansion coefficients read
\begin{align}
  \BareBeam[(0)]{ij}
    ={}& \BareBeamUnren[(0)]{ij}
  \,, \\
  \BareBeam[(1)]{ij}
    ={}& \left[R(\mu^2) \BareBeamUnren[(1)]{ij}\right]
  \,, \\
  \BareBeam[(2)]{ij}
    ={}& \left[R(\mu^2)^2 \BareBeamUnren[(2)]{ij}\right]
         -\frac{\beta_0}{\ep} \left[R(\mu^2) \BareBeamUnren[(1)]{ij}\right]
  \,, \\
  \BareBeam[(3)]{ij}
    ={}& \left[R(\mu^2)^3 \BareBeamUnren[(3)]{ij}\right]
         -\frac{2 \beta_0}{\ep} \left[R(\mu^2)^2 \BareBeamUnren[(2)]{ij}\right]
  \notag \\ &
         +\left(\frac{\beta_0^2}{\ep^2} - \frac{\beta_1}{2\ep}\right)
           \left[R(\mu^2) \BareBeamUnren[(1)]{ij}\right]
  \,.
\end{align}

The bare partonic beam function is related to the partonic beam function by a
renormalisation \cite{Stewart:2009yx}
\begin{align}
  \BareBeam{ij}(t,z)
    &= Z_i(t,\mu) \underset{t}{\otimes} \Beam{ij}(t,z,\mu)
  \label{eq:beam-renormalisation}
  \,.
\end{align}
In the above equation the convolution with respect to $t$ is defined through
the following equation
\begin{align}
  f_1(t) \underset{t}{\otimes} f_2(t)
    &= \int\limits_0^\infty \rmd t_1 \, \rmd t_2 \, f(t_1) f(t_2) \,
       \delta(t-t_1-t_2)
  \,.
\end{align}

We write expansions for the partonic beam function and the
renormalisation constant in $\alpha_s$
\begin{align}
  \Beam{ij}(t,z,\mu)
    &= \sum_{k=0}^\infty \left(\frac{\alpha_s}{4 \pi}\right)^k
       \Beam[(k)]{ij}(t,z,\mu)
  \,, \\
  Z_i(t,\mu)
    &= \sum_{k=0}^{\infty} \left(\frac{\alpha_s}{4 \pi}\right)^n
       Z_i^{(k)}(t,\mu)
  \,,
\end{align}
make use of the fact that the leading-order coefficients are given by
\begin{align}
  \BareBeam[(0)]{ij}(t,z)
     &= \Beam[(0)]{ij}(t,z,\mu)
      = \delta_{ij} \delta(t) \delta(1-z)
  \,, &
  Z_i^{(0)}(t,\mu) &= \delta(t)
  \,,
\end{align}
insert those expansions into Eq.~\eqref{eq:beam-renormalisation} and obtain
the expansion coefficients of the beam function in terms of those of
the bare beam function,
\begin{align}
  \Beam[(1)]{ij}(t,z,\mu)
    ={}& \BareBeam[(1)]{ij}(t,z)
         -\delta_{ij} \delta(1-z) Z_i^{(1)}(t,\mu)
  \label{eq:BBtoB-nlo}
  \,, \\
  \Beam[(2)]{ij}(t,z,\mu)
    ={}& \BareBeam[(2)]{ij}(t,z)
         -\delta_{ij} \delta(1-z) Z_i^{(2)}(t,\mu)
         -\Beam[(1)]{ij}(t,z,\mu) \underset{t}{\otimes} Z_i^{(1)}(t,\mu)
  \label{eq:BBtoB-nnlo}
  \,, \\
  \Beam[(3)]{ij}(t,z,\mu)
    ={}& \BareBeam[(3)]{ij}(t,z)
         -\delta_{ij} \delta(1-z) Z_i^{(3)}(t,\mu)
         -\Beam[(2)]{ij}(t,z,\mu) \underset{t}{\otimes} Z_i^{(1)}(t,\mu)
  \notag \\ &
         -\Beam[(1)]{ij}(t,z,\mu) \underset{t}{\otimes} Z_i^{(2)}(t,\mu)
  \label{eq:BBtoB-n3lo}
  \,.
\end{align}

To compute the matching coefficients $\MC{ij}$, we use the matching relation
for the partonic beam function
\begin{align}
  \Beam{ij}(t,z,\mu)
    &= \sum_{k \in \{g,u,\bar{u},d,\bar{d},\dots\}}
       \MC{ik}(t,z,\mu) \underset{z}{\otimes} \PDFs{kj}(z,\mu)
  \label{eq:partonic-matching-relation-arg}
  \,.
\end{align}
In Eq.~\eqref{eq:partonic-matching-relation-arg}, $\underset{z}{\otimes}$
stands for the Mellin convolution with respect to the variable $z$. It is
defined as follows
\begin{align}
  f_1(z) \underset{z}{\otimes} f_2(z)
    &= \int \limits_{0}^{1} \rmd z_1 \rmd z_2 \,
       f_1(z_1) f_2(z_2) \delta(z - z_1 z_2)
  \,.
\end{align}
We again expand all quantities in the strong coupling
\begin{align}
  \MC{ij}(t,z,\mu)
    &= \sum_{n} \left(\frac{\alpha_s}{4\pi}\right)^n
       \MC[(n)]{ij}(t,z,\mu)
  \,, &
  \PDFs{ij}(z,\mu)
    &= \sum_{n} \left(\frac{\alpha_s}{2\pi}\right)^n \PDFs[(n)]{ij}(z)
  \,,
\end{align}
use the leading-order coefficients
\begin{align}
  \MC[(0)]{ij}
    &= \delta_{ij} \delta(t) \delta(1-z)
  \,, &
  \PDFs[(0)]{ij} &= \delta_{ij} \delta(1-z)
  \label{eq:leading-order-If}
  \,,
\end{align}
insert these expansions in Eq.~\eqref{eq:partonic-matching-relation-arg}, and
obtain
\begin{align}
  \MC[(1)]{ij}(t,z,\mu)
    ={}& \Beam[(1)]{ij}(t,z,\mu) - 2 \delta(t) \PDFs[(1)]{ij}(z)
  \label{eq:matching-coeff-nlo}
  \,, \\
  \MC[(2)]{ij}(t,z,\mu)
    ={}& \Beam[(2)]{ij}(t,z,\mu) - 4 \delta(t) \PDFs[(2)]{ij}(z)
         -2 \sum_{k} \MC[(1)]{ik}(t,z,\mu)
           \underset{z}{\otimes} \PDFs[(1)]{kj}(z)
  \label{eq:matching-coeff-nnlo}
  \,, \\
  \MC[(3)]{ij}(t,z,\mu)
    ={}& \Beam[(3)]{ij}(t,z,\mu) - 8 \delta(t) \PDFs[(3)]{ij}(z)
         -4 \sum_{k} \MC[(1)]{ik}(t,z,\mu)
           \underset{z}{\otimes} \PDFs[(2)]{kj}(z)
  \notag \\ &
         -2 \sum_{k} \MC[(2)]{ik}(t,z,\mu)
           \underset{z}{\otimes} \PDFs[(1)]{kj}(z)
  \label{eq:matching-coeff-n3lo}
  \,.
\end{align}

Finally, we combine Eqs.~(\ref{eq:BBtoB-nlo},\ref{eq:BBtoB-nnlo},%
\ref{eq:BBtoB-n3lo}) with Eqs.~(\ref{eq:matching-coeff-nlo},%
\ref{eq:matching-coeff-nnlo},\ref{eq:matching-coeff-n3lo}), and obtain the
expressions for the matching coefficients in terms of the bare beam function
\begin{align}
  \MC[(1)]{ij}(t,z,\mu)
    ={}& \BareBeam[(1)]{ij}(t,z)
         -\delta_{ij}\delta(1-z) \, Z^{(1)}_{i}(t,\mu)
         -2 \, \delta(t) \, \PDFs[(1)]{ij}(z)
  \label{eq:matching-coeff-nlo-barebeam}
  \,, \\
  \MC[(2)]{ij}(t,z,\mu)
    ={}& \BareBeam[(2)]{ij}(t,z)
         -\delta_{ij}\delta(1-z) \, Z^{(2)}_{i}(t,\mu)
         -2 Z^{(1)}_{i}(t,\mu) \, \PDFs[(1)]{ij}(z)
  \notag \\ &
         -4 \, \delta(t) \, \PDFs[(2)]{ij}(z)
         -Z^{(1)}_{i}(t,\mu)
           \underset{t}{\otimes} \MC[(1)]{ij}(t,z,\mu)
  \notag \\ &
         -2 \sum_k \MC[(1)]{ik}(t,z,\mu)
           \underset{z}{\otimes} \PDFs[(1)]{kj}(z)
  \label{eq:matching-coeff-nnlo-barebeam}
  \,, \\
  \MC[(3)]{ij}(t,z,\mu)
    ={}& \BareBeam[(3)]{ij}(t,z)
         -\delta_{ij}\delta(1-z) \, Z^{(3)}_{i}(t,\mu)
         -2 Z^{(2)}_{i}(t,\mu) \, \PDFs[(1)]{ij}(z)
  \notag \\ &
         -4 Z^{(1)}_{i}(t,\mu) \, \PDFs[(2)]{ij}(z)
         -Z^{(2)}_{i}(t,\mu)
           \underset{t}{\otimes} \MC[(1)]{ij}(t,z,\mu)
         -8 \, \delta(t) \, \PDFs[(3)]{ij}(z)
  \notag \\ &
         -2 \sum_k Z^{(1)}_{i}(t,\mu)
           \underset{t}{\otimes} \MC[(1)]{ik}(t,z,\mu)
           \underset{z}{\otimes} \PDFs[(1)]{kj}(z)
  \notag \\ &
         -Z^{(1)}_{i}(t,\mu)
           \underset{t}{\otimes} \MC[(2)]{ij}(t,z,\mu)
         -4 \sum_k \MC[(1)]{ik}(t,z,\mu)
           \underset{z}{\otimes} \PDFs[(2)]{kj}(z)
  \notag \\ &
         -2 \sum_k \MC[(2)]{ik}(t,z,\mu)
           \underset{z}{\otimes} \PDFs[(1)]{kj}(z).
  \label{eq:matching-coeff-n3lo-barebeam}
\end{align}

As Eqs.~(\ref{eq:matching-coeff-nlo-barebeam},%
\ref{eq:matching-coeff-nnlo-barebeam},\ref{eq:matching-coeff-n3lo-barebeam})
demonstrate, we have to compute convolutions of the renormalisation constants
and divergent partonic distribution functions with lower-order matching
coefficients. For this reason, both NLO and NNLO partonic beam functions are
required to higher orders in the expansion in the dimensional regularisation
parameter $\ep$. The corresponding computations for all relevant partonic beam
functions were performed in Ref.~\cite{Baranowski:2020xlp} and we use the
results of that reference in the current calculation.

It remains to explain how to compute the expansion coefficients of the
renormalisation constants and of partonic distribution functions. We begin
with the renormalisation constants $Z_i$. It is known \cite{Stewart:2009yx}
that these constants satisfy the renormalisation group equation
\begin{align}
  \mu \frac{\rmd Z_i}{\rmd \mu}
    &= - Z_i(t,\mu) \underset{t}{\otimes} \gamma_i(t,\mu)
  \label{eq:adim-to-Zi}
  \,,
\end{align}
where the anomalous dimensions read
\begin{align}
  \gamma_i(t,\mu)
    &= \gamma_{\text{B}}^{i}\delta(t)
       -2 \Gamma_{\text{cusp}}^{i} \,
         \mathcal{L}_0 \left(\frac{t}{\mu^2}\right)
  \label{eq:adim-structure}
  \,.
\end{align}
The anomalous dimensions $\Gamma_{\text{cusp}}$ and $\gamma_{\text{B}}^i$ are
known through $\mathcal{O}(\alpha_s^4)$~\cite{Korchemsky:1987wg,Moch:2004pa,%
Vogt:2004mw,Grozin:2014hna,Grozin:2015kna,Henn:2016men,Henn:2016wlm,%
Davies:2016jie,Lee:2017mip,Moch:2017uml,Grozin:2018vdn,Moch:2018wjh,%
Lee:2019zop,Henn:2019rmi,vonManteuffel:2019wbj,Bruser:2019auj,Henn:2019swt,%
vonManteuffel:2020vjv,Stewart:2010qs,Duhr:2022cob}. In
Eq.~\eqref{eq:adim-structure}, $\mathcal{L}_0(t/\mu^2)$ is a plus-distribution
defined in Eq.~\eqref{eq:plus-distributions}. We integrate
Eq.~\eqref{eq:adim-structure} to determine the renormalisation constants
$Z_{q,g}$. The analytic expressions for these constants through
$\mathcal{O}(\alpha_s^3)$ are provided in Appendix~\ref{sec:reno} and in an
ancillary file.

The perturbative expansion of the partonic distribution functions is obtained
by solving the Altarelli-Parisi equation
\begin{align}
  \mu^2 \frac{\rmd \PDFs{ij}(z,\mu^2) }{\rmd \mu^2}
    &= \frac{\alpha_s(\mu^2)}{2 \pi} \sum_{k} P_{ik}(z)
         \underset{z}{\otimes} \PDFs{kj}(z,\mu^2)
  \label{eq:DGLAP}
  \,,
\end{align}
order by order in $\alpha_s$ and using the boundary conditions in
Eq.~\eqref{eq:leading-order-If}. Since the parton distribution functions in
Eq.~\eqref{eq:partonic-matching-relation-arg} are defined in the $\MSbar$
scheme, the perturbative parton distribution functions should only contain
poles in $\ep$. Explicit results for the functions $\PDFs[(n)]{kj}(z)$, with
$n=1,2,3$, in terms of Altarelli-Parisi splitting functions are given in
Ref.~\cite{Behring:2019quf}; we present them in Appendix~\ref{sec:reno} for
completeness.

We note in passing that the cancellation of infra-red poles that occurs once
the renormalisation procedure described in this section is performed, provides
a direct cross-check on the three-loop Altarelli-Parisi splitting functions
computed in Refs.~\cite{Moch:2004pa,Vogt:2004mw,Ablinger:2014lka,%
Ablinger:2014vwa,Ablinger:2014nga,Ablinger:2017tan,Blumlein:2021enk}.

\section{Results}
\label{sec:res}
We used the procedure described in Sections~\ref{sec:calc} and
\ref{sec:matching-coeff} to compute the N$^3$LO matching coefficients
$\MC[(3)]{ij}(t,z,\mu)$ for all partonic combinations $(ij) \in \{q_i q_j, q g,
g q, g g, q_i \bar{q}_j\}$.

We observe a significant simplification of the alphabet of the iterated
integrals that appear in the final result for the matching coefficients
compared to the alphabet for the individual master integrals given in
Eq.~\eqref{eq:letters}. Already when inserting the master integrals into the
$\RRR$ and $\RRV$ squared amplitudes some letters cancel completely. In
particular, the square root $\sqrt{z}\sqrt{4+z} = \sqrt{1-\zb}\sqrt{5-\zb}$
only appears in integrals entering the $\RRR$ contribution and cancels at the
level of squared amplitudes. Moreover, upon combining the $\RRR$ and $\RRV$
contributions, the two letters with the square root $\sqrt{4+z^2} =
\sqrt{\zb^2-2\zb+5}$ drop out. The iterated integrals in the final results for
the matching coefficients depend only on the following letters
\begin{align}
  f_i(\zb) &\in \left\{
    \frac{1}{\zb-1},
    \frac{1}{\zb},
    \frac{1}{\zb+1},
    \frac{1}{\zb-2},
    \frac{1}{\sqrt{1-\zb}\sqrt{3+\zb}}
  \right\}
  \,.
\end{align}
The remaining square root corresponds to $\sqrt{z}\sqrt{4-z}$ when written in
terms of the variable $\zb$. We note that in total only twelve different
iterated integrals that depend on square roots appear in the results and that
they only occur in four distinct combinations. We illustrate this below.

It is instructive to discuss the general structure of the matching coefficients.
For all partonic channels the $n$-th order matching coefficient can be written
as
\begin{align}
  \MC[(n)]{ij}
    &= \sum_{k=0}^{2n-1} \mathcal{L}_k\left(\frac{t}{\mu^2}\right)
         F_{ij,+}^{(n,k)}(z)
       +\delta(t) F_{ij,\delta}^{(n)}(z)
  \,.
\end{align}
The function $F_{ij,\delta}^{(n)}$ can be further decomposed as
\begin{align}
  F_{ij,\delta}^{(n)}
    &= C^{(n)}_{ij,-1} \delta(1-z)
       +\sum_{k=0}^{2n-1} C_{ij,k}^{(n)} \mathcal{D}_k(z)
       +F_{ij,\delta,h}^{(n)}(z)
  \,.
\end{align}

The above decomposition is important because it isolates all the contributions
to N$^3$LO matching coefficients which can be predicted without explicit
computation. Indeed, the functions $F_{ij,+}^{(n,k)}(z)$ can be obtained from
the renormalisation group equation for the beam function \cite{Stewart:2009yx}.
Similarly, the coefficients $C_{ij,k}^{(n)}$ for $k=-1,\dots,5$, can be derived
by analysing soft contributions to the beam function \cite{Billis:2019vxg}. The
genuinely new result that cannot be obtained without a dedicated calculation is
the hard contribution $F_{ij,\delta,h}^{(n)}(z)$. Clearly, the above
predictions for certain contributions to the matching coefficients are
important since they allow us to check our calculation in a non-trivial way.
Finally, we note that our results for the matching coefficients agree with the
results published in Ref.~\cite{Ebert:2020unb}.

Unfortunately, the results for the matching coefficients are quite sizeable.
Because of that, we refrain from displaying them in the paper and, instead,
provide them in electronically readable form in an ancillary file. However, to
illustrate their structure, we will describe a particular contribution to the
matching coefficient $\MC{q_i q_j}$.

In addition to being a function of the energy fraction $z$, the hard
contribution $F_{q_i q_j,\delta,h}^{(3)}$ is also a function of the number of
colours $N_c$ and the number of fermion species $n_f$. Explicitly, this
dependence reads\footnote{For better readability, we suppress flavour- and
order-related indices and the argument $z$ on the right-hand side in
Eq.~\eqref{eq:mc-structure}.}
\begin{align}
  F_{q_i q_j,\delta,h}^{(3)}(z)
    ={}& \delta_{ij} \Biggl[
           n_f^2 \Biggl(
             N_c F_{n_f^2 N_c}
             +\frac{F_{n_f^2 N_c^{-1}}}{N_c}
           \Biggr)
           +n_f \Biggl(
             N_c^2 F_{n_f N_c^2}
             +F_{n_f}
             +\frac{F_{n_f N_c^{-2}}}{N_c^2}
           \Biggr)
  \notag \\ &
           +N_c^3 F_{N_c^3}
           +N_c F_{N_c}
           +\frac{F_{N_c^{-1}}}{N_c}
           +\frac{F_{N_c^{-3}}}{N_c^3}
         \Biggr]
         +n_f \Biggl(
           N_c F_{n_f N_c}
           +\frac{F_{n_f N_c^{-1}}}{N_c}
         \Biggr)
  \notag \\ &
         +N_c^2 F_{N_c^2}
         +F_{1}
         +\frac{F_{N_c^{-2}}}{N_c^2}
  \label{eq:mc-structure}
  \,.
\end{align}
In writing this decomposition we have used $T_F = \frac{1}{2}$. The
coefficients $F_{N_c^3}$, $F_{n_f N_c^2}$ and $F_{n_f^2 N_c}$ were published in
Ref.~\cite{Behring:2019quf}. To illustrate the structure of the new results, we
show the coefficient $F_{N_c^{-3}}$, which only contributes for equal flavours,
$i=j$, below.

The function $F_{N_c^{-3}}$ features iterated integrals with
square-root-valued letters as well as square roots in the coefficients of the
iterated integrals. We use the notation
\begin{align}
  L_{a,\vec{b}}(\zb)
    &= \int\limits_0^{\zb} \rmd t \, f_a(t) L_{\vec{b}}(t)
  \,, &
  L_{a}(\zb)
    &= \int\limits_0^{\zb} \rmd t \, f_a(t)
  \,, &
  L_{\underbrace{0,\dots,0}_{n}}(\zb)
    &= \frac{1}{n!} \log^n(\zb)
\end{align}
for the iterated integrals and the letters
\begin{align}
  f_{-1}(\zb) &= \frac{1}{\zb+1} \,, &
  f_{0}(\zb)  &= \frac{1}{\zb}   \,, &
  f_{1}(\zb)  &= \frac{1}{\zb-1} \,, \notag \\
  f_{2}(\zb)  &= \frac{1}{\zb-2} \,, &
  f_{r}(\zb)  &= \frac{1}{\sqrt{1-\zb}\sqrt{3+\zb}}
  \,.
\end{align}
Following the standard practice, we suppress the argument $\zb$ of the iterated
integrals for brevity. We note that the iterated integrals that depend on
square roots only occur in four distinct combinations throughout all matching
coefficients. They are given by
\begin{align}
  R_1 &= L_{r,0,1} + \frac{2}{3} L_{r,1,1} - \frac{\pi^2}{6} L_r
  \,, \\
  R_2 &= L_{r,r,0,1} + \frac{2}{3} L_{r,r,1,1} - \frac{\pi^2}{6} L_{r,r}
  \,, \\
  R_3 &= L_{0,r,r,0,1} + \frac{2}{3} L_{0,r,r,1,1} - \frac{\pi^2}{6} L_{0,r,r}
  \,, \\
  R_4 &= L_{1,r,r,0,1} + \frac{2}{3} L_{1,r,r,1,1} - \frac{\pi^2}{6} L_{1,r,r}
  \,.
\end{align}
With these abbreviations the coefficient $F_{N_c^{-3}}$ reads
\begin{align}
  F_{N_c^{-3}}
    ={}&
        \frac{1}{12} \big(-2116 \zb+15\big)
        +\frac{1}{12} \big(1344 \zb+31\big) L_0
        +\frac{1}{48} \big(-947 \zb-2390\big) L_1
  \notag \\ &
        +\frac{1}{3} \big(5 \zb-12\big) L_{0,0}
        +\frac{1}{6} \big(143 \zb-68\big) L_{0,1}
        +\frac{1}{6} \big(-253 \zb+569\big) L_{1,0}
  \notag \\ &
        +\frac{1}{24} \big(16 \zb^2-1147 \zb+482\big) L_{1,1}
        +3 \big(9 \zb+1\big) L_{0,0,0}
        +\frac{1}{6} \big(-113 \zb+34\big) L_{0,0,1}
  \notag \\ &
        +\frac{1}{2} \big(-61 \zb-26\big) L_{0,1,0}
        +\frac{1}{6} \big(95 \zb-6\big) L_{0,1,1}
        -\frac{19}{6} \big(19 \zb-14\big) L_{1,0,0}
  \notag \\ &
        +\frac{1}{2} \big(60 \zb-31\big) L_{1,0,1}
        +\frac{1}{12} \big(1307 \zb-936\big) L_{1,1,0}
        +\frac{1}{6} \big(118 \zb-197\big) L_{1,1,1}
  \notag \\ &
        -\frac{2}{3} \big(24 \zb-17\big) L_{1,2,1}
        +12 \zb L_{0,0,0,0}
        +\frac{1}{3} \big(-41 \zb+2\big) L_{0,0,0,1}
  \notag \\ &
        +\frac{1}{3} \big(-57 \zb+26\big) L_{0,0,1,0}
        +\frac{4}{3} \big(10 \zb+3\big) L_{0,0,1,1}
        -\frac{2}{3} \big(11 \zb+2\big) L_{0,1,0,0}
  \notag \\ &
        +\frac{1}{3} \big(47 \zb-2\big) L_{0,1,0,1}
        +\frac{2}{3} \big(3 \zb-10\big) L_{0,1,1,0}
        -\frac{4}{3} \big(41 \zb-6\big) L_{0,1,1,1}
  \notag \\ &
        +\frac{4}{3} \big(7 \zb-2\big) L_{0,1,2,1}
        -3 \big(13 \zb-14\big) L_{1,0,0,0}
        +\frac{1}{6} \big(143 \zb-150\big) L_{1,0,0,1}
  \notag \\ &
        +\frac{1}{2} \big(99 \zb-130\big) L_{1,0,1,0}
        +\frac{1}{2} \big(-83 \zb+108\big) L_{1,0,1,1}
        +\frac{1}{6} \big(409 \zb-442\big) L_{1,1,0,0}
  \notag \\ &
        +\frac{1}{3} \big(-101 \zb+113\big) L_{1,1,0,1}
        +\frac{1}{6} \big(-463 \zb+568\big) L_{1,1,1,0}
        -\frac{1}{2} \big(\zb+9\big) L_{1,1,1,1}
  \notag \\ &
        +4 \big(13 \zb-14\big) L_{1,1,2,1}
        -4 \big(5 \zb-4\big) L_{1,2,1,0}
        +\frac{4}{3} \big(42 \zb-37\big) L_{1,2,1,1}
  \notag \\ &
        -\frac{4}{3} \big(53 \zb-54\big) L_{1,2,2,1}
        +\big(1+\zb\big) \Biggl(
                \frac{1}{3} \big(2 \zb-11\big) L_{2,1}
                -\frac{8}{3} L_{2,1,0}
                +\frac{94}{3} L_{2,1,1}
  \notag \\ &
                -\frac{44}{3} L_{2,2,1}
                +16 L_{2,1,0,1}
                +\frac{28}{3} L_{2,1,1,0}
                -24 L_{2,1,1,1}
                +40 L_{2,1,2,1}
                -\frac{56}{3} L_{2,2,1,0}
  \notag \\ &
                +32 L_{2,2,1,1}
                -\frac{152}{3} L_{2,2,2,1}
                +60 L_{0,0,0,0,0}
                -12 L_{1,0,0,0,0}
                +16 L_{1,0,0,0,1}
  \notag \\ &
                +14 L_{1,0,0,1,0}
                -\frac{55}{3} L_{1,0,0,1,1}
                +12 L_{1,0,1,0,0}
                -\frac{50}{3} L_{1,0,1,0,1}
                -27 L_{1,0,1,1,0}
  \notag \\ &
                +29 L_{1,0,1,1,1}
                +24 L_{1,1,0,0,0}
                -28 L_{1,1,0,0,1}
                -\frac{64}{3} L_{1,1,0,1,0}
                +\frac{70}{3} L_{1,1,0,1,1}
  \notag \\ &
                -\frac{73}{3} L_{1,1,1,0,0}
                +\frac{58}{3} L_{1,1,1,0,1}
                +\frac{97}{6} L_{1,1,1,1,0}
                +\frac{5}{2} L_{1,1,1,1,1}
                +\frac{28}{3} L_{1,1,1,2,1}
  \notag \\ &
                -4 L_{1,1,2,1,0}
                +\frac{20}{3} L_{1,1,2,1,1}
                -4 L_{1,1,2,2,1}
                +\biggl(
                        -\frac{22}{9} L_{2,1}
                        -9 L_{0,0,0}
                        +\frac{11}{6} L_{1,0,0}
  \notag \\ &
                        -\frac{35}{18} L_{1,0,1}
                        -\frac{29}{9} L_{1,1,0}
                        +\frac{31}{12} L_{1,1,1}
                \biggr) \pi^2
                +\biggl(
                        40 L_{0,0}
                        -8 L_{1,0}
                        +11 L_{1,1}
                \biggr) \zeta_3
  \notag \\ &
                +\biggl(
                        \frac{25}{72} L_0
                        -\frac{13}{180} L_1
                \biggr) \pi^4
                -6 \pi^2 \zeta_3
                +24 \zeta_5
        \Biggr)
        +\frac{\zb^2-2\zb+2}{\zb} \Biggl(
                \frac{83}{48} L_1
                +L_{0,1}
  \notag \\ &
                -\frac{1}{3} L_{1,0}
                -\frac{29}{24} L_{1,1}
                -4 L_{0,0,1}
                -\frac{11}{3} L_{0,1,0}
                +14 L_{0,1,1}
                -\frac{13}{3} L_{1,0,0}
                +15 L_{1,0,1}
  \notag \\ &
                -\frac{9}{4} L_{1,1,0}
                +\frac{11}{3} L_{1,1,1}
                +\frac{5}{3} L_{1,2,1}
                +3 L_{0,0,1,0}
                -\frac{9}{2} L_{0,0,1,1}
                +6 L_{0,1,0,0}
  \notag \\ &
                -8 L_{0,1,0,1}
                -\frac{41}{2} L_{0,1,1,0}
                +\frac{63}{2} L_{0,1,1,1}
                +9 L_{1,0,0,0}
                -\frac{27}{2} L_{1,0,0,1}
                -\frac{23}{2} L_{1,0,1,0}
  \notag \\ &
                +23 L_{1,0,1,1}
                -\frac{23}{2} L_{1,1,0,0}
                +18 L_{1,1,0,1}
                +\frac{7}{2} L_{1,1,1,0}
                +4 L_{1,2,1,1}
                +84 L_{0,0,0,0,1}
  \notag \\ &
                +72 L_{0,0,0,1,0}
                -114 L_{0,0,0,1,1}
                +66 L_{0,0,1,0,0}
                -80 L_{0,0,1,0,1}
                -68 L_{0,0,1,1,0}
  \notag \\ &
                +\frac{296}{3} L_{0,0,1,1,1}
                +\frac{32}{3} L_{0,0,1,2,1}
                +66 L_{0,1,0,0,0}
                -\frac{236}{3} L_{0,1,0,0,1}
                -68 L_{0,1,0,1,0}
  \notag \\ &
                +\frac{226}{3} L_{0,1,0,1,1}
                -\frac{248}{3} L_{0,1,1,0,0}
                +\frac{164}{3} L_{0,1,1,0,1}
                +74 L_{0,1,1,1,0}
                +14 L_{0,1,1,1,1}
  \notag \\ &
                -\frac{128}{3} L_{0,1,1,2,1}
                +24 L_{0,1,2,1,0}
                -\frac{140}{3} L_{0,1,2,1,1}
                +\frac{112}{3} L_{0,1,2,2,1}
                +4 L_{1,0,-1,0,1}
  \notag \\ &
                +72 L_{1,0,0,0,0}
                -82 L_{1,0,0,0,1}
                -\frac{206}{3} L_{1,0,0,1,0}
                +\frac{220}{3} L_{1,0,0,1,1}
                -\frac{208}{3} L_{1,0,1,0,0}
  \notag \\ &
                +\frac{212}{3} L_{1,0,1,0,1}
                +\frac{182}{3} L_{1,0,1,1,0}
                -50 L_{1,0,1,1,1}
                -\frac{32}{3} L_{1,0,1,2,1}
                -74 L_{1,1,0,0,0}
  \notag \\ &
                +68 L_{1,1,0,0,1}
                +\frac{152}{3} L_{1,1,0,1,0}
                -\frac{58}{3} L_{1,1,0,1,1}
                +\frac{170}{3} L_{1,1,1,0,0}
                -44 L_{1,1,1,0,1}
  \notag \\ &
                -4 L_{1,1,1,1,0}
                -4 L_{1,1,1,1,1}
                -\frac{40}{3} L_{1,1,1,2,1}
                -\frac{16}{3} L_{1,1,2,1,0}
                -10 L_{1,1,2,1,1}
  \notag \\ &
                +28 L_{1,1,2,2,1}
                +16 L_{1,2,1,0,1}
                +\frac{28}{3} L_{1,2,1,1,0}
                -24 L_{1,2,1,1,1}
                +40 L_{1,2,1,2,1}
  \notag \\ &
                -\frac{56}{3} L_{1,2,2,1,0}
                +32 L_{1,2,2,1,1}
                -\frac{152}{3} L_{1,2,2,2,1}
                +\biggl(
                        \frac{35}{36} L_1
                        -\frac{5}{6} L_{0,1}
                        -\frac{17}{12} L_{1,0}
  \notag \\ &
                        +\frac{21}{8} L_{1,1}
                        -\frac{28}{3} L_{0,0,1}
                        -\frac{82}{9} L_{0,1,0}
                        +\frac{21}{2} L_{0,1,1}
                        -\frac{1}{3} L_{1,0,-1}
                        -\frac{59}{6} L_{1,0,0}
  \notag \\ &
                        +\frac{64}{9} L_{1,0,1}
                        +\frac{73}{9} L_{1,1,0}
                        -\frac{14}{3} L_{1,1,1}
                        -\frac{22}{9} L_{1,2,1}
                \biggr) \pi^2
                +\biggl(
                        13 L_1
                        +34 L_{0,1}
  \notag \\ &
                        +42 L_{1,0}
                        -\frac{76}{3} L_{1,1}
                \biggr) \zeta_3
                +\frac{371}{1080} L_1 \pi^4
        \Biggr)
        +\biggl(
                \frac{1}{18} \big(-\zb^2-22 \zb+26\big)
  \notag \\ &
                +\frac{1}{36} \big(-108 \zb+11\big) L_0
                +\frac{1}{18} \big(55 \zb-47\big) L_1
                -\frac{16}{9} \zb L_{0,0}
                +\frac{1}{9} \big(18 \zb-1\big) L_{0,1}
  \notag \\ &
                +\frac{1}{12} \big(55 \zb-52\big) L_{1,0}
                +\frac{1}{72} \big(-553 \zb+586\big) L_{1,1}
        \biggr) \pi^2
        +\biggl(
                \frac{1}{6} \big(235 \zb-18\big)
  \notag \\ &
                -\frac{2}{3} \big(29 \zb-64\big) L_1
        \biggr) \zeta_3
        +\frac{1}{1080} \big(187 \zb+92\big) \pi^4
        -\frac{3 \sqrt{1-\zb} \big(5 \zb+14\big)}{\sqrt{3+\zb}} R_{1}
  \notag \\ &
        -15 \zb R_{2}
        +6 \big(\zb-2\big) R_{4}
        +\frac{\zb^2-2\zb+2}{\zb} \left(
                18 R_{3}
                -6 R_{4}
        \right)
  \label{eq:Fqq3Ncm3}
  \,.
\end{align}
We note that the expressions that contain square roots either as coefficients
or as letters of the iterated integrals are confined to the \emph{last four
terms} of the above formula and are very compact. The structure of the results
for other colour factors and also for the other matching coefficients is quite
similar to what is shown in Eq.~\eqref{eq:Fqq3Ncm3}.

It is obvious that, for practical computations, it is important to be able to
evaluate the matching coefficients numerically. From this perspective, the
result shown above is not optimal as the numerical evaluation of iterated
integrals with square-root-valued letters is complicated. It is possible to get
around this problem by providing the expansion of the matching coefficient in
powers of $z$ and/or $\zb$ as was done in Ref.~\cite{Ebert:2020unb}.

Here, we note that it is also possible to obtain an expression for the matching
coefficient which is suitable for numerical evaluation, by rationalising the
only square-root-valued letter that appears in the final result. Indeed, since
the matching coefficients only depend on a single square root, we express those
iterated integrals which actually depend on this square root in terms of
Goncharov polylograrithms (GPLs) using the rationalising variable
transformation from Eq.~\eqref{eq:sqrt-rat-x}. These GPLs are then evaluated
at the argument
\begin{align}
  x &= \frac{\sqrt{z} + \I \sqrt{4-z}}{\sqrt{z} - \I \sqrt{4-z}}
  \,.
\end{align}
The remaining iterated integrals only depend on linear letters and therefore
also fall into the class of GPLs, but they are still evaluated at the argument
$\zb$. Also the algebraic coefficients in front of the integrals are expressed
in terms of $\zb$. The use of these mixed arguments in the iterated integrals
allows for a relatively compact representation of the matching coefficients
which is straightforward to evaluate numerically. The corresponding expressions
for the matching coefficients are also included in an ancillary file.

\section{Conclusions}
\label{sec:conc}

In this paper, we have described the computation of the matching coefficients
for $N$-jettiness beam functions through third order in perturbative QCD. This
computation extends our previous results reported in
Ref.~\cite{Behring:2019quf}, where only the generalised leading-colour
contribution to the $q \to q$ matching coefficient was presented.

Although beam functions were originally defined in soft-collinear effective
theory, we benefited from the observation made in Ref.~\cite{Ritzmann:2014mka}
that matching coefficients can be computed by integrating collinear splitting
functions over the $N$-jettiness phase space simplified in the collinear
limits. This observation together with the known fact that collinear splitting
functions can be calculated in a process-independent way if physical gauges are
used, opened up a way to compute the matching coefficients as described in
this paper.

To simplify the actual computation, we made use of reverse unitarity which
allowed us to map the various real emission integrals onto loop-like integrals.
We then employed the (by now) standard machinery for computing loop integrals,
such as integration by parts, differential equations, canonical bases etc.\ to
make the computation manageable. It is worth noting that, although we were able
to construct a canonical basis for all the master integrals involved in the
computation, this could not be always done with publicly-available programs
because multiple square roots appeared in an alphabet associated with the
differential equations. Because of this, we had to resort to a manual
construction of candidate integrals for the canonical basis by studying their
leading singularities in the Baikov representation.

Another non-trivial aspect of the calculation involves the computation of the
boundary conditions for the master integrals. We calculated the required
boundary conditions by studying the soft $z \to 1$ limit of the master
integrals. Interestingly, it turns out that many boundary constants that we
require for this computation are related to the boundary constants which appear
in the calculation of the N$^3$LO QCD corrections to Higgs boson production
cross section obtained in Refs.~\cite{Anastasiou:2013srw}. This relation allows
us to compare the boundary constants that we computed in this and earlier
papers \cite{Melnikov:2018jxb,Melnikov:2019pdm} with the results of
Refs.~\cite{Anastasiou:2013srw}, offering a welcome cross-check at the
intermediate stages of the calculation.

We computed the matching coefficients for the $q_i q_j$, $qg$, $gq$, $gg$ and
$q_i \bar{q}_j$ partonic channels. The results are written as linear
combinations of plus-distributions and regular functions given by linear
combinations of iterated integrals with mostly rational $z$-dependent
coefficients. As we pointed out in Section~\ref{sec:calc}, the differential
equations for individual master integrals involve various square roots of
second degree $z$-polynomials. It is interesting that many of these square
roots disappear from the final answer, when all the master integrals are
combined. In fact, \emph{all} such square roots should disappear in the
divergent parts of the partonic beam functions since these divergences are
removed, e.g., by collinear counter-terms that involve Altarelli-Parisi
splitting functions. These functions were first calculated through N$^3$LO in
Refs.~\cite{Moch:2004pa,Vogt:2004mw} and are known to contain no square roots
of $z$-polynomials.

The matching coefficients computed in this paper have been already calculated
in Ref.~\cite{Ebert:2020unb}. Our calculation agrees with these results and
provides a completely independent check of the matching coefficients reported
in that reference. It goes without saying that for computations of such a
complexity, an independent confirmation is always welcome.

The application of the $N$-jettiness slicing scheme to collider processes
requires calculation of beam functions, jet functions and soft functions. The
beam functions are universal in that they do not depend on the number of jets
in the final state; the same applies to jet functions and, by now, both are
known through N$^3$LO in perturbative QCD.\footnote{The N$^3$LO QCD
contributions to jet functions were computed in
Refs.~\cite{Bruser:2018rad,Banerjee:2018ozf}} The soft functions, on the other
hand, are not known at that perturbative order even for the simplest
zero-jettiness case, in spite of the interesting progress in recent years
\cite{Baranowski:2022khd,Baranowski:2021gxe,Chen:2022cvz}. Hence, to fully
unlock the potential of the $N$-jettiness slicing scheme for N$^3$LO QCD
computations, further progress with computing $N$-jettiness soft-functions at
N$^3$LO is needed.

\section*{Acknowledgement}
We are grateful to B.~Mistlberger for useful conversations and comparison of
some contributions at the intermediate stage of the calculation, as well as to
C.~Duhr for providing the results of Ref.~\cite{Duhr:2014nda} in
computer-readable form. A.~B. would like to thank T.~Becher, F.~Herren and
B.~Page for helpful conversations. We are particularly grateful to R.~Rietkerk
for the collaboration in the earlier stages of this project.

Our work on this project was supported by various funding agencies. In
particular, A.~B., D.~B. and K.~M. were partially supported by the German
Research Foundation (DFG, Deutsche Forschungsgemeinschaft) under grant
396021762-TRR 257. L.~T.~was supported by the Excellence Cluster ORIGINS
funded by the DFG under Germany's Excellence Strategy - EXC-2094 - 390783311,
by the ERC Starting Grant 949279 HighPHun and, in the initial phase of this
work, by the Royal Society through grant URF/R1/191125. During his work on
this project, Ch.~W. was supported by the BMBF project No.~05H18WOCA1. The
research of D.~B. was also supported by Karlsruhe School of Particle and
Astroparticle Physics (KSETA).

Some work of this paper was performed during the program ``Gearing up for
high-precision LHC physics'' at the Munich Institute for Astro-, Particle and
BioPhysics (MIAPbP) whose support we gratefully acknowlege. MIAPbP is funded by
the DFG under Germany's Excellence Strategy - EXC-2094 - 390783311. The
diagrams in this article have been drawn using
\progname{TikZ-Feynman}~\cite{Ellis:2016jkw}.

\appendix

\section{Example of deriving partial fraction relations}
\label{sec:parfrac}

In this appendix, we discuss an example to illustrate the algorithm that was
used to find partial fraction relations between different integrals. This
algorithm was first proposed in Ref.~\cite{Pak:2011xt}; a clear and concise
description was provided in Ref.~\cite{Hoff:2015kub}. More details about the
relevant mathematical concepts can be found in standard textbooks on the
subject (see, for example, Ref.~\cite{Cox:IVA}).

\begin{table}
\renewcommand*{\arraystretch}{1.2}
  \caption{Integral family definitions for the example on partial fraction
           relations.}
  \label{tab:intfam-def}
  \centering
  \begin{tabular}{ccccc}
    \toprule
    $f$          & T1              & T28                & T29                & T30                \\
    \midrule
    $D^{f}_{1}$  & \multicolumn{4}{c}{$k_1^2$}                                                    \\
    $D^{f}_{2}$  & \multicolumn{4}{c}{$k_2^2$}                                                    \\
    $D^{f}_{3}$  & \multicolumn{4}{c}{$k_3^2$}                                                    \\
    $D^{f}_{4}$  & \multicolumn{4}{c}{$2 k_{123} \cdot p - \frac{t}{z}$}                          \\
    $D^{f}_{5}$  & \multicolumn{4}{c}{$\frac{2 k_{123} \cdot \pb}{s} - (1-z)$}                    \\
    $D^{f}_{6}$  & $(k_1-p)^2$     & $(k_1-p)^2$        & $(k_1-p)^2$        & $k_{12}^2$         \\
    $D^{f}_{7}$  & $(k_2-p)^2$     & $(k_2-p)^2$        & $(k_2-p)^2$        & $(k_1-p)^2$        \\
    $D^{f}_{8}$  & $(k_{12}-p)^2$  & $(k_{12}-p)^2$     & $(k_{12}-p)^2$     & $(k_{12}-p)^2$     \\
    $D^{f}_{9}$  & $(k_{13}-p)^2$  & $(k_{13}-p)^2$     & $(k_{13}-p)^2$     & $(k_{13}-p)^2$     \\
    $D^{f}_{10}$ & $(k_{123}-p)^2$ & $(k_{123}-p)^2$    & $(k_{123}-p)^2$    & $(k_{123}-p)^2$    \\
    $D^{f}_{11}$ & $k_1 \cdot \pb$ & $k_1 \cdot \pb$    & $k_{13} \cdot \pb$ & $k_1 \cdot \pb$    \\
    $D^{f}_{12}$ & $k_2 \cdot \pb$ & $k_{13} \cdot \pb$ & $k_3 \cdot \pb$    & $k_{12} \cdot \pb$ \\
    \bottomrule
  \end{tabular}
\end{table}

We consider the following problem. Suppose that there exists a number of
integral families constructed in some way starting from relevant Feynman
diagrams. An example is given in Table~\ref{tab:intfam-def} where four families
are displayed. The integrals described by these families read
\begin{align}
  I^f_{a_1 \dots a_{12}}
    &= \frac{1}{(2\pi)^{3d}} \int\frac{\rmd^d k_1 \, \rmd^d k_2 \, \rmd^d k_3}{%
       \bigl[D^f_1\bigr]_c^{a_1} \dots \bigl[D^f_5\bigr]_c^{a_5}
       \bigl(D^f_6\bigr)^{a_6} \dots \bigl(D^f_{12}\bigr)^{a_{12}}}
  \,,
\end{align}
where $[\dots]_c$ denotes a cut propagator which implements a delta-function
constraint via reverse unitarity. We would like to find if all integrals that
belong to these four families are independent, and to derive linear relations
between them if they are not.

Since each family contains enough inverse propagators to describe all
independent scalar products between $p$, $\pb$ and $k_{1,2,3}$, there exist
linear relations between inverse propagators of different families shown in
Table~\ref{tab:intfam-def}, and it is these linear relations that, potentially,
lead to useful partial fraction relations. To find these relations, the first
step is to combine all inverse propagators that belong to different families
into an overcomplete set. For the sake of simplicity, we consider the topology
T1 and add $\pfD{T30}{12}$ to its inverse propagators, so that a toy version of
the overcomplete set reads
\begin{align}
  \left\{\pfD{T1}{1}, \dots, \pfD{T1}{12}, \pfD{T30}{12}\right\}
  \label{eq:toy-superfam}
  \,.
\end{align}
Since there are twelve independent scalar products and thirteen inverse
propagators, there is one linear relation between elements of the list in
Eq.~\eqref{eq:toy-superfam}; it reads
\begin{align}
  0 &= \pfD{T1}{11} + \pfD{T11}{12} - \pfD{T30}{12}
  \label{eq:pf-seed-linrel}
  \,.
\end{align}
To find a partial fraction relation, we divide Eq.~\eqref{eq:pf-seed-linrel} by
the product of the three propagators and obtain
\begin{align}
  0 &=  \frac{1}{\pfD{T1}{12} \pfD{T30}{12}}
       +\frac{1}{\pfD{T1}{11} \pfD{T30}{12}}
       -\frac{1}{\pfD{T1}{11} \pfD{T1}{12}}
  \label{eq:pfrel-den}
  \,.
\end{align}
Our goal is to treat both linear relations for the numerators and partial
fraction relations for the denominators on the same footing and systematically
apply them to integrands of Feynman integrals. To unify the description of
numerator and denominator relations, we describe the integrands not as rational
functions of the $D^f_i$, but as polynomials in $D^f_i$ and $\tilde{D}^f_i =
1/D^f_i$. We treat the $D^f_i$ and $\tilde{D}^f_i$ as a priori independent
variables which are then subject to the relations $D^f_i \tilde{D}^f_i - 1 =
0$, that describe cancelling numerators against denominators. Moreover, in the
example at hand the variables are also subject to the linear relation in
Eq.~\eqref{eq:pf-seed-linrel} and the partial fraction relation in
Eq.~\eqref{eq:pfrel-den} which becomes
\begin{align}
  0 &=  \pfDt{T1}{12} \pfDt{T30}{12}
       +\pfDt{T1}{11} \pfDt{T30}{12}
       -\pfDt{T1}{11} \pfDt{T1}{12}
  \label{eq:pfrel-D}
  \,.
\end{align}
In this language, the relations between the variables $D^f_i$ and
$\tilde{D}^f_i$ are encoded by polynomials that we equate to zero and applying
one of these relations can be thought of as extracting one of the polynomials,
setting it to zero and keeping the remainder. A trivial example would be a
cancellation of some factors in a numerator and a denominator using
\begin{align}
  \pfD{T1}{7} \pfDt{T1}{7} \pfDt{T1}{8} \pfDt{T1}{9}
    &= \underbrace{\left(\pfD{T1}{7} \pfDt{T1}{7} - 1\right)}_{=0}
         \pfDt{T1}{8} \pfDt{T1}{9}
       +\pfDt{T1}{8} \pfDt{T1}{9}
     = \pfDt{T1}{8} \pfDt{T1}{9}
  \label{eq:pf-example-cancel}
  \,.
\end{align}
Similarly, extracting $\pfD{T1}{11} + \pfD{T11}{12} - \pfD{T30}{12}$
corresponds to eliminating a linear relation between numerators in
Eq.~\eqref{eq:pf-seed-linrel}, and extracting $\pfDt{T1}{12} \pfDt{T30}{12}
+\pfDt{T1}{11} \pfDt{T30}{12}-\pfDt{T1}{11} \pfDt{T1}{12}$ corresponds to
eliminating a partial fraction relation in Eq.~\eqref{eq:pfrel-den}.

To phrase this more generally, given an integrand of a Feynman integral
$\mathcal{I}_D$ written as a polynomial in the variables $D^f_i$ and
$\tilde{D}^f_i$ and a set of polynomials $\{p_i\}$ that encode relations
between the variables, we would like to systematically construct a
decomposition
\begin{align}
  \mathcal{I}_D &= \sum_i c_{D,i} \, p_i + r_D
  \label{eq:pf-decomposition}
  \,,
\end{align}
where the coefficients $c_{D,i}$ and the remainder $r_D$ are also polynomials
in $D^f_i$ and $\tilde{D}^f_i$. The remainder $r_D$ is equivalent to
$\mathcal{I}_D$ because if we consider the case where all relations encoded by
the polynomials $p_i$ hold, i.e. where all the $p_i$'s vanish simultaneously,
then Eq.~\eqref{eq:pf-decomposition} becomes
\begin{align}
  \left.\mathcal{I}_D\right|_{\{p_i=0\}} &= r_D
  \,.
\end{align}
Moreover, the remainder should be unique so that if we decompose two or more
integrands in this way and find that their remainders are linearly independent
over functions of the kinematic variables, the integrands have to be
independent. Thus, such a procedure allows us to remove all linear and partial
fraction relations between integrands.

In order to make this approach systematic, we have to specify which monomials
are preferred over others. For example, in Eq.~\eqref{eq:pf-example-cancel} we
preferred terms with a lower total degree, i.e. we preferred the term $1$ over
$\pfD{T1}{7} \pfDt{T1}{7}$. Intuitively, this ensures that numerators are
cancelled against denominators whenever possible. For the linear and partial
fraction relations in Eqs.~\eqref{eq:pf-seed-linrel} and \eqref{eq:pfrel-D} all
terms have same total degree, so we must use other additional criteria to
specify preferred terms. In general, this is done by specifying a
\emph{monomial ordering}, which defines an ordering of the exponents of the
variables.\footnote{We follow the suggestion from
Refs.~\cite{Pak:2011xt,Hoff:2015kub} and use a monomial ordering where we first
compare the total degree of two monomials for the variables
$\{\tilde{D}^f_1,\dots,\tilde{D}^f_{12},D^f_1,\dots,D^f_{12}\}$. If that is the
same, we successively compare the total degrees considering fewer variables,
i.e. $\{\tilde{D}^f_2,\dots,\tilde{D}^f_{12},D^f_1,\dots,D^f_{12}\}$, then
$\{\tilde{D}^f_3,\dots,\tilde{D}^f_{12},D^f_1,\dots,D^f_{12}\}$, etc.}

The problem we just described is a classic problem in the field of algebraic
geometry and corresponds to \emph{finding a canonical representative modulo
polynomial ideal}. The set of all linear combinations $\langle \{p_i\} \rangle
= \sum_i c_{D,i} \, p_i$, where the $c_{D,i}$ are polynomials in the $D^f_i$
and $\tilde{D}^f_i$, is called the \emph{polynomial ideal} spanned by the
polynomials $p_i$. Decomposing a polynomial $\mathcal{I}_D$ into an element of
the ideal $\langle \{p_i\} \rangle$ and a remainder $r$ can be achieved using a
procedure called \emph{polynomial reduction}. In general, this decomposition is
not unique for an arbitrary set of polynomials $\{p_i\}$. However, what is
important for us is the set of simultaneous zeros of the polynomials $\{p_i\}$.
Therefore, we can choose different sets of polynomials $\{\tilde{p}_i\}$ which
have the same set of simultaneous zeros and generate the same polynomial ideal.
For each given set of polynomials there exists a particular choice of
polynomials, called a \emph{Gr\"obner basis}, for which the remainder $r_D$
becomes unique. Here, we will not go into detail of how to construct a
Gr\"obner basis and only note that the construction of Gr\"obner bases is
implemented in many computer algebra systems. We have used the implementations
in \progname{Mathematica} and \progname{Singular}~\cite{Singular} for our
calculation.

To apply this method to find relations between integrands of Feynman integrals,
the idea is to start from the polynomials that encode cancellations between
numerators $D^f_i$ and denominators $\tilde{D}^f_i$, i.e. $D^f_i \tilde{D}^f_i
- 1$, as well as the linear relations between the numerators. The partial
  fraction relations between denominators automatically arise as a consequence
of the numerator relations.\footnote{As pointed out in
Refs.~\cite{Abreu:2019odu,Heller:2021qkz} it is also possible to directly use
the independent scalar products as the variables for the numerators. In that
case one does not even have to specify the linear relations between numerators
since they will automatically be found though the construction of the Gr\"obner
basis.} For the toy overcomplete set from Eq.~\eqref{eq:toy-superfam} we start
with
\begin{align}
   L_D &= \{
         \pfD{T1}{1} \pfDt{T1}{1} - 1, \;
         \dots, \;
         \pfD{T1}{12} \pfDt{T1}{12} - 1, \;
         \pfD{T30}{12} \pfDt{T30}{12} - 1, \;
         \pfD{T1}{11} + \pfD{T1}{12} - \pfD{T30}{12}
       \}
  \,.
  \label{eq:pf-rel-start}
\end{align}
We now compute the Gr\"obner basis for this set and find
\begin{align}
  L_{D}^{G}
    ={}& \Bigl\{
           \pfD{T1}{1} \pfDt{T1}{1} - 1, \;
           \dots, \;
           \pfD{T1}{12} \pfDt{T1}{12} - 1, \;
           (\pfD{T1}{12} + \pfD{T1}{11}) \pfDt{T30}{12} - 1,
  \notag \\ &
           \pfD{T30}{12} - \pfD{T1}{12} - \pfD{T1}{11}, \;
           \pfDt{T1}{12} \pfDt{T30}{12}
             +\pfDt{T1}{11} \pfDt{T30}{12}
             -\pfDt{T1}{11} \pfDt{T1}{12}
         \Bigr\}
  \label{eq:pf-groebner-basis}
  \,.
\end{align}
Compared to Eq.~\eqref{eq:pf-rel-start}, the next-to-last polynomial in
Eq.~\eqref{eq:pf-rel-start} war reexpressed and the last polynomial in
Eq.~\eqref{eq:pf-groebner-basis} was added. Finally, this Gr\"obner basis can
be used to compute the remainders of the integrands of Feynman integrals via
polynomial reduction.

We illustrate this procedure by considering an integral from the topology T1
defined by the integrand
\begin{align}
  I^{\text{T1}}_{11111;1010011}
    &= \pfDt{T1}{1} \dots \pfDt{T1}{5}
       \pfDt{T1}{6} \pfDt{T1}{8}
       \pfDt{T1}{11} \pfDt{T1}{12}
  \label{eq:pf-seed-int-D}
  \,.
\end{align}
Decomposing $I^{\text{T1}}_{11111;1010011}$ with respect to the Gr\"obner basis
$L_D^G$ in Eq.~\eqref{eq:pf-groebner-basis} yields
\begin{align}
  I^{\text{T1}}_{11111;1010011}
    ={}& \left[\pfDt{T1}{11} \pfDt{T1}{12}
           -\pfDt{T1}{11} \pfDt{T30}{12}
           -\pfDt{T1}{12} \pfDt{T30}{12}\right]
         \pfDt{T1}{1} \dots \pfDt{T1}{5}
         \pfDt{T1}{6} \pfDt{T1}{8}
  \notag \\ &
         +\pfDt{T1}{1} \dots \pfDt{T1}{5}
           \pfDt{T1}{6} \pfDt{T1}{8}
           \pfDt{T1}{11} \pfDt{T30}{12}
         +\pfDt{T1}{1} \dots \pfDt{T1}{5}
           \pfDt{T1}{6} \pfDt{T1}{8}
           \pfDt{T1}{12} \pfDt{T30}{12}
  \label{eq:pf-poly-red-D}
  \,.
\end{align}
The first term on the right hand side of Eq.~\eqref{eq:pf-poly-red-D} is
proportional to the last element of the Gr\"obner basis and therefore vanishes;
the last two terms represent the remainder. Hence, we obtain
\begin{align}
  I^{\text{T1}}_{11111;1010011}
    ={}& \pfDt{T1}{1} \dots \pfDt{T1}{5}
           \pfDt{T1}{6} \pfDt{T1}{8}
           \pfDt{T1}{11} \pfDt{T30}{12}
         +\pfDt{T1}{1} \dots \pfDt{T1}{5}
           \pfDt{T1}{6} \pfDt{T1}{8}
           \pfDt{T1}{12} \pfDt{T30}{12}
  \label{eq:pf-result-D}
  \,.
\end{align}

Upon inspecting the two integrals on the right hand side of
Eq.~\eqref{eq:pf-result-D}, we find that we can map them to the integral
families shown in Table~\ref{tab:intfam-def} provided that we redefine the loop
momenta $k_2 \leftrightarrow k_3$. We obtain
\begin{align}
  I^{\text{T1}}_{11111;1010011}
    &= I^{\text{T28}}_{11111;1001011} + I^{\text{T29}}_{11111;1001011}
  \,.
\end{align}
This is the partial fraction relation that was displayed in
Eq.~\eqref{eq:pfrel}. We note that, although we added the linearly-dependent
propagator from a family T30 to the twelve propagators of the family T1, the
linear relation that we found involves integrals from the families T1, T28,
T29. As a final comment we note that, in general, it may still be necessary to
reduce integrals found in the last step to master integrals using the IBP
relations.

\section{Leading-singularity analysis: an example}
\label{sec:leading-sing}

In this appendix, we give an explicit example of finding a candidate for a
canonical integral based on the analysis of the leading singularities in the
Baikov representation \cite{Baikov:1996rk,Baikov:1996iu}. For this example, we
consider a family of RRV integrals defined by the inverse
propagators\footnote{The first four propagators correspond to delta-function
constraints re-written as propagators using reverse unitarity.}
\begin{align}
  \begin{aligned}
    D_1 &= k_1^2                            \,, &
    D_2 &= k_2^2                            \,, &
    D_3 &= 2 k_{12} \cdot p - \frac{t}{z}   \,, &
    D_4 &= \frac{2 k_{12} \cdot p}{s} - \zb \,, \\
    D_5 &= k_3^2                            \,, &
    D_6 &= k_{12}^2                         \,, &
    D_7 &= k_{13}^2                         \,, &
    D_8 &= k_{123}^2                        \,, \\
    D_9 &= (k_1-p)^2                        \,, &
    D_{10} &= (k_{123}-p)^2                 \,, &
    D_{11} &= k_2 \cdot \pb                 \,, &
    D_{12} &= k_3 \cdot \pb                 \,.
  \end{aligned}
  \label{eq:A3-intfam}
\end{align}
We focus on the following integral
\begin{align}
  I^{\text{A3}}_{1111;10111111}
    &= \frac{1}{(2 \pi)^{3d}}
       \int \frac{\rmd^d k_1 \, \rmd^d k_2 \, \rmd^d k_3}{%
       [D_1]_c [D_2]_2 [D_3]_c [D_4]_c D_5 D_7 D_8 D_9 D_{10} D_{11} D_{12}}
  \,.
  \label{eq:A3-4063-integrand}
\end{align}
In Eq.~\eqref{eq:A3-4063-integrand} $[\dots]_c$ denotes a cut propagator
introduced via reverse unitarity. Note that the propagator $D_6$ is absent in
the integrand in Eq.~\eqref{eq:A3-4063-integrand}. We assume that $s = 2 p
\cdot \pb = 1$ and $t = 1$ since, as we explained in the main text of the
paper, the dependence of all integrals on $s$ and $t$ is uniform and can be
easily restored. The differential equation in $z$, that
$I^{\text{A3}}_{1111;10111111}$ satisfies, is not in canonical form. Hence, our
goal is to find a new master integral which is directly related to
$I^{\text{A3}}_{1111;10111111}$ and is canonical.

To proceed, we start by employing the Baikov
representation~\cite{Baikov:1996rk,Baikov:1996iu} to render the integrand into
a convenient form where the propagators $D_i$ take the role of the integration
variables. Up to an irrelevant overall prefactor, we therefore write
\begin{align}
  I^{\text{A3}}_{1111;10111111}
    &\sim \BaikCut_{D_1,D_2,D_3,D_4} \int \frac{\rmd D_1 \dots \rmd D_{12}}{%
          D_1 \dots D_5 \, D_7 \dots D_{12}} P(D_1,\dots,D_{12})^{(d-6)/2}
  \label{eq:A3-4063-baikov}
  \,.
\end{align}
In Eq.~\eqref{eq:A3-4063-baikov}
$P(D_1,\dots,D_{12}) = G(k_1,k_2,k_3,p,\pb)|_{p_i \cdot p_j = \sum_k c_k D_k}$
is the so-called Baikov polynomial, i.e. the Gram determinant of the momenta
$k_{1,2,3}$ and external momenta $p$ and $\pb$ where all scalar products have
been expressed in terms of inverse propagators. The advantage of the Baikov
representation for the problem at hand is that it allows us to derive a
convenient starting point for the analysis of the cut integrals. Indeed, cutting
propagators corresponds to taking residues at $D_i = 0$, $i=1,2,3,4$, in
Eq.~\eqref{eq:A3-4063-baikov} which, in turn, amounts to simply evaluating the
Baikov polynomial at the point $D_1 = D_2 = D_3 = D_4 = 0$.

For the cut integral in Eq.~\eqref{eq:A3-4063-baikov}, the Baikov polynomial
reads
\begin{align}
  \MoveEqLeft{P(0,0,0,0,D_5,\dots,D_{12})}
  \notag \\ ={}&
    \frac{1}{16 z^2} \bigl(
      D_5 \zb^2
      -D_6 \zb^2
      -D_7 \zb^2
      +D_8 \zb^2
      -D_{5,6} z \zb
      +D_6^2 z \zb
      +D_{5,7} z \zb
      -D_7^2 z \zb
  \notag \\ &
      -D_{5,8} z \zb
      -D_{6,8} z^2 \zb
      +D_{7,8} (2-z) z \zb
      -D_8^2 z \zb^2
      +2 D_{5,9} z \zb^2
      -D_{6,9} z \zb^2
      -D_{7,9} z \zb^2
  \notag \\ &
      +D_{8,9} z \zb^2
      -D_{6,10} z \zb^2
      -D_{7,10} z \zb^2
      +D_{8,10} z \zb^2
      -2 D_{5,11} \zb
      +2 D_{6,11} \zb
      +2 D_{7,11} \zb
  \notag \\ &
      -4 D_{8,11} \zb
      +2 D_{5,12} \zb
      -2 D_{6,12} \zb
      -2 D_{7,12} \zb
      -D_{5,6,7} z^2
      +D_6^2 D_7 z^2
      +D_6 D_7^2 z^2
  \notag \\ &
      +D_{5,6,8} z^2
      -D_6^2 D_8 z^2 \zb
      -D_{6,7,8} (2-z) z^2
      +D_6 D_8^2 z^2 \zb
      -D_{5,6,9} z^2 \zb
      +D_6^2 D_9 z^2 \zb
  \notag \\ &
      +D_{5,7,9} z^2 \zb
      +D_{6,7,9} z^2 \zb
      -D_{5,8,9} z^2 \zb
      -D_{6,8,9} z^2 (z+1) \zb
      -D_{7,8,9} z^3 \zb
      +D_8^2 D_9 z^3 \zb
  \notag \\ &
      +D_5 D_9^2 z^2 \zb^2
      +D_6^2 D_{10} z^2 \zb
      +D_{6,7,10} z^2 \zb
      -D_{6,8,10} z^2 \zb
      -D_{6,9,10} z^2 \zb^2
      -D_{7,9,10} z^2 \zb^2
  \notag \\ &
      +D_{8,9,10} z^2 \zb^2
      -2 D_5^2 D_{11} z
      +4 D_{5,6,11} z
      -2 D_6^2 D_{11} z
      +2 D_{5,7,11} z
      -2 D_{6,7,11} z
  \notag \\ &
      +2 D_{5,8,11} z^2
      +2 D_{6,8,11} z
      -2 D_{7,8,11} (2-z) z
      +4 D_8^2 D_{11} z \zb
      -2 D_{5,9,11} z \zb
  \notag \\ &
      +2 D_{6,9,11} z \zb
      -2 D_{8,9,11} z \zb
      +2 D_{5,10,11} z \zb
      +2 D_{7,10,11} z \zb
      -4 D_{8,10,11} z \zb
      -2 D_{5,6,12} z
  \notag \\ &
      +2 D_6^2 D_{12} z
      +2 D_{6,7,12} z
      +D_8 D_{11}^2 4
      +2 D_{6,8,12} z \zb
      +4 D_{5,9,12} z \zb
      -2 D_{7,9,12} z \zb
  \notag \\ &
      -2 D_{8,9,12} z \zb
      -2 D_{6,10,12} z \zb
      +D_{5,11,12} -4
      +D_{6,11,12} 4
      +D_{7,11,12} 4
      -2 D_{5,6,8,11} z^2
  \notag \\ &
      +2 D_6^2 D_{8,11} z^2
      +4 D_{6,7,8,11} z^2
      -2 D_8^2 D_{6,11} (2-z) z^2
      -2 D_5^2 D_{9,11} z^2
      +4 D_{5,6,9,11} z^2
  \notag \\ &
      -2 D_6^2 D_{9,11} z^2
      +2 D_{5,8,9,11} z^2 (z+1)
      +2 D_{6,8,9,11} z^2 (z+1)
      -2 D_8^2 D_{9,11} z^3
  \notag \\ &
      +2 D_{5,6,10,11} z^2
      -2 D_6^2 D_{10,11} z^2
      -4 D_{6,7,10,11} z^2
      +2 D_{6,8,10,11} (3-2 z) z^2
  \notag \\ &
      +2 D_{5,9,10,11} z^2 \zb
      +2 D_{6,9,10,11} z^2 \zb
      -2 D_{8,9,10,11} z^2 \zb
      -2 D_{10}^2 D_{6,11} z^2 \zb
      +4 D_{11}^2 D_{5,8} z
  \notag \\ &
      -4 D_{11}^2 D_{6,8} z
      -2 D_6^2 D_{8,12} z^2
      -4 D_{8,11}^2 z
      -2 D_{5,6,9,12} z^2
      +2 D_6^2 D_{9,12} z^2
      +4 D_{6,7,9,12} z^2
  \notag \\ &
      -2 D_{6,8,9,12} z^3
      -4 D_{11}^2 D_{5,10} z
      +2 D_9^2 D_{5,12} z^2 \zb
      +4 D_{11}^2 D_{6,10} z
      +2 D_9^2 D_{6,12} z^2 \zb
  \notag \\ &
      +2 D_6^2 D_{10,12} z^2
      +4 D_{11}^2 D_{8,10} z
      -2 D_9^2 D_{8,12} z^2 \zb
      -2 D_{6,9,10,12} z^2 \zb
      -4 D_{6,8,11,12} z
  \notag \\ &
      -4 D_{5,9,11,12} z
      +4 D_{6,9,11,12} z
      +4 D_{8,9,11,12} z
      +4 D_{6,10,11,12} z
      +4 D_{12}^2 D_{6,9} z
  \notag \\ &
      +4 D_6 D_{8,11}^2 z^2
      -8 D_{11}^2 D_{6,8,10} z^2
      +4 D_6 D_{10,11}^2 z^2
      -8 D_{6,8,9,11,12} z^2
      +8 D_{6,9,10,11,12} z^2
  \notag \\ &
      +4 D_6 D_{9,12}^2 z^2
    \bigr)
  \,,
\end{align}
where we have used the abbreviations $D_{a,b,c,\dots} = D_a D_b D_c \dots$.

Calculation of the leading singularities requires us to compute (iterated)
residues in all integration variables at all possible locations of the poles,
eventually including those induced by the Baikov polynomial
$P(D_1,\dots,D_{12})$. Conjecturally~\cite{Henn:2020lye}, this analysis allows
us to determine candidates for canonical integrals by searching for linear
combinations of integrals
\begin{enumerate}
  \item which are ultra-violet (UV) finite;%
  \item which do not have any double poles in any of the integration variables;
  \item whose leading singularities, i.e. the maximally iterated residues at
        all single poles, are constant in the sense that they do not depend on
        the kinematic variable $z$.
\end{enumerate}

To find candidate integrals, we work in $d=4$ dimensions. As we already
mentioned, the integral in Eq.~\eqref{eq:A3-4063-integrand} is not canonical and
has to be modified appropriately. Hence, we make a general ansatz for the
numerator and write
\begin{align}
  &I_\text{can}
    \sim \BaikCut_{D1,D2,D3,D4} \int \frac{\rmd D_1 \dots \rmd D_{12}}{%
          D_1 \dots D_5 \, D_7 \dots D_{12}}
          \frac{N_\text{can}(D_1,\dots,D_{12})}{P(D_1,\dots,D_{12})^2}
  \label{eq:cutrep}
  \,, \\
  &N_\text{can}(D_1,\dots,D_{12})
    = a_0 + \sum\limits_{k=5}^{12} a_{k} D_k
         +\sum\limits_{k=5}^{12} \sum\limits_{l=5}^{k} a_{k,l} D_k D_l
  \label{eq:num-ansatz}
  \,,
\end{align}
where the degree of $N_\text{can}(D_1,\dots,D_{12})$ is bounded by the
requirement of UV finiteness.\footnote{This requirement can be relaxed in some
cases allowing for logarithmic UV divergences.} This formula implies that the
candidate integral is a linear combination of $I^{\text{A3}}_{1111;10111111}$
and simpler integrals that belong to the same integral family. Our goal is to
find the coefficients $a_0$, $a_k$ and $a_{k,l}$ that make $I_{\text{can}}$
satisfy the three requirements listed above.

We note that the analysis of the leading singularities of the full integral,
which necessarily involves studying the pole structure of the Baikov polynomial
$P(D_1,\dots,D_{12})$, is quite demanding. To simplify it, we organise the
analysis iteratively, starting with the computation of simplest residues, and
then moving to more complex ones. Experience shows that we do not need to carry
this analysis to the very end and that after several iterations a clear
candidate for a canonical integral emerges. Below we illustrate how this is
done in practice.

A glance at Eq.~\eqref{eq:cutrep} reveals that the simplest residue to compute
(the maximal cut) is the one at $D_i=0$, with $i=1,\dots,5,7,\dots,12$.
Computing this residue, we find
\begin{align}
  \BaikCut_{D_1,\dots,\widehat{D_6},\dots,D_{12}} \left[ I_\text{can} \right]
    &\sim \frac{16 z^2}{1-z} \int \rmd D_6 \,
          \frac{a_0 + a_6 D_6 + a_{6,6} D_6^2}{D_6 (z D_6 - (1-z))}
  \label{eq:A3-4063-maxcut}
  \,,
\end{align}
where $\widehat{D_6}$ means that no residue in $D_6$ has been calculated. The
denominator of the integrand in Eq.~\eqref{eq:A3-4063-maxcut} comes from the
Baikov polynomial which simplifies once all the other propagators are set to
zero.

For the maximal-cut computation we need to analyse residues in $D_6$ in
Eq.~\eqref{eq:A3-4063-maxcut}. We note that absence of double poles at infinity
immediately implies $a_{6,6} = 0$. It is then clear that there are two simple
poles in Eq.~\eqref{eq:A3-4063-maxcut}, one at $D_6 = 0$ and another one at
$D_6 = (1-z)/z$. Computing the two residues we obtain
\begin{align}
  \BaikCut_{D_1,\dots,D_6,\dots,D_{12}} \left[ I_\text{can} \right]
    &\sim \left\{
            -\frac{16 z^2}{(1-z)^2} a_0
            \,, \quad
            \frac{16 z^2}{(1-z)^2} \left(a_0 + \frac{1-z}{z} a_6\right)
          \right\}
  \,.
\end{align}
These residues are called the ``leading singularities''. According to the third
requirement mentioned above, a candidate for a canonical master integral should
have constant (i.e. $z$-independent) leading singularities. We accomplish this
by choosing
\begin{align}
  a_0 &= \frac{(1-z)^2}{16 z^2}
  \,, \quad
  a_6 = 0
  \,.
\end{align}

Computing the maximal cut allows us to determine some, but not all coefficients
in the general ansatz shown in Eq.~\eqref{eq:num-ansatz}. To fix more
coefficients, we need to inspect the next-to-maximal cuts. This means that,
instead of computing the residues at $D_i=0$ for all $i=5,7,8,\dots,12$, we do
this for all but one of them. It is clear that the number of next-to-maximal
cuts that need to be considered is seven in this case.

As an example, we consider the next-to-maximal cut where we do not take the
residue at $D_7=0$. The corresponding integrand reads
\begin{align}
  \BaikCut_{D_1,\dots,\widehat{D_6},\widehat{D_7},\dots,D_{12}} I_\text{can}
    &\sim 16 z^2 \int \rmd D_6 \, \rmd D_7
          \frac{\frac{(1-z)^2}{16 z^2} + a_7 D_7 + a_{7,6} D_6 D_7
            +a_{7,7} D_7^2}{D_7 (D_6 + D_7) ((1-z)+z D_7) (z D_6 - (1-z))}
  \label{eq:A3-4063-nmaxcut-D7}
  \,.
\end{align}
To avoid poles at $D_7 \to \infty$, we set $a_{7,7} = 0$. Computing the
remaining poles in $D_7$, we find that this induces double poles in $D_6$ at
$D_6 = (1-z)/z$ unless we choose
\begin{align}
  a_{7,6} = 0
  \,, \quad
  a_7 = \frac{1-z}{16 z}
  \,.
\end{align}
Calculating all the remaining residues in $D_6$ and $D_7$ in
Eq.~\eqref{eq:A3-4063-nmaxcut-D7}, we find that with this choice of constants
the leading singularities are $\{-1,1\}$.

As another example, we consider the next-to-maximal cut where the $D_8=0$
residue is not taken. We then find
\begin{align}
  \BaikCut_{D_1,\dots,\widehat{D_6},D_7,\widehat{D_8},\dots,D_{12}} I_\text{can}
    &\sim \frac{16 z^2}{1-z} \int \rmd D_6 \, \rmd D_8
          \frac{\frac{(1-z)^2}{16 z^2} + a_8 D_8 + a_{8,6} D_6 D_8
            +a_{8,8} D_8^2}{D_8 (D_6 - D_8) (1-z D_8) (z D_6 - (1-z))}
  \,,
\end{align}
We can again set $a_{8,8} = 0$ to avoid poles at infinity. Computing residues at
$D_8$, we find that the result does not have double poles in the variable $D_6$.
However, since our goal is to find candidates for canonical integrals, we can
choose the simplest option whenever possible. Thus, we choose $a_{8,6} = 0$ and
find the following leading singularities
\begin{align}
  \left\{
    \pm 1,
    \pm \left(\frac{1}{z} + \frac{16 a_8}{1-z}\right),
    \pm \left(1 - \frac{1}{z} + \frac{16 a_8}{1-z}\right)
  \right\}
  \,.
\end{align}
They become $z$-independent if we choose
\begin{align}
  a_8 &= -\frac{(1-z)^2}{16 z}
  \,.
\end{align}

We perform the analysis of all next-to-maximal and next-to-next-to-maximal cuts
and find that with the choice of the following numerator polynomial
\begin{align}
  N_\text{can}(D_1,\dots,D_{12})
    &= \frac{1-z}{16 z} \left(
         \frac{1-z}{z}
         +D_7
         -(1-z) D_8
         +\frac{2}{z} D_{12}
         +2 D_{10} D_{11}
       \right)
  \,,
\end{align}
all requirements mentioned above are satisfied.

The presence of the integration variables $D_i$ in the numerator of the
integrand removes the corresponding propagators in the original
integral. Therefore, the candidate for a canonical integral to replace
$I^{\text{A3}}_{1111;1011 1111}$ reads
\begin{align}
  I_\text{can}
    ={}& \frac{1-z}{16 z} \biggl(
           \frac{1-z}{z} I^{\text{A3}}_{1111;1011 1111}
           +             I^{\text{A3}}_{1111;1001 1111}
           -(1-z)        I^{\text{A3}}_{1111;1010 1111}
  \notag \\ &
           +\frac{2}{z}  I^{\text{A3}}_{1111;1011 1110}
           +2            I^{\text{A3}}_{1111;1011 1001}
         \biggr)
  \,.
  \label{eq:canonical-candidate}
\end{align}

We can easily check whether or not this candidate integral is indeed canonical
since we know the differential equations for all integrals that appear in
Eq.~\eqref{eq:canonical-candidate}. Although in this particular case
$I_\text{can}$ turns out to be canonical, in general this does not happen since
we terminated the cut analysis once the next-to-next-to-maximal cut was
computed.

Nevertheless, even if the candidate integral turns out to be not fully
canonical after the next-to-leading cut analysis, knowing a good candidate is
extremely helpful. Indeed, we note that once the differential equations for
the integrals that $I_\text{can}$ couples to have been brought to a canonical
form, also the differential equation that $I_\text{can}$ satisfies becomes
partially canonical. Since the integral we started from has eleven propagators
and we analysed the leading singularities up to the next-to-next-to-maximal
cuts, all blocks in the differential equation corresponding to sectors with at
least nine propagators are canonical.
To deal with the rest, we resorted to methods described in
Ref.~\cite{Gehrmann:2014bfa} where a bottom-up construction of the canonical
basis is described.
As a final remark we note that an analysis of leading singularities combined
with the methods of Ref.~\cite{Gehrmann:2014bfa} allowed us to find canonical
bases for all integrals that appear in the computation of the beam function at
N$^3$LO in perturbative QCD.

\section{Building blocks for the beam-function renormalisation}
\label{sec:reno}
In this appendix, we collect formulas that are needed for the extraction of the
matching coefficients from the bare partonic beam functions.

First, we describe how to construct the $\MSbar$ parton distribution functions
in perturbation theory. The starting point is the Altarelli-Parisi equation,
Eq.~\eqref{eq:DGLAP}, and the perturbative expansion of the splitting functions
\begin{align}
  P_{ij}(z)
    &= \sum_{n=0}^{\infty} \left(\frac{\alpha_s}{2\pi}\right)^n P_{ij}^{(n)}(z)
  \,.
\end{align}
To construct the parton distribution functions $\PDFs{ij}$, we integrate the
Altarelli-Parisi equation with the boundary condition $f_{ij}^{(0)}(z) =
\delta(1-z)$. We employ the evolution equation for the strong coupling constant
\begin{align}
  \mu^2 \frac{\rmd}{\rmd \mu^2} \alpha_s(\mu^2)
    &= \beta(\alpha_s) - \ep \,\alpha_s(\mu^2)
  \,,
\end{align}
where
\begin{align}
  \beta(\alpha_s)
    &= -\frac{\alpha_s^2}{4\pi} \beta_0 - \frac{\alpha_s^3}{(4\pi)^2} \beta_1
       + \mathcal{O}(\alpha_s^4)
  \,,
\end{align}
and
\begin{align}
  \beta_0 &= \frac{11}{3} C_A - \frac{4}{3} T_F N_f
  \,, &
  \beta_1 &= \frac{34}{3} C_A^2 - \left(\frac{20}{3} C_A + 4 C_F \right) T_F N_f
  \,,
\end{align}
are the well-known expansion coefficients of the beta function.
We write the result for the partonic PDFs as
\begin{align}
  \PDFs[(1)]{ij}
    ={}& -\frac{1}{\ep} P_{ij}^{(0)}
  \label{eq:f_1}
  \,, \\
  \PDFs[(2)]{ij}
    ={}& \frac{1}{2\ep^2} \sum_k P_{ik}^{(0)}
           \underset{z}{\otimes} P_{kj}^{(0)}
         +\frac{\beta_0}{4 \ep^2} P_{ij}^{(0)}
         -\frac{1}{2\ep} P_{ij}^{(1)}
  \label{eq:f_2}
  \,, \\
  \PDFs[(3)]{ij}
    ={}& -\frac{1}{6\ep^3} \sum_{k,\ell} P_{ik}^{(0)}
           \underset{z}{\otimes} P_{k\ell}^{(0)}
           \underset{z}{\otimes} P_{\ell j}^{(0)}
         -\frac{\beta_0}{4\ep^3}
           \sum_k P_{ik}^{(0)} \underset{z}{\otimes} P_{kj}^{(0)}
         -\frac{\beta_0^2}{12\ep^3} P_{ij}^{(0)}
  \notag \\ &
         +\frac{1}{3\ep^2} \sum_k P_{ik}^{(1)}
           \underset{z}{\otimes} P_{kj}^{(0)}
         +\frac{\beta_0}{6\ep^2} P_{ij}^{(1)}
         +\frac{1}{6\ep^2} \sum_k P_{ik}^{(0)}
           \underset{z}{\otimes} P_{kj}^{(1)}
  \notag \\ &
         +\frac{\beta_1}{12\ep^2} P_{ij}^{(0)}
         -\frac{1}{3\ep} P_{ij}^{(2)}\,,
\end{align}
where the dependence of $\PDFs{ij}$'s and $P_{ij}$'s on $z$ has been
suppressed.

The expansion coefficients of the renormalisation constants for the quark case
($i=q$) read
\begin{align}
  Z_q^{(1)} ={}&
        C_F \Biggl[
                -\mathcal{L}_0\left(\frac{t}{\mu^2}\right) \frac{4}{\ep}
                +\delta(t) \biggl[
                        \frac{4}{\ep^2}
                        +\frac{3}{\ep}
                \biggr]
        \Biggr]
  \,, \\
  Z_q^{(2)} ={}&
        C_A C_F \Biggl[
                \mathcal{L}_0\left(\frac{t}{\mu^2}\right) \biggl[
                      \frac{22}{3 \ep^2}
                        +\frac{1}{\ep} \biggl(
                                -\frac{134}{9}
                                +\frac{2}{3} \pi^2
                        \biggr)
                \biggr]
                +\delta(t) \biggl[
                        -\frac{11}{\ep^3}
  \notag \\ &
                        +\frac{1}{\ep^2} \biggl(
                                \frac{35}{18}
                                -\frac{1}{3} \pi^2
                        \biggr)
                        +\frac{1}{\ep} \biggl(
                                \frac{1769}{108}
                                +\frac{11}{18} \pi^2
                                -20 \zeta_3
                        \biggr)
                \biggr]
        \Biggr]
        +C_F^2 \Biggl[
                \mathcal{L}_1\left(\frac{t}{\mu^2}\right) \frac{16}{\ep^2}
  \notag \\ &
                +\mathcal{L}_0\left(\frac{t}{\mu^2}\right) \biggl[
                        -\frac{16}{\ep^3}
                        -\frac{12}{\ep^2}
                \biggr]
                +\delta(t) \biggl[
                        \frac{8}{\ep^4}
                        +\frac{12}{\ep^3}
                        +\frac{1}{\ep^2} \biggl(
                                \frac{9}{2}
                                -\frac{4}{3} \pi^2
                        \biggr)
                        +\frac{1}{\ep} \biggl(
                                \frac{3}{4}
                                -\pi^2
  \notag \\ &
                                +12 \zeta_3
                        \biggr)
                \biggr]
        \Biggr]
        +C_F N_F T_F \Biggl[
                \mathcal{L}_0\left(\frac{t}{\mu^2}\right) \biggl[
                        -\frac{1}{\ep^2} \frac{8}{3}
                        +\frac{1}{\ep} \frac{40}{9}
                \biggr]
                +\delta(t) \biggl[
                        \frac{4}{\ep^3}
                        -\frac{2}{9\ep^2}
  \notag \\ &
                        +\frac{1}{\ep} \biggl(
                                -\frac{121}{27}
                                -\frac{2}{9} \pi^2
                        \biggr)
                \biggr]
        \Biggr]
  \,,
\end{align}
and
\begin{align}
  Z_q^{(3)} ={}&
        C_A^2 C_F \Biggl[
                \mathcal{L}_0\left(\frac{t}{\mu^2}\right) \biggl[
                        -\frac{484}{27 \ep^3}
                        +\frac{1}{\ep^2} \biggl(
                                \frac{4172}{81}
                                -\frac{44}{27} \pi^2
                        \biggr)
                        +\frac{1}{\ep} \biggl(
                                -\frac{490}{9}
                                +\frac{536}{81} \pi^2
  \notag \\ &
                                -\frac{88}{9} \zeta_3
                                -\frac{44}{135} \pi^4
                        \biggr)
                \biggr]
                +\delta(t) \biggl[
                        \frac{2662}{81 \ep^4}
                        +\frac{1}{\ep^3} \biggl(
                                -\frac{8999}{243}
                                +\frac{110}{81} \pi^2
                        \biggr)
  \notag \\ &
                        +\frac{1}{\ep^2} \biggl(
                                -\frac{16147}{486}
                                -\frac{899}{243} \pi^2
                                +\frac{1408}{27} \zeta_3
                                +\frac{44}{405} \pi^4
                        \biggr)
                        +\frac{1}{\ep} \biggl(
                                \frac{412907}{8748}
                                +\frac{419}{729} \pi^2
  \notag \\ &
                                -\frac{5500}{27} \zeta_3
                                +\frac{19}{30} \pi^4
                                +\frac{88}{27} \pi^2 \zeta_3
                                +\frac{232}{3} \zeta_5
                        \biggr)
                \biggr]
        \Biggr]
  \notag \\ &
        +C_A C_F^2 \Biggl[
                \mathcal{L}_1\left(\frac{t}{\mu^2}\right) \biggl[
                        -\frac{1}{\ep^3} \frac{176}{3}
                        +\frac{1}{\ep^2} \biggl(
                                \frac{1072}{9}
                                -\frac{16}{3} \pi^2
                        \biggr)
                \biggr]
  \notag \\ &
                +\mathcal{L}_0\left(\frac{t}{\mu^2}\right) \biggl[
                        \frac{220}{3 \ep^4}
                        +\frac{1}{\ep^3} \biggl(
                                -\frac{136}{3}
                                +4 \pi^2
                        \biggr)
                        -\frac{1}{\ep^2} \biggl(
                                \frac{2975}{27}
                                +\frac{4}{9} \pi^2
  \notag \\ &
                                -80 \zeta_3
                        \biggr)
                \biggr]
                +\delta(t) \biggl[
                        -\frac{44}{\ep^5}
                        -\frac{1}{\ep^4} \biggl(
                                \frac{227}{9}
                                +\frac{4}{3} \pi^2
                        \biggr)
                        +\frac{1}{\ep^3} \biggl(
                                \frac{3853}{54}
                                +\frac{19}{3} \pi^2
  \notag \\ &
                                -80 \zeta_3
                        \biggr)
                        +\frac{1}{\ep^2} \biggl(
                                \frac{1703}{36}
                                -\frac{305}{54} \pi^2
                                -\frac{268}{3} \zeta_3
                                +\frac{4}{9} \pi^4
                        \biggr)
                        +\frac{1}{\ep} \biggl(
                                \frac{151}{12}
                                -\frac{205}{27} \pi^2
  \notag \\ &
                                +\frac{844}{9} \zeta_3
                                -\frac{247}{405} \pi^4
                                +\frac{8}{9} \pi^2 \zeta_3
                                +40 \zeta_5
                        \biggr)
                \biggr]
        \Biggr]
        +C_F^3 \Biggl[
                -\mathcal{L}_2\left(\frac{t}{\mu^2}\right) \frac{32}{\ep^3}
  \notag \\ &
                +\mathcal{L}_1\left(\frac{t}{\mu^2}\right) \biggl[
                        \frac{64}{\ep^4}
                        +\frac{48}{\ep^3}
                \biggr]
                +\mathcal{L}_0\left(\frac{t}{\mu^2}\right) \biggl[
                        -\frac{32}{\ep^5}
                        -\frac{48}{\ep^4}
                        -\frac{1}{\ep^3} \biggl(
                                18
                                -\frac{16}{3} \pi^2
                        \biggr)
  \notag \\ &
                        -\frac{1}{\ep^2} \biggl(
                                3
                                -4 \pi^2
                                +48 \zeta_3
                        \biggr)
                \biggr]
                +\delta(t) \biggl[
                        \frac{32}{3 \ep^6}
                        +\frac{24}{\ep^5}
                        +\frac{1}{\ep^4} \biggl(
                                18
                                -\frac{16}{3} \pi^2
                        \biggr)
  \notag \\ &
                        +\frac{1}{\ep^3} \biggl(
                                \frac{15}{2}
                                -8 \pi^2
                                +\frac{80}{3} \zeta_3
                        \biggr)
                        +\frac{1}{\ep^2} \biggl(
                                \frac{9}{4}
                                -3 \pi^2
                                +36 \zeta_3
                        \biggr)
                        +\frac{1}{\ep} \biggl(
                                \frac{29}{6}
                                +\pi^2
  \notag \\ &
                                +\frac{68}{3} \zeta_3
                                +\frac{8}{15} \pi^4
                                -\frac{16}{9} \pi^2 \zeta_3
                                -80 \zeta_5
                        \biggr)
                \biggr]
        \Biggr]
  \notag \\ &
        +C_A C_F T_F N_F \Biggl[
                \mathcal{L}_0\left(\frac{t}{\mu^2}\right) \biggl[
                        \frac{352}{27 \ep^3}
                        -\frac{1}{\ep^2} \biggl(
                                \frac{2672}{81}
                                -\frac{16}{27} \pi^2
                        \biggr)
                        +\frac{1}{\ep} \biggl(
                                \frac{1672}{81}
  \notag \\ &
                                -\frac{160}{81} \pi^2
                                +\frac{224}{9} \zeta_3
                        \biggr)
                \biggr]
                +\delta(t) \biggl[
                        -\frac{1}{\ep^4} \frac{1936}{81}
                        +\frac{1}{\ep^3} \biggl(
                                \frac{5384}{243}
                                -\frac{40}{81} \pi^2
                        \biggr)
  \notag \\ &
                        +\frac{1}{\ep^2} \biggl(
                                \frac{6148}{243}
                                +\frac{424}{243} \pi^2
                                -\frac{704}{27} \zeta_3
                        \biggr)
                        +\frac{1}{\ep} \biggl(
                                \frac{5476}{2187}
                                -\frac{1180}{729} \pi^2
                                +\frac{2656}{81} \zeta_3
  \notag \\ &
                                -\frac{46}{135} \pi^4
                        \biggr)
                \biggr]
        \Biggr]
        +C_F^2 T_F N_F \Biggl[
                \mathcal{L}_1\left(\frac{t}{\mu^2}\right) \biggl[
                        \frac{64}{3 \ep^3}
                        -\frac{1}{\ep^2} \frac{320}{9}
                \biggr]
  \notag \\ &
                +\mathcal{L}_0\left(\frac{t}{\mu^2}\right) \biggl[
                        -\frac{80}{3 \ep^4}
                        +\frac{32}{3 \ep^3}
                        +\frac{1}{\ep^2} \biggl(
                                \frac{700}{27}
                                +\frac{8}{9} \pi^2
                        \biggr)
                        +\frac{1}{\ep} \biggl(
                                \frac{220}{9}
                                -\frac{64}{3} \zeta_3
                        \biggr)
                \biggr]
  \notag \\ &
                +\delta(t) \biggl[
                        \frac{16}{\ep^5}
                        +\frac{100}{9 \ep^4}
                        -\frac{1}{\ep^3} \biggl(
                                \frac{310}{27}
                                +\frac{8}{3} \pi^2
                        \biggr)
                        -\frac{1}{\ep^2} \biggl(
                                \frac{457}{27}
                                -\frac{38}{27} \pi^2
                                -\frac{160}{9} \zeta_3
                        \biggr)
  \notag \\ &
                        -\frac{1}{\ep} \biggl(
                                \frac{4664}{81}
                                -\frac{32}{27} \pi^2
                                +\frac{208}{27} \zeta_3
                                -\frac{164}{405} \pi^4
                        \biggr)
                \biggr]
        \Biggr]
  \notag \\ &
        +C_F T_F^2 N_F^2 \Biggl[
                \mathcal{L}_0\left(\frac{t}{\mu^2}\right) \biggl[
                        -\frac{64}{27 \ep^3}
                        +\frac{320}{81 \ep^2}
                        +\frac{64}{81 \ep}
                \biggr]
                +\delta(t) \biggl[
                        \frac{352}{81 \ep^4}
  \notag \\ &
                        -\frac{368}{243 \ep^3}
                        -\frac{1}{\ep^2} \biggl(
                                \frac{344}{81}
                                +\frac{16}{81} \pi^2
                        \biggr)
                        -\frac{1}{\ep} \biggl(
                                \frac{13828}{2187}
                                -\frac{80}{243} \pi^2
                                -\frac{256}{81} \zeta_3
                        \biggr)
                \biggr]
        \Biggr]
  \,.
\end{align}

For gluons ($i=g$) we find
\begin{align}
  Z_g^{(1)} ={}&
        C_A \Biggl[
                -\mathcal{L}_0\left(\frac{t}{\mu^2}\right) \frac{4}{\ep}
                +\delta(t) \biggl[
                        \frac{4}{\ep^2}
                        +\frac{11}{3 \ep}
                \biggr]
        \Biggr]
        -N_F T_F\delta(t) \frac{4}{3 \ep}
  \,, \\
  Z_g^{(2)} ={}&
        C_A^2 \Biggl[
                \mathcal{L}_1\left(\frac{t}{\mu^2}\right) \frac{16}{\ep^2}
                +\mathcal{L}_0\left(\frac{t}{\mu^2}\right) \biggl[
                        -\frac{16}{\ep^3}
                        -\frac{22}{3 \ep^2}
                        +\frac{1}{\ep} \biggl(
                                -\frac{134}{9}
                                +\frac{2}{3} \pi^2
                        \biggr)
                \biggr]
  \notag \\ &
                +\delta(t) \biggl[
                        \frac{8}{\ep^4}
                        +\frac{11}{3 \ep^3}
                        +\frac{1}{\ep^2} \biggl(
                                \frac{67}{9}
                                -\frac{5}{3} \pi^2
                        \biggr)
                        +\frac{1}{\ep} \biggl(
                                \frac{548}{27}
                                -\frac{11}{18} \pi^2
                                -8 \zeta_3
                        \biggr)
                \biggr]
        \Biggr]
  \notag \\ &
        +C_A N_F T_F \Biggl[
                \mathcal{L}_0\left(\frac{t}{\mu^2}\right) \biggl[
                        \frac{8}{3\ep^2}
                        +\frac{40}{9 \ep}
                \biggr]
                +\delta(t) \biggl[
                        -\frac{4}{3 \ep^3}
                        -\frac{20}{9 \ep^2}
  \notag \\ &
                        +\frac{1}{\ep} \biggl(
                                -\frac{184}{27}
                                +\frac{2}{9} \pi^2
                        \biggr)
                \biggr]
        \Biggr]
        -C_F N_F T_F\delta(t) \frac{2}{\ep}
  \,,
\end{align}
and
\begin{align}
  Z_g^{(3)} ={}&
        C_A^3 \Biggl[
                -\mathcal{L}_2\left(\frac{t}{\mu^2}\right) \frac{32}{\ep^3}
                +\mathcal{L}_1\left(\frac{t}{\mu^2}\right) \biggl[
                        \frac{64}{\ep^4}
                        +\frac{1}{\ep^2} \biggl(
                                \frac{1072}{9}
                                -\frac{16}{3} \pi^2
                        \biggr)
                \biggr]
  \notag \\ &
                +\mathcal{L}_0\left(\frac{t}{\mu^2}\right) \biggl[
                        -\frac{32}{\ep^5}
                        +\frac{44}{3 \ep^4}
                        -\frac{1}{\ep^3} \biggl(
                                \frac{2170}{27}
                                -\frac{28}{3} \pi^2
                        \biggr)
                        -\frac{1}{\ep^2} \biggl(
                                \frac{6826}{81}
  \notag \\ &
                                -\frac{88}{27} \pi^2
                                -32 \zeta_3
                        \biggr)
                        +\frac{1}{\ep} \biggl(
                                -\frac{490}{9}
                                +\frac{536}{81} \pi^2
                                -\frac{88}{9} \zeta_3
                                -\frac{44}{135} \pi^4
                        \biggr)
                \biggr]
  \notag \\ &
                +\delta(t) \biggl[
                        \frac{32}{3\ep^6}
                        -\frac{44}{3 \ep^5}
                        +\frac{1}{\ep^4} \biggl(
                                \frac{1807}{81}
                                -\frac{20}{3} \pi^2
                        \biggr)
                        +\frac{1}{\ep^3} \biggl(
                                \frac{14095}{243}
                                -\frac{187}{81} \pi^2
  \notag \\ &
                                -\frac{160}{3} \zeta_3
                        \biggr)
                        +\frac{1}{\ep^2} \biggl(
                                \frac{7072}{243}
                                -\frac{6259}{486} \pi^2
                                -\frac{176}{27} \zeta_3
                                +\frac{224}{405} \pi^4
                        \biggr)
  \notag \\ &
                        +\frac{1}{\ep} \biggl(
                                \frac{331153}{4374}
                                -\frac{6217}{729} \pi^2
                                -\frac{260}{3} \zeta_3
                                +\frac{583}{810} \pi^4
                                +\frac{64}{27} \pi^2 \zeta_3
                                +\frac{112}{3} \zeta_5
                        \biggr)
                \biggr]
        \Biggr]
  \notag \\ &
        +C_A^2 T_F N_F \Biggl[
                -\mathcal{L}_1\left(\frac{t}{\mu^2}\right) \frac{320}{9 \ep^2}
                +\mathcal{L}_0\left(\frac{t}{\mu^2}\right) \biggl[
                        -\frac{16}{3 \ep^4}
                        +\frac{544}{27 \ep^3}
  \notag \\ &
                        +\frac{1}{\ep^2} \biggl(
                                \frac{2464}{81}
                                -\frac{32}{27} \pi^2
                        \biggr)
                        +\frac{1}{\ep} \biggl(
                                \frac{1672}{81}
                                -\frac{160}{81} \pi^2
                                +\frac{224}{9} \zeta_3
                        \biggr)
                \biggr]
  \notag \\ &
                +\delta(t) \biggl[
                       \frac{16}{3\ep^5}
                        -\frac{280}{81 \ep^4}
                        +\frac{1}{\ep^3} \biggl(
                                -\frac{3256}{243}
                                +\frac{68}{81} \pi^2
                        \biggr)
                        -\frac{1}{\ep^2} \biggl(
                                \frac{2684}{243}
  \notag \\ &
                                -\frac{1012}{243} \pi^2
                                +\frac{128}{27} \zeta_3
                        \biggr)
                        +\frac{1}{\ep} \biggl(
                                -\frac{42557}{2187}
                                +\frac{2612}{729} \pi^2
                                +\frac{16}{81} \zeta_3
                                -\frac{154}{405} \pi^4
                        \biggr)
                \biggr]
        \Biggr]
  \notag \\ &
        +C_A C_F T_F N_F \Biggl[
                \mathcal{L}_0\left(\frac{t}{\mu^2}\right) \biggl[
                        \frac{8}{3\ep^2}
                        +\frac{1}{\ep} \biggl(
                                \frac{220}{9}
                                -\frac{64}{3} \zeta_3
                        \biggr)
                \biggr]
                +\delta(t) \biggl[
                        -\frac{1}{\ep^3} \frac{8}{9}
  \notag \\ &
                        +\frac{1}{\ep^2} \biggl(
                                -\frac{154}{27}
                                +\frac{64}{9} \zeta_3
                        \biggr)
                        +\frac{1}{\ep} \biggl(
                                -\frac{4145}{81}
                                +\frac{4}{9} \pi^2
                                +\frac{608}{27} \zeta_3
                                +\frac{16}{135} \pi^4
                        \biggr)
                \biggr]
        \Biggr]
  \notag \\ &
        +C_F^2 T_F N_F \delta(t) \frac{1}{\ep} \frac{2}{3}
        +C_A T_F^2 N_F^2 \Biggl[
                \mathcal{L}_0\left(\frac{t}{\mu^2}\right) \biggl[
                       \frac{32}{27 \ep^3}
                        -\frac{160}{81 \ep^2}
                        +\frac{64}{81 \ep}
                \biggr]
  \notag \\ &
                +\delta(t) \biggl[
                        -\frac{80}{81 \ep^4}
                        -\frac{80}{243 \ep^3}
                        -\frac{1}{\ep^2} \biggl(
                                \frac{16}{81}
                                +\frac{8}{81} \pi^2
                        \biggr)
                        -\frac{1}{\ep} \biggl(
                                \frac{3622}{2187}
                                +\frac{80}{243} \pi^2
  \notag \\ &
                                -\frac{448}{81} \zeta_3
                        \biggr)
                \biggr]
        \Biggr]
        +C_F N_F^2 T_F^2\delta(t) \biggl[
                -\frac{8}{9 \ep^2}
                +\frac{44}{27 \ep}
        \biggr].
\end{align}
The NLO and NNLO coefficients agree with Ref.~\cite{Ritzmann:2014mka}.

\bibliographystyle{utphys}
\bibliography{paper}{}

\providecommand{\href}[2]{#2}\begingroup\raggedright\begin{thebibliography}{100}

\bibitem{Ebert:2020unb}
M.~A. Ebert, B.~Mistlberger, and G.~Vita, ``{$N$-jettiness beam functions at
  N$^{3}$LO},'' \href{http://dx.doi.org/10.1007/JHEP09(2020)143}{{\em JHEP}
  {\bfseries 09} (2020) 143}, \href{http://arxiv.org/abs/2006.03056}{{\ttfamily
  arXiv:2006.03056 [hep-ph]}}.

\bibitem{Catani:1996vz}
S.~Catani and M.~H. Seymour, ``{A General algorithm for calculating jet
  cross-sections in NLO QCD},''
  \href{http://dx.doi.org/10.1016/S0550-3213(96)00589-5}{{\em Nucl. Phys. B}
  {\bfseries 485} (1997) 291--419},
  \href{http://arxiv.org/abs/hep-ph/9605323}{{\ttfamily arXiv:hep-ph/9605323}}.
  [Erratum: Nucl. Phys. B \textbf{510} (1998) 503].

\bibitem{Frixione:1995ms}
S.~Frixione, Z.~Kunszt, and A.~Signer, ``{Three jet cross-sections to
  next-to-leading order},''
  \href{http://dx.doi.org/10.1016/0550-3213(96)00110-1}{{\em Nucl. Phys. B}
  {\bfseries 467} (1996) 399--442},
  \href{http://arxiv.org/abs/hep-ph/9512328}{{\ttfamily arXiv:hep-ph/9512328}}.

\bibitem{Somogyi:2005xz}
G.~Somogyi, Z.~Trocsanyi, and V.~Del~Duca, ``{Matching of singly- and
  doubly-unresolved limits of tree-level QCD squared matrix elements},''
  \href{http://dx.doi.org/10.1088/1126-6708/2005/06/024}{{\em JHEP} {\bfseries
  06} (2005) 024}, \href{http://arxiv.org/abs/hep-ph/0502226}{{\ttfamily
  arXiv:hep-ph/0502226}}.

\bibitem{Gehrmann-DeRidder:2005btv}
A.~Gehrmann-De~Ridder, T.~Gehrmann, and E.~W.~N. Glover, ``{Antenna subtraction
  at NNLO},'' \href{http://dx.doi.org/10.1088/1126-6708/2005/09/056}{{\em JHEP}
  {\bfseries 09} (2005) 056},
  \href{http://arxiv.org/abs/hep-ph/0505111}{{\ttfamily arXiv:hep-ph/0505111}}.

\bibitem{Czakon:2010td}
M.~Czakon, ``{A novel subtraction scheme for double-real radiation at NNLO},''
  \href{http://dx.doi.org/10.1016/j.physletb.2010.08.036}{{\em Phys. Lett. B}
  {\bfseries 693} (2010) 259--268},
  \href{http://arxiv.org/abs/1005.0274}{{\ttfamily arXiv:1005.0274 [hep-ph]}}.

\bibitem{Caola:2017dug}
F.~Caola, K.~Melnikov, and R.~R\"ontsch, ``{Nested soft-collinear subtractions
  in NNLO QCD computations},''
  \href{http://dx.doi.org/10.1140/epjc/s10052-017-4774-0}{{\em Eur. Phys. J. C}
  {\bfseries 77} no.~4, (2017) 248},
  \href{http://arxiv.org/abs/1702.01352}{{\ttfamily arXiv:1702.01352
  [hep-ph]}}.

\bibitem{Herzog:2018ily}
F.~Herzog, ``{Geometric IR subtraction for final state real radiation},''
  \href{http://dx.doi.org/10.1007/JHEP08(2018)006}{{\em JHEP} {\bfseries 08}
  (2018) 006}, \href{http://arxiv.org/abs/1804.07949}{{\ttfamily
  arXiv:1804.07949 [hep-ph]}}.

\bibitem{Magnea:2018hab}
L.~Magnea, E.~Maina, G.~Pelliccioli, C.~Signorile-Signorile, P.~Torrielli, and
  S.~Uccirati, ``{Local analytic sector subtraction at NNLO},''
  \href{http://dx.doi.org/10.1007/JHEP12(2018)107}{{\em JHEP} {\bfseries 12}
  (2018) 107}, \href{http://arxiv.org/abs/1806.09570}{{\ttfamily
  arXiv:1806.09570 [hep-ph]}}. [Erratum: JHEP \textbf{06} (2019) 013].

\bibitem{Catani:2007vq}
S.~Catani and M.~Grazzini, ``{An NNLO subtraction formalism in hadron
  collisions and its application to Higgs boson production at the LHC},''
  \href{http://dx.doi.org/10.1103/PhysRevLett.98.222002}{{\em Phys. Rev. Lett.}
  {\bfseries 98} (2007) 222002},
  \href{http://arxiv.org/abs/hep-ph/0703012}{{\ttfamily arXiv:hep-ph/0703012}}.

\bibitem{Stewart:2009yx}
I.~W. Stewart, F.~J. Tackmann, and W.~J. Waalewijn, ``{Factorization at the
  LHC: From PDFs to Initial State Jets},''
  \href{http://dx.doi.org/10.1103/PhysRevD.81.094035}{{\em Phys. Rev. D}
  {\bfseries 81} (2010) 094035},
  \href{http://arxiv.org/abs/0910.0467}{{\ttfamily arXiv:0910.0467 [hep-ph]}}.

\bibitem{Stewart:2010tn}
I.~W. Stewart, F.~J. Tackmann, and W.~J. Waalewijn, ``{N-Jettiness: An
  Inclusive Event Shape to Veto Jets},''
  \href{http://dx.doi.org/10.1103/PhysRevLett.105.092002}{{\em Phys. Rev.
  Lett.} {\bfseries 105} (2010) 092002},
  \href{http://arxiv.org/abs/1004.2489}{{\ttfamily arXiv:1004.2489 [hep-ph]}}.

\bibitem{Bauer:2001ct}
C.~W. Bauer and I.~W. Stewart, ``{Invariant operators in collinear effective
  theory},'' \href{http://dx.doi.org/10.1016/S0370-2693(01)00902-9}{{\em Phys.
  Lett. B} {\bfseries 516} (2001) 134--142},
  \href{http://arxiv.org/abs/hep-ph/0107001}{{\ttfamily arXiv:hep-ph/0107001}}.

\bibitem{Bauer:2001yt}
C.~W. Bauer, D.~Pirjol, and I.~W. Stewart, ``{Soft collinear factorization in
  effective field theory},''
  \href{http://dx.doi.org/10.1103/PhysRevD.65.054022}{{\em Phys. Rev. D}
  {\bfseries 65} (2002) 054022},
  \href{http://arxiv.org/abs/hep-ph/0109045}{{\ttfamily arXiv:hep-ph/0109045}}.

\bibitem{Bauer:2002nz}
C.~W. Bauer, S.~Fleming, D.~Pirjol, I.~Z. Rothstein, and I.~W. Stewart, ``{Hard
  scattering factorization from effective field theory},''
  \href{http://dx.doi.org/10.1103/PhysRevD.66.014017}{{\em Phys. Rev. D}
  {\bfseries 66} (2002) 014017},
  \href{http://arxiv.org/abs/hep-ph/0202088}{{\ttfamily arXiv:hep-ph/0202088}}.

\bibitem{Beneke:2002ph}
M.~Beneke, A.~P. Chapovsky, M.~Diehl, and T.~Feldmann, ``{Soft collinear
  effective theory and heavy to light currents beyond leading power},''
  \href{http://dx.doi.org/10.1016/S0550-3213(02)00687-9}{{\em Nucl. Phys. B}
  {\bfseries 643} (2002) 431--476},
  \href{http://arxiv.org/abs/hep-ph/0206152}{{\ttfamily arXiv:hep-ph/0206152}}.

\bibitem{Beneke:2002ni}
M.~Beneke and T.~Feldmann, ``{Multipole expanded soft collinear effective
  theory with non-abelian gauge symmetry},''
  \href{http://dx.doi.org/10.1016/S0370-2693(02)03204-5}{{\em Phys. Lett. B}
  {\bfseries 553} (2003) 267--276},
  \href{http://arxiv.org/abs/hep-ph/0211358}{{\ttfamily arXiv:hep-ph/0211358}}.

\bibitem{Stewart:2010qs}
I.~W. Stewart, F.~J. Tackmann, and W.~J. Waalewijn, ``{The Quark Beam Function
  at NNLL},'' \href{http://dx.doi.org/10.1007/JHEP09(2010)005}{{\em JHEP}
  {\bfseries 09} (2010) 005}, \href{http://arxiv.org/abs/1002.2213}{{\ttfamily
  arXiv:1002.2213 [hep-ph]}}.

\bibitem{Gaunt:2014cfa}
J.~Gaunt, M.~Stahlhofen, and F.~J. Tackmann, ``{The Gluon Beam Function at Two
  Loops},'' \href{http://dx.doi.org/10.1007/JHEP08(2014)020}{{\em JHEP}
  {\bfseries 08} (2014) 020}, \href{http://arxiv.org/abs/1405.1044}{{\ttfamily
  arXiv:1405.1044 [hep-ph]}}.

\bibitem{Gaunt:2014xga}
J.~R. Gaunt, M.~Stahlhofen, and F.~J. Tackmann, ``{The Quark Beam Function at
  Two Loops},'' \href{http://dx.doi.org/10.1007/JHEP04(2014)113}{{\em JHEP}
  {\bfseries 04} (2014) 113}, \href{http://arxiv.org/abs/1401.5478}{{\ttfamily
  arXiv:1401.5478 [hep-ph]}}.

\bibitem{Boughezal:2017tdd}
R.~Boughezal, F.~Petriello, U.~Schubert, and H.~Xing, ``{Spin-dependent quark
  beam function at NNLO},''
  \href{http://dx.doi.org/10.1103/PhysRevD.96.034001}{{\em Phys. Rev. D}
  {\bfseries 96} no.~3, (2017) 034001},
  \href{http://arxiv.org/abs/1704.05457}{{\ttfamily arXiv:1704.05457
  [hep-ph]}}.

\bibitem{Melnikov:2019pdm}
K.~Melnikov, R.~Rietkerk, L.~Tancredi, and C.~Wever, ``{Triple-real
  contribution to the quark beam function in QCD at
  next-to-next-to-next-to-leading order},''
  \href{http://dx.doi.org/10.1007/JHEP06(2019)033}{{\em JHEP} {\bfseries 06}
  (2019) 033}, \href{http://arxiv.org/abs/1904.02433}{{\ttfamily
  arXiv:1904.02433 [hep-ph]}}.

\bibitem{Melnikov:2018jxb}
K.~Melnikov, R.~Rietkerk, L.~Tancredi, and C.~Wever, ``{Double-real
  contribution to the quark beam function at N$^{3}$LO QCD},''
  \href{http://dx.doi.org/10.1007/JHEP02(2019)159}{{\em JHEP} {\bfseries 02}
  (2019) 159}, \href{http://arxiv.org/abs/1809.06300}{{\ttfamily
  arXiv:1809.06300 [hep-ph]}}.

\bibitem{Behring:2019quf}
A.~Behring, K.~Melnikov, R.~Rietkerk, L.~Tancredi, and C.~Wever, ``{Quark beam
  function at next-to-next-to-next-to-leading order in perturbative QCD in the
  generalized large-$N_c$ approximation},''
  \href{http://dx.doi.org/10.1103/PhysRevD.100.114034}{{\em Phys. Rev. D}
  {\bfseries 100} no.~11, (2019) 114034},
  \href{http://arxiv.org/abs/1910.10059}{{\ttfamily arXiv:1910.10059
  [hep-ph]}}.

\bibitem{Ebert:2020lxs}
M.~A. Ebert, B.~Mistlberger, and G.~Vita, ``{Collinear expansion for color
  singlet cross sections},''
  \href{http://dx.doi.org/10.1007/JHEP09(2020)181}{{\em JHEP} {\bfseries 09}
  (2020) 181}, \href{http://arxiv.org/abs/2006.03055}{{\ttfamily
  arXiv:2006.03055 [hep-ph]}}.

\bibitem{Ritzmann:2014mka}
M.~Ritzmann and W.~J. Waalewijn, ``{Fragmentation in Jets at NNLO},''
  \href{http://dx.doi.org/10.1103/PhysRevD.90.054029}{{\em Phys. Rev. D}
  {\bfseries 90} no.~5, (2014) 054029},
  \href{http://arxiv.org/abs/1407.3272}{{\ttfamily arXiv:1407.3272 [hep-ph]}}.

\bibitem{Catani:2022sgr}
S.~Catani and P.~K. Dhani, ``{Collinear functions for QCD resummations},''
  \href{http://arxiv.org/abs/2208.05840}{{\ttfamily arXiv:2208.05840
  [hep-ph]}}.

\bibitem{Catani:1999ss}
S.~Catani and M.~Grazzini, ``{Infrared factorization of tree level QCD
  amplitudes at the next-to-next-to-leading order and beyond},''
  \href{http://dx.doi.org/10.1016/S0550-3213(99)00778-6}{{\em Nucl. Phys. B}
  {\bfseries 570} (2000) 287--325},
  \href{http://arxiv.org/abs/hep-ph/9908523}{{\ttfamily arXiv:hep-ph/9908523}}.

\bibitem{Bern:2004cz}
Z.~Bern, L.~J. Dixon, and D.~A. Kosower, ``{Two-loop $g \to gg$ splitting
  amplitudes in QCD},''
  \href{http://dx.doi.org/10.1088/1126-6708/2004/08/012}{{\em JHEP} {\bfseries
  08} (2004) 012}, \href{http://arxiv.org/abs/hep-ph/0404293}{{\ttfamily
  arXiv:hep-ph/0404293}}.

\bibitem{Badger:2004uk}
S.~D. Badger and E.~W.~N. Glover, ``{Two loop splitting functions in QCD},''
  \href{http://dx.doi.org/10.1088/1126-6708/2004/07/040}{{\em JHEP} {\bfseries
  07} (2004) 040}, \href{http://arxiv.org/abs/hep-ph/0405236}{{\ttfamily
  arXiv:hep-ph/0405236}}.

\bibitem{DelDuca:2019ggv}
V.~Del~Duca, C.~Duhr, R.~Haindl, A.~Lazopoulos, and M.~Michel, ``{Tree-level
  splitting amplitudes for a quark into four collinear partons},''
  \href{http://dx.doi.org/10.1007/JHEP02(2020)189}{{\em JHEP} {\bfseries 02}
  (2020) 189}, \href{http://arxiv.org/abs/1912.06425}{{\ttfamily
  arXiv:1912.06425 [hep-ph]}}.

\bibitem{DelDuca:2020vst}
V.~Del~Duca, C.~Duhr, R.~Haindl, A.~Lazopoulos, and M.~Michel, ``{Tree-level
  splitting amplitudes for a gluon into four collinear partons},''
  \href{http://dx.doi.org/10.1007/JHEP10(2020)093}{{\em JHEP} {\bfseries 10}
  (2020) 093}, \href{http://arxiv.org/abs/2007.05345}{{\ttfamily
  arXiv:2007.05345 [hep-ph]}}.

\bibitem{Catani:2003vu}
S.~Catani, D.~de~Florian, and G.~Rodrigo, ``{The Triple collinear limit of one
  loop QCD amplitudes},''
  \href{http://dx.doi.org/10.1016/j.physletb.2004.02.039}{{\em Phys. Lett. B}
  {\bfseries 586} (2004) 323--331},
  \href{http://arxiv.org/abs/hep-ph/0312067}{{\ttfamily arXiv:hep-ph/0312067}}.

\bibitem{Badger:2015cxa}
S.~Badger, F.~Buciuni, and T.~Peraro, ``{One-loop triple collinear splitting
  amplitudes in QCD},'' \href{http://dx.doi.org/10.1007/JHEP09(2015)188}{{\em
  JHEP} {\bfseries 09} (2015) 188},
  \href{http://arxiv.org/abs/1507.05070}{{\ttfamily arXiv:1507.05070
  [hep-ph]}}.

\bibitem{Sborlini:2014mpa}
G.~F.~R. Sborlini, D.~de~Florian, and G.~Rodrigo, ``{Triple collinear splitting
  functions at NLO for scattering processes with photons},''
  \href{http://dx.doi.org/10.1007/JHEP10(2014)161}{{\em JHEP} {\bfseries 10}
  (2014) 161}, \href{http://arxiv.org/abs/1408.4821}{{\ttfamily arXiv:1408.4821
  [hep-ph]}}.

\bibitem{Czakon:2022fqi}
M.~Czakon and S.~Sapeta, ``{Complete collection of one-loop triple-collinear
  splitting operators for dimensionally-regulated QCD},''
  \href{http://dx.doi.org/10.1007/JHEP07(2022)052}{{\em JHEP} {\bfseries 07}
  (2022) 052}, \href{http://arxiv.org/abs/2204.11801}{{\ttfamily
  arXiv:2204.11801 [hep-ph]}}.

\bibitem{Anastasiou:2002yz}
C.~Anastasiou and K.~Melnikov, ``{Higgs boson production at hadron colliders in
  NNLO QCD},'' \href{http://dx.doi.org/10.1016/S0550-3213(02)00837-4}{{\em
  Nucl. Phys. B} {\bfseries 646} (2002) 220--256},
  \href{http://arxiv.org/abs/hep-ph/0207004}{{\ttfamily arXiv:hep-ph/0207004}}.

\bibitem{Tkachov:1981wb}
F.~V. Tkachov, ``{A Theorem on Analytical Calculability of Four Loop
  Renormalization Group Functions},''
  \href{http://dx.doi.org/10.1016/0370-2693(81)90288-4}{{\em Phys. Lett. B}
  {\bfseries 100} (1981) 65--68}.

\bibitem{Chetyrkin:1981qh}
K.~G. Chetyrkin and F.~V. Tkachov, ``{Integration by Parts: The Algorithm to
  Calculate beta Functions in 4 Loops},''
  \href{http://dx.doi.org/10.1016/0550-3213(81)90199-1}{{\em Nucl. Phys. B}
  {\bfseries 192} (1981) 159--204}.

\bibitem{Kotikov:1990kg}
A.~V. Kotikov, ``{Differential equations method: New technique for massive
  Feynman diagrams calculation},''
  \href{http://dx.doi.org/10.1016/0370-2693(91)90413-K}{{\em Phys. Lett. B}
  {\bfseries 254} (1991) 158--164}.

\bibitem{Bern:1993kr}
Z.~Bern, L.~J. Dixon, and D.~A. Kosower, ``{Dimensionally regulated pentagon
  integrals},'' \href{http://dx.doi.org/10.1016/0550-3213(94)90398-0}{{\em
  Nucl. Phys. B} {\bfseries 412} (1994) 751--816},
  \href{http://arxiv.org/abs/hep-ph/9306240}{{\ttfamily arXiv:hep-ph/9306240}}.

\bibitem{Remiddi:1997ny}
E.~Remiddi, ``{Differential equations for Feynman graph amplitudes},''
  \href{http://dx.doi.org/10.1007/BF03185566}{{\em Nuovo Cim. A} {\bfseries
  110} (1997) 1435--1452},
  \href{http://arxiv.org/abs/hep-th/9711188}{{\ttfamily arXiv:hep-th/9711188}}.

\bibitem{Gehrmann:1999as}
T.~Gehrmann and E.~Remiddi, ``{Differential equations for two loop four point
  functions},'' \href{http://dx.doi.org/10.1016/S0550-3213(00)00223-6}{{\em
  Nucl. Phys. B} {\bfseries 580} (2000) 485--518},
  \href{http://arxiv.org/abs/hep-ph/9912329}{{\ttfamily arXiv:hep-ph/9912329}}.

\bibitem{Nogueira:1991ex}
P.~Nogueira, ``{Automatic Feynman graph generation},''
\href{http://dx.doi.org/10.1006/jcph.1993.1074}{{\em J. Comput. Phys.}
  {\bfseries 105} (1993) 279--289}.

\bibitem{Vermaseren:2000nd}
J.~Vermaseren, ``{New features of FORM},''
  \href{http://arxiv.org/abs/math-ph/0010025}{{\ttfamily
  arXiv:math-ph/0010025}}.

\bibitem{Kuipers:2012rf}
J.~Kuipers, T.~Ueda, J.~Vermaseren, and J.~Vollinga, ``{FORM version 4.0}''
  \href{http://dx.doi.org/10.1016/j.cpc.2012.12.028}{{\em Comput. Phys.
  Commun.} {\bfseries 184} (2013) 1453--1467},
  \href{http://arxiv.org/abs/1203.6543}{{\ttfamily arXiv:1203.6543 [cs.SC]}}.

\bibitem{Kuipers:2013pba}
J.~Kuipers, T.~Ueda, and J.~Vermaseren, ``{Code Optimization in FORM},''
  \href{http://dx.doi.org/10.1016/j.cpc.2014.08.008}{{\em Comput. Phys.
  Commun.} {\bfseries 189} (2015) 1--19},
  \href{http://arxiv.org/abs/1310.7007}{{\ttfamily arXiv:1310.7007 [cs.SC]}}.

\bibitem{Ruijl:2017dtg}
B.~Ruijl, T.~Ueda, and J.~Vermaseren, ``{FORM version 4.2}''
  \href{http://arxiv.org/abs/1707.06453}{{\ttfamily arXiv:1707.06453
  [hep-ph]}}.

\bibitem{vanRitbergen:1998pn}
T.~van Ritbergen, A.~Schellekens, and J.~Vermaseren, ``{Group theory factors
  for Feynman diagrams},''
  \href{http://dx.doi.org/10.1142/S0217751X99000038}{{\em Int. J. Mod. Phys. A}
  {\bfseries 14} (1999) 41--96},
  \href{http://arxiv.org/abs/hep-ph/9802376}{{\ttfamily arXiv:hep-ph/9802376}}.

\bibitem{vonManteuffel:2012np}
A.~von Manteuffel and C.~Studerus, ``{Reduze 2 - Distributed Feynman Integral
  Reduction},'' \href{http://arxiv.org/abs/1201.4330}{{\ttfamily
  arXiv:1201.4330 [hep-ph]}}.

\bibitem{Maierhofer:2017gsa}
P.~Maierh\"ofer, J.~Usovitsch, and P.~Uwer, ``{Kira\textemdash{}A Feynman
  integral reduction program},''
  \href{http://dx.doi.org/10.1016/j.cpc.2018.04.012}{{\em Comput. Phys.
  Commun.} {\bfseries 230} (2018) 99--112},
  \href{http://arxiv.org/abs/1705.05610}{{\ttfamily arXiv:1705.05610
  [hep-ph]}}.

\bibitem{Maierhofer:2018gpa}
P.~Maierh\"ofer and J.~Usovitsch, ``{Kira 1.2 Release Notes},''
  \href{http://arxiv.org/abs/1812.01491}{{\ttfamily arXiv:1812.01491
  [hep-ph]}}.

\bibitem{Maierhofer:2019goc}
P.~Maierh\"ofer and J.~Usovitsch, ``{Recent developments in Kira},''
  \href{http://dx.doi.org/10.23731/CYRM-2020-003.201}{{\em CERN Yellow Reports:
  Monographs} {\bfseries 3} (2020) 201--204}.

\bibitem{Klappert:2020nbg}
J.~Klappert, F.~Lange, P.~Maierh\"ofer, and J.~Usovitsch, ``{Integral reduction
  with Kira 2.0 and finite field methods},''
  \href{http://dx.doi.org/10.1016/j.cpc.2021.108024}{{\em Comput. Phys.
  Commun.} {\bfseries 266} (2021) 108024},
  \href{http://arxiv.org/abs/2008.06494}{{\ttfamily arXiv:2008.06494
  [hep-ph]}}.

\bibitem{Pak:2011xt}
A.~Pak, ``{The Toolbox of modern multi-loop calculations: novel analytic and
  semi-analytic techniques},''
  \href{http://dx.doi.org/10.1088/1742-6596/368/1/012049}{{\em J. Phys. Conf.
  Ser.} {\bfseries 368} (2012) 012049},
  \href{http://arxiv.org/abs/1111.0868}{{\ttfamily arXiv:1111.0868 [hep-ph]}}.

\bibitem{Hoff:2015kub}
J.~S. Hoff, \href{http://dx.doi.org/10.5445/IR/1000047447}{{\em {Methods for
  multiloop calculations and Higgs boson production at the LHC}}}.
\newblock PhD thesis, KIT, Karlsruhe, 2015.

\bibitem{Meyer:2017joq}
C.~Meyer, ``{Algorithmic transformation of multi-loop master integrals to a
  canonical basis with CANONICA},''
  \href{http://dx.doi.org/10.1016/j.cpc.2017.09.014}{{\em Comput. Phys.
  Commun.} {\bfseries 222} (2018) 295--312},
  \href{http://arxiv.org/abs/1705.06252}{{\ttfamily arXiv:1705.06252
  [hep-ph]}}.

\bibitem{Abreu:2019odu}
S.~Abreu, J.~Dormans, F.~Febres~Cordero, H.~Ita, B.~Page, and V.~Sotnikov,
  ``{Analytic Form of the Planar Two-Loop Five-Parton Scattering Amplitudes in
  QCD},'' \href{http://dx.doi.org/10.1007/JHEP05(2019)084}{{\em JHEP}
  {\bfseries 05} (2019) 084}, \href{http://arxiv.org/abs/1904.00945}{{\ttfamily
  arXiv:1904.00945 [hep-ph]}}.

\bibitem{Heller:2021qkz}
M.~Heller and A.~von Manteuffel, ``{MultivariateApart: Generalized partial
  fractions},'' \href{http://dx.doi.org/10.1016/j.cpc.2021.108174}{{\em Comput.
  Phys. Commun.} {\bfseries 271} (2022) 108174},
  \href{http://arxiv.org/abs/2101.08283}{{\ttfamily arXiv:2101.08283 [cs.SC]}}.

\bibitem{Henn:2013pwa}
J.~M. Henn, ``{Multiloop integrals in dimensional regularization made
  simple},'' \href{http://dx.doi.org/10.1103/PhysRevLett.110.251601}{{\em Phys.
  Rev. Lett.} {\bfseries 110} (2013) 251601},
  \href{http://arxiv.org/abs/1304.1806}{{\ttfamily arXiv:1304.1806 [hep-th]}}.

\bibitem{Chen:1977oja}
K.-T. Chen, ``{Iterated path integrals},''
  \href{http://dx.doi.org/10.1090/S0002-9904-1977-14320-6}{{\em Bull. Am. Math.
  Soc.} {\bfseries 83} (1977) 831--879}.

\bibitem{Gehrmann:2014bfa}
T.~Gehrmann, A.~von Manteuffel, L.~Tancredi, and E.~Weihs, ``{The two-loop
  master integrals for $q\overline{q} \to VV$},''
  \href{http://dx.doi.org/10.1007/JHEP06(2014)032}{{\em JHEP} {\bfseries 06}
  (2014) 032}, \href{http://arxiv.org/abs/1404.4853}{{\ttfamily arXiv:1404.4853
  [hep-ph]}}.

\bibitem{Lee:2014ioa}
R.~N. Lee, ``{Reducing differential equations for multiloop master
  integrals},'' \href{http://dx.doi.org/10.1007/JHEP04(2015)108}{{\em JHEP}
  {\bfseries 04} (2015) 108}, \href{http://arxiv.org/abs/1411.0911}{{\ttfamily
  arXiv:1411.0911 [hep-ph]}}.

\bibitem{Argeri:2014qva}
M.~Argeri, S.~Di~Vita, P.~Mastrolia, E.~Mirabella, J.~Schlenk, U.~Schubert, and
  L.~Tancredi, ``{Magnus and Dyson Series for Master Integrals},''
  \href{http://dx.doi.org/10.1007/JHEP03(2014)082}{{\em JHEP} {\bfseries 03}
  (2014) 082}, \href{http://arxiv.org/abs/1401.2979}{{\ttfamily arXiv:1401.2979
  [hep-ph]}}.

\bibitem{Gituliar:2017vzm}
O.~Gituliar and V.~Magerya, ``{Fuchsia: a tool for reducing differential
  equations for Feynman master integrals to epsilon form},''
  \href{http://dx.doi.org/10.1016/j.cpc.2017.05.004}{{\em Comput. Phys.
  Commun.} {\bfseries 219} (2017) 329--338},
  \href{http://arxiv.org/abs/1701.04269}{{\ttfamily arXiv:1701.04269
  [hep-ph]}}.

\bibitem{Meyer:2016slj}
C.~Meyer, ``{Transforming differential equations of multi-loop Feynman
  integrals into canonical form},''
  \href{http://dx.doi.org/10.1007/JHEP04(2017)006}{{\em JHEP} {\bfseries 04}
  (2017) 006}, \href{http://arxiv.org/abs/1611.01087}{{\ttfamily
  arXiv:1611.01087 [hep-ph]}}.

\bibitem{Prausa:2017ltv}
M.~Prausa, ``{epsilon: A tool to find a canonical basis of master integrals},''
  \href{http://dx.doi.org/10.1016/j.cpc.2017.05.026}{{\em Comput. Phys.
  Commun.} {\bfseries 219} (2017) 361--376},
  \href{http://arxiv.org/abs/1701.00725}{{\ttfamily arXiv:1701.00725
  [hep-ph]}}.

\bibitem{Lee:2020zfb}
R.~N. Lee, ``{Libra: A package for transformation of differential systems for
  multiloop integrals},''
  \href{http://dx.doi.org/10.1016/j.cpc.2021.108058}{{\em Comput. Phys.
  Commun.} {\bfseries 267} (2021) 108058},
  \href{http://arxiv.org/abs/2012.00279}{{\ttfamily arXiv:2012.00279
  [hep-ph]}}.

\bibitem{Arkani-Hamed:2010pyv}
N.~Arkani-Hamed, J.~L. Bourjaily, F.~Cachazo, and J.~Trnka, ``{Local Integrals
  for Planar Scattering Amplitudes},''
  \href{http://dx.doi.org/10.1007/JHEP06(2012)125}{{\em JHEP} {\bfseries 06}
  (2012) 125}, \href{http://arxiv.org/abs/1012.6032}{{\ttfamily arXiv:1012.6032
  [hep-th]}}.

\bibitem{Henn:2020lye}
J.~Henn, B.~Mistlberger, V.~A. Smirnov, and P.~Wasser, ``{Constructing d-log
  integrands and computing master integrals for three-loop four-particle
  scattering},'' \href{http://dx.doi.org/10.1007/JHEP04(2020)167}{{\em JHEP}
  {\bfseries 04} (2020) 167}, \href{http://arxiv.org/abs/2002.09492}{{\ttfamily
  arXiv:2002.09492 [hep-ph]}}.

\bibitem{Hoschele:2014qsa}
M.~H\"oschele, J.~Hoff, and T.~Ueda, ``{Adequate bases of phase space master
  integrals for gg $\to$ h at NNLO and beyond},''
  \href{http://dx.doi.org/10.1007/JHEP09(2014)116}{{\em JHEP} {\bfseries 09}
  (2014) 116}, \href{http://arxiv.org/abs/1407.4049}{{\ttfamily arXiv:1407.4049
  [hep-ph]}}.

\bibitem{Dlapa:2020cwj}
C.~Dlapa, J.~Henn, and K.~Yan, ``{Deriving canonical differential equations for
  Feynman integrals from a single uniform weight integral},''
  \href{http://dx.doi.org/10.1007/JHEP05(2020)025}{{\em JHEP} {\bfseries 05}
  (2020) 025}, \href{http://arxiv.org/abs/2002.02340}{{\ttfamily
  arXiv:2002.02340 [hep-ph]}}.

\bibitem{Chen:2020uyk}
J.~Chen, X.~Jiang, X.~Xu, and L.~L. Yang, ``{Constructing canonical Feynman
  integrals with intersection theory},''
  \href{http://dx.doi.org/10.1016/j.physletb.2021.136085}{{\em Phys. Lett. B}
  {\bfseries 814} (2021) 136085},
  \href{http://arxiv.org/abs/2008.03045}{{\ttfamily arXiv:2008.03045
  [hep-th]}}.

\bibitem{Chen:2022lzr}
J.~Chen, X.~Jiang, C.~Ma, X.~Xu, and L.~L. Yang, ``{Baikov representations,
  intersection theory, and canonical Feynman integrals},''
  \href{http://dx.doi.org/10.1007/JHEP07(2022)066}{{\em JHEP} {\bfseries 07}
  (2022) 066}, \href{http://arxiv.org/abs/2202.08127}{{\ttfamily
  arXiv:2202.08127 [hep-th]}}.

\bibitem{Melnikov:2016qoc}
K.~Melnikov, L.~Tancredi, and C.~Wever, ``{Two-loop $gg \to Hg$ amplitude
  mediated by a nearly massless quark},''
  \href{http://dx.doi.org/10.1007/JHEP11(2016)104}{{\em JHEP} {\bfseries 11}
  (2016) 104}, \href{http://arxiv.org/abs/1610.03747}{{\ttfamily
  arXiv:1610.03747 [hep-ph]}}.

\bibitem{Usovitsch:2020jrk}
J.~Usovitsch, ``{Factorization of denominators in integration-by-parts
  reductions},'' \href{http://arxiv.org/abs/2002.08173}{{\ttfamily
  arXiv:2002.08173 [hep-ph]}}.

\bibitem{Smirnov:2020quc}
A.~V. Smirnov and V.~A. Smirnov, ``{How to choose master integrals},''
  \href{http://dx.doi.org/10.1016/j.nuclphysb.2020.115213}{{\em Nucl. Phys. B}
  {\bfseries 960} (2020) 115213},
  \href{http://arxiv.org/abs/2002.08042}{{\ttfamily arXiv:2002.08042
  [hep-ph]}}.

\bibitem{Baikov:1996rk}
P.~A. Baikov, ``{Explicit solutions of the three loop vacuum integral
  recurrence relations},''
  \href{http://dx.doi.org/10.1016/0370-2693(96)00835-0}{{\em Phys. Lett. B}
  {\bfseries 385} (1996) 404--410},
  \href{http://arxiv.org/abs/hep-ph/9603267}{{\ttfamily arXiv:hep-ph/9603267}}.

\bibitem{Baikov:1996iu}
P.~A. Baikov, ``{Explicit solutions of the multiloop integral recurrence
  relations and its application},''
  \href{http://dx.doi.org/10.1016/S0168-9002(97)00126-5}{{\em Nucl. Instrum.
  Meth. A} {\bfseries 389} (1997) 347--349},
  \href{http://arxiv.org/abs/hep-ph/9611449}{{\ttfamily arXiv:hep-ph/9611449}}.

\bibitem{Goncharov:1998kja}
A.~B. Goncharov, ``{Multiple polylogarithms, cyclotomy and modular
  complexes},'' \href{http://dx.doi.org/10.4310/MRL.1998.v5.n4.a7}{{\em Math.
  Res. Lett.} {\bfseries 5} (1998) 497--516},
  \href{http://arxiv.org/abs/1105.2076}{{\ttfamily arXiv:1105.2076 [math.AG]}}.

\bibitem{Remiddi:1999ew}
E.~Remiddi and J.~A.~M. Vermaseren, ``{Harmonic polylogarithms},''
  \href{http://dx.doi.org/10.1142/S0217751X00000367}{{\em Int. J. Mod. Phys.}
  {\bfseries A15} (2000) 725--754},
\href{http://arxiv.org/abs/hep-ph/9905237}{{\ttfamily arXiv:hep-ph/9905237
  [hep-ph]}}.

\bibitem{Vollinga:2004sn}
J.~Vollinga and S.~Weinzierl, ``{Numerical evaluation of multiple
  polylogarithms},'' \href{http://dx.doi.org/10.1016/j.cpc.2004.12.009}{{\em
  Comput. Phys. Commun.} {\bfseries 167} (2005) 177},
  \href{http://arxiv.org/abs/hep-ph/0410259}{{\ttfamily arXiv:hep-ph/0410259}}.

\bibitem{Gehrmann:2000zt}
T.~Gehrmann and E.~Remiddi, ``{Two loop master integrals for $\gamma^* \to 3
  \text{~jets}$: The Planar topologies},''
  \href{http://dx.doi.org/10.1016/S0550-3213(01)00057-8}{{\em Nucl. Phys. B}
  {\bfseries 601} (2001) 248--286},
  \href{http://arxiv.org/abs/hep-ph/0008287}{{\ttfamily arXiv:hep-ph/0008287}}.

\bibitem{Aglietti:2004tq}
U.~Aglietti and R.~Bonciani, ``{Master integrals with 2 and 3 massive
  propagators for the 2 loop electroweak form-factor - planar case},''
  \href{http://dx.doi.org/10.1016/j.nuclphysb.2004.07.018}{{\em Nucl. Phys. B}
  {\bfseries 698} (2004) 277--318},
  \href{http://arxiv.org/abs/hep-ph/0401193}{{\ttfamily arXiv:hep-ph/0401193}}.

\bibitem{Weinzierl:2004bn}
S.~Weinzierl, ``{Expansion around half integer values, binomial sums and
  inverse binomial sums},'' \href{http://dx.doi.org/10.1063/1.1758319}{{\em J.
  Math. Phys.} {\bfseries 45} (2004) 2656--2673},
  \href{http://arxiv.org/abs/hep-ph/0402131}{{\ttfamily arXiv:hep-ph/0402131}}.

\bibitem{Ablinger:2011te}
J.~Ablinger, J.~Bl{\"u}mlein, and C.~Schneider, ``{Harmonic Sums and
  Polylogarithms Generated by Cyclotomic Polynomials},''
  \href{http://dx.doi.org/10.1063/1.3629472}{{\em J. Math. Phys.} {\bfseries
  52} (2011) 102301},
\href{http://arxiv.org/abs/1105.6063}{{\ttfamily arXiv:1105.6063 [math-ph]}}.

\bibitem{Bonciani:2010ms}
R.~Bonciani, G.~Degrassi, and A.~Vicini, ``{On the Generalized Harmonic
  Polylogarithms of One Complex Variable},''
  \href{http://dx.doi.org/10.1016/j.cpc.2011.02.011}{{\em Comput. Phys.
  Commun.} {\bfseries 182} (2011) 1253--1264},
  \href{http://arxiv.org/abs/1007.1891}{{\ttfamily arXiv:1007.1891 [hep-ph]}}.

\bibitem{Ablinger:2014bra}
J.~Ablinger, J.~Bl\"umlein, C.~G. Raab, and C.~Schneider, ``{Iterated Binomial
  Sums and their Associated Iterated Integrals},''
  \href{http://dx.doi.org/10.1063/1.4900836}{{\em J. Math. Phys.} {\bfseries
  55} (2014) 112301}, \href{http://arxiv.org/abs/1407.1822}{{\ttfamily
  arXiv:1407.1822 [hep-th]}}.

\bibitem{Ablinger:2021fnc}
J.~Ablinger, J.~Bl\"umlein, and C.~Schneider, ``{Iterated integrals over
  letters induced by quadratic forms},''
  \href{http://dx.doi.org/10.1103/PhysRevD.103.096025}{{\em Phys. Rev. D}
  {\bfseries 103} no.~9, (2021) 096025},
  \href{http://arxiv.org/abs/2103.08330}{{\ttfamily arXiv:2103.08330
  [hep-th]}}.

\bibitem{Ablinger:2010kw}
J.~Ablinger, {\em {A Computer Algebra Toolbox for Harmonic Sums Related to
  Particle Physics}}.
\newblock Diploma thesis, J. Kepler University Linz, 2009.
\newblock
\href{http://arxiv.org/abs/1011.1176}{{\ttfamily arXiv:1011.1176 [math-ph]}}.
\newblock

\bibitem{Ablinger:2013hcp}
J.~Ablinger, {\em {Computer Algebra Algorithms for Special Functions in
  Particle Physics}}.
\newblock PhD thesis, J. Kepler University Linz, 2012.
\newblock
\href{http://arxiv.org/abs/1305.0687}{{\ttfamily arXiv:1305.0687 [math-ph]}}.
\newblock

\bibitem{Vermaseren:1998uu}
J.~A.~M. Vermaseren, ``{Harmonic sums, Mellin transforms and integrals},''
  \href{http://dx.doi.org/10.1142/S0217751X99001032}{{\em Int. J. Mod. Phys.}
  {\bfseries A14} (1999) 2037--2076},
\href{http://arxiv.org/abs/hep-ph/9806280}{{\ttfamily arXiv:hep-ph/9806280
  [hep-ph]}}.

\bibitem{Blumlein:2009ta}
J.~Bl{\"u}mlein, ``{Structural Relations of Harmonic Sums and Mellin Transforms
  up to Weight $w = 5$},''
  \href{http://dx.doi.org/10.1016/j.cpc.2009.07.004}{{\em Comput. Phys.
  Commun.} {\bfseries 180} (2009) 2218--2249},
\href{http://arxiv.org/abs/0901.3106}{{\ttfamily arXiv:0901.3106 [hep-ph]}}.

\bibitem{Ablinger:2013cf}
J.~Ablinger, J.~Bl{\"u}mlein, and C.~Schneider, ``{Analytic and Algorithmic
  Aspects of Generalized Harmonic Sums and Polylogarithms},''
  \href{http://dx.doi.org/10.1063/1.4811117}{{\em J. Math. Phys.} {\bfseries
  54} (2013) 082301},
\href{http://arxiv.org/abs/1302.0378}{{\ttfamily arXiv:1302.0378 [math-ph]}}.

\bibitem{Ablinger:2014rba}
J.~Ablinger, ``{The package HarmonicSums: Computer Algebra and Analytic aspects
  of Nested Sums},'' {\em PoS} {\bfseries LL2014} (2014) 019,
\href{http://arxiv.org/abs/1407.6180}{{\ttfamily arXiv:1407.6180 [cs.SC]}}.

\bibitem{Ablinger:2016ll}
J.~Ablinger, ``{Inverse Mellin Transform of Holonomic Sequences},'' {\em PoS}
  {\bfseries LL2016} (2016) 067,
\href{http://arxiv.org/abs/1606.02845}{{\ttfamily arXiv:1606.02845 [cs.SC]}}.

\bibitem{Ablinger:2017rad}
J.~Ablinger, ``{Computing the Inverse Mellin Transform of Holonomic Sequences
  using Kovacic's Algorithm},'' {\em PoS} {\bfseries RADCOR2017} (2017) 069,
\href{http://arxiv.org/abs/1801.01039}{{\ttfamily arXiv:1801.01039 [cs.SC]}}.

\bibitem{Ablinger:2019mkx}
J.~Ablinger, ``{Discovering and Proving Infinite Pochhammer Sum Identities},''
  \href{http://dx.doi.org/10.1080/10586458.2019.1627254}{{\em {Experimental
  Mathematics}} {\bfseries 31} no.~1, (2019) 309--323},
\href{http://arxiv.org/abs/1902.11001}{{\ttfamily arXiv:1902.11001 [math.CO]}}.

\bibitem{Maitre:2005uu}
D.~Maitre, ``{HPL, a mathematica implementation of the harmonic
  polylogarithms},'' \href{http://dx.doi.org/10.1016/j.cpc.2005.10.008}{{\em
  Comput. Phys. Commun.} {\bfseries 174} (2006) 222--240},
  \href{http://arxiv.org/abs/hep-ph/0507152}{{\ttfamily arXiv:hep-ph/0507152}}.

\bibitem{Maitre:2007kp}
D.~Maitre, ``{Extension of HPL to complex arguments},''
  \href{http://dx.doi.org/10.1016/j.cpc.2011.11.015}{{\em Comput. Phys.
  Commun.} {\bfseries 183} (2012) 846},
  \href{http://arxiv.org/abs/hep-ph/0703052}{{\ttfamily arXiv:hep-ph/0703052}}.

\bibitem{Duhr:2019tlz}
C.~Duhr and F.~Dulat, ``{PolyLogTools --- polylogs for the masses},''
  \href{http://dx.doi.org/10.1007/JHEP08(2019)135}{{\em JHEP} {\bfseries 08}
  (2019) 135}, \href{http://arxiv.org/abs/1904.07279}{{\ttfamily
  arXiv:1904.07279 [hep-th]}}.

\bibitem{Bauer:2000cp}
C.~W. Bauer, A.~Frink, and R.~Kreckel, ``{Introduction to the GiNaC framework
  for symbolic computation within the C++ programming language},''
  \href{http://dx.doi.org/10.1006/jsco.2001.0494}{{\em J. Symb. Comput.}
  {\bfseries 33} (2000) 1},
\href{http://arxiv.org/abs/cs/0004015}{{\ttfamily arXiv:cs/0004015 [cs-sc]}}.

\bibitem{Anastasiou:2013srw}
C.~Anastasiou, C.~Duhr, F.~Dulat, and B.~Mistlberger, ``{Soft triple-real
  radiation for Higgs production at N3LO},''
  \href{http://dx.doi.org/10.1007/JHEP07(2013)003}{{\em JHEP} {\bfseries 07}
  (2013) 003}, \href{http://arxiv.org/abs/1302.4379}{{\ttfamily arXiv:1302.4379
  [hep-ph]}}.

\bibitem{Li:2014bfa}
Y.~Li, A.~von Manteuffel, R.~M. Schabinger, and H.~X. Zhu, ``{N$^3$LO Higgs
  boson and Drell-Yan production at threshold: The one-loop two-emission
  contribution},'' \href{http://dx.doi.org/10.1103/PhysRevD.90.053006}{{\em
  Phys. Rev. D} {\bfseries 90} no.~5, (2014) 053006},
  \href{http://arxiv.org/abs/1404.5839}{{\ttfamily arXiv:1404.5839 [hep-ph]}}.

\bibitem{Zhu:2014fma}
H.~X. Zhu, ``{On the calculation of soft phase space integral},''
  \href{http://dx.doi.org/10.1007/JHEP02(2015)155}{{\em JHEP} {\bfseries 02}
  (2015) 155}, \href{http://arxiv.org/abs/1501.00236}{{\ttfamily
  arXiv:1501.00236 [hep-ph]}}.

\bibitem{Duhr:2022cob}
C.~Duhr, B.~Mistlberger, and G.~Vita, ``{Soft Integrals and Soft Anomalous
  Dimensions at N$^3$LO and Beyond},''
  \href{http://arxiv.org/abs/2205.04493}{{\ttfamily arXiv:2205.04493
  [hep-ph]}}.

\bibitem{Beneke:1997zp}
M.~Beneke and V.~A. Smirnov, ``{Asymptotic expansion of Feynman integrals near
  threshold},'' \href{http://dx.doi.org/10.1016/S0550-3213(98)00138-2}{{\em
  Nucl. Phys. B} {\bfseries 522} (1998) 321--344},
  \href{http://arxiv.org/abs/hep-ph/9711391}{{\ttfamily arXiv:hep-ph/9711391}}.

\bibitem{Anastasiou:2015yha}
C.~Anastasiou, C.~Duhr, F.~Dulat, E.~Furlan, F.~Herzog, and B.~Mistlberger,
  ``{Soft expansion of double-real-virtual corrections to Higgs production at
  N$^{3}$LO},'' \href{http://dx.doi.org/10.1007/JHEP08(2015)051}{{\em JHEP}
  {\bfseries 08} (2015) 051}, \href{http://arxiv.org/abs/1505.04110}{{\ttfamily
  arXiv:1505.04110 [hep-ph]}}.

\bibitem{Duhr:2014nda}
C.~Duhr, T.~Gehrmann, and M.~Jaquier, ``{Two-loop splitting amplitudes and the
  single-real contribution to inclusive Higgs production at N$^3$LO},''
  \href{http://dx.doi.org/10.1007/JHEP02(2015)077}{{\em JHEP} {\bfseries 02}
  (2015) 077}, \href{http://arxiv.org/abs/1411.3587}{{\ttfamily arXiv:1411.3587
  [hep-ph]}}.

\bibitem{Duhr:2013msa}
C.~Duhr and T.~Gehrmann, ``{The two-loop soft current in dimensional
  regularization},''
  \href{http://dx.doi.org/10.1016/j.physletb.2013.10.063}{{\em Phys. Lett. B}
  {\bfseries 727} (2013) 452--455},
  \href{http://arxiv.org/abs/1309.4393}{{\ttfamily arXiv:1309.4393 [hep-ph]}}.

\bibitem{Baranowski:2020xlp}
D.~Baranowski, ``{NNLO zero-jettiness beam and soft functions to higher orders
  in the dimensional-regularization parameter $\epsilon$},''
  \href{http://dx.doi.org/10.1140/epjc/s10052-020-8047-y}{{\em Eur. Phys. J. C}
  {\bfseries 80} no.~6, (2020) 523},
  \href{http://arxiv.org/abs/2004.03285}{{\ttfamily arXiv:2004.03285
  [hep-ph]}}.

\bibitem{Korchemsky:1987wg}
G.~P. Korchemsky and A.~V. Radyushkin, ``{Renormalization of the Wilson Loops
  Beyond the Leading Order},''
  \href{http://dx.doi.org/10.1016/0550-3213(87)90277-X}{{\em Nucl. Phys. B}
  {\bfseries 283} (1987) 342--364}.

\bibitem{Moch:2004pa}
S.~Moch, J.~A.~M. Vermaseren, and A.~Vogt, ``{The Three loop splitting
  functions in QCD: The Nonsinglet case},''
  \href{http://dx.doi.org/10.1016/j.nuclphysb.2004.03.030}{{\em Nucl. Phys. B}
  {\bfseries 688} (2004) 101--134},
  \href{http://arxiv.org/abs/hep-ph/0403192}{{\ttfamily arXiv:hep-ph/0403192}}.

\bibitem{Vogt:2004mw}
A.~Vogt, S.~Moch, and J.~A.~M. Vermaseren, ``{The Three-loop splitting
  functions in QCD: The Singlet case},''
  \href{http://dx.doi.org/10.1016/j.nuclphysb.2004.04.024}{{\em Nucl. Phys. B}
  {\bfseries 691} (2004) 129--181},
  \href{http://arxiv.org/abs/hep-ph/0404111}{{\ttfamily arXiv:hep-ph/0404111}}.

\bibitem{Grozin:2014hna}
A.~Grozin, J.~M. Henn, G.~P. Korchemsky, and P.~Marquard, ``{Three Loop Cusp
  Anomalous Dimension in QCD},''
  \href{http://dx.doi.org/10.1103/PhysRevLett.114.062006}{{\em Phys. Rev.
  Lett.} {\bfseries 114} no.~6, (2015) 062006},
  \href{http://arxiv.org/abs/1409.0023}{{\ttfamily arXiv:1409.0023 [hep-ph]}}.

\bibitem{Grozin:2015kna}
A.~Grozin, J.~M. Henn, G.~P. Korchemsky, and P.~Marquard, ``{The three-loop
  cusp anomalous dimension in QCD and its supersymmetric extensions},''
  \href{http://dx.doi.org/10.1007/JHEP01(2016)140}{{\em JHEP} {\bfseries 01}
  (2016) 140}, \href{http://arxiv.org/abs/1510.07803}{{\ttfamily
  arXiv:1510.07803 [hep-ph]}}.

\bibitem{Henn:2016men}
J.~M. Henn, A.~V. Smirnov, V.~A. Smirnov, and M.~Steinhauser, ``{A planar
  four-loop form factor and cusp anomalous dimension in QCD},''
  \href{http://dx.doi.org/10.1007/JHEP05(2016)066}{{\em JHEP} {\bfseries 05}
  (2016) 066}, \href{http://arxiv.org/abs/1604.03126}{{\ttfamily
  arXiv:1604.03126 [hep-ph]}}.

\bibitem{Henn:2016wlm}
J.~Henn, A.~V. Smirnov, V.~A. Smirnov, M.~Steinhauser, and R.~N. Lee,
  ``{Four-loop photon quark form factor and cusp anomalous dimension in the
  large-$N_c$ limit of QCD},''
  \href{http://dx.doi.org/10.1007/JHEP03(2017)139}{{\em JHEP} {\bfseries 03}
  (2017) 139}, \href{http://arxiv.org/abs/1612.04389}{{\ttfamily
  arXiv:1612.04389 [hep-ph]}}.

\bibitem{Davies:2016jie}
J.~Davies, A.~Vogt, B.~Ruijl, T.~Ueda, and J.~A.~M. Vermaseren, ``{Large-$n_f$
  contributions to the four-loop splitting functions in QCD},''
  \href{http://dx.doi.org/10.1016/j.nuclphysb.2016.12.012}{{\em Nucl. Phys. B}
  {\bfseries 915} (2017) 335--362},
  \href{http://arxiv.org/abs/1610.07477}{{\ttfamily arXiv:1610.07477
  [hep-ph]}}.

\bibitem{Lee:2017mip}
R.~N. Lee, A.~V. Smirnov, V.~A. Smirnov, and M.~Steinhauser, ``{The $n_f^2$
  contributions to fermionic four-loop form factors},''
  \href{http://dx.doi.org/10.1103/PhysRevD.96.014008}{{\em Phys. Rev. D}
  {\bfseries 96} no.~1, (2017) 014008},
  \href{http://arxiv.org/abs/1705.06862}{{\ttfamily arXiv:1705.06862
  [hep-ph]}}.

\bibitem{Moch:2017uml}
S.~Moch, B.~Ruijl, T.~Ueda, J.~A.~M. Vermaseren, and A.~Vogt, ``{Four-Loop
  Non-Singlet Splitting Functions in the Planar Limit and Beyond},''
  \href{http://dx.doi.org/10.1007/JHEP10(2017)041}{{\em JHEP} {\bfseries 10}
  (2017) 041}, \href{http://arxiv.org/abs/1707.08315}{{\ttfamily
  arXiv:1707.08315 [hep-ph]}}.

\bibitem{Grozin:2018vdn}
A.~Grozin, ``{Four-loop cusp anomalous dimension in QED},''
  \href{http://dx.doi.org/10.1007/JHEP01(2019)134}{{\em JHEP} {\bfseries 06}
  (2018) 073}, \href{http://arxiv.org/abs/1805.05050}{{\ttfamily
  arXiv:1805.05050 [hep-ph]}}. [Addendum: JHEP \textbf{01} (2019) 134].

\bibitem{Moch:2018wjh}
S.~Moch, B.~Ruijl, T.~Ueda, J.~A.~M. Vermaseren, and A.~Vogt, ``{On quartic
  colour factors in splitting functions and the gluon cusp anomalous
  dimension},'' \href{http://dx.doi.org/10.1016/j.physletb.2018.06.017}{{\em
  Phys. Lett. B} {\bfseries 782} (2018) 627--632},
  \href{http://arxiv.org/abs/1805.09638}{{\ttfamily arXiv:1805.09638
  [hep-ph]}}.

\bibitem{Lee:2019zop}
R.~N. Lee, A.~V. Smirnov, V.~A. Smirnov, and M.~Steinhauser, ``{Four-loop quark
  form factor with quartic fundamental colour factor},''
  \href{http://dx.doi.org/10.1007/JHEP02(2019)172}{{\em JHEP} {\bfseries 02}
  (2019) 172}, \href{http://arxiv.org/abs/1901.02898}{{\ttfamily
  arXiv:1901.02898 [hep-ph]}}.

\bibitem{Henn:2019rmi}
J.~M. Henn, T.~Peraro, M.~Stahlhofen, and P.~Wasser, ``{Matter dependence of
  the four-loop cusp anomalous dimension},''
  \href{http://dx.doi.org/10.1103/PhysRevLett.122.201602}{{\em Phys. Rev.
  Lett.} {\bfseries 122} no.~20, (2019) 201602},
  \href{http://arxiv.org/abs/1901.03693}{{\ttfamily arXiv:1901.03693
  [hep-ph]}}.

\bibitem{vonManteuffel:2019wbj}
A.~von Manteuffel and R.~M. Schabinger, ``{Quark and gluon form factors in four
  loop QCD: The $N_f^2$ and $N_{q\gamma} N_f$ contributions},''
  \href{http://dx.doi.org/10.1103/PhysRevD.99.094014}{{\em Phys. Rev. D}
  {\bfseries 99} no.~9, (2019) 094014},
  \href{http://arxiv.org/abs/1902.08208}{{\ttfamily arXiv:1902.08208
  [hep-ph]}}.

\bibitem{Bruser:2019auj}
R.~Br\"user, A.~Grozin, J.~M. Henn, and M.~Stahlhofen, ``{Matter dependence of
  the four-loop QCD cusp anomalous dimension: from small angles to all
  angles},'' \href{http://dx.doi.org/10.1007/JHEP05(2019)186}{{\em JHEP}
  {\bfseries 05} (2019) 186}, \href{http://arxiv.org/abs/1902.05076}{{\ttfamily
  arXiv:1902.05076 [hep-ph]}}.

\bibitem{Henn:2019swt}
J.~M. Henn, G.~P. Korchemsky, and B.~Mistlberger, ``{The full four-loop cusp
  anomalous dimension in $\mathcal{N}=4$ super Yang-Mills and QCD},''
  \href{http://dx.doi.org/10.1007/JHEP04(2020)018}{{\em JHEP} {\bfseries 04}
  (2020) 018}, \href{http://arxiv.org/abs/1911.10174}{{\ttfamily
  arXiv:1911.10174 [hep-th]}}.

\bibitem{vonManteuffel:2020vjv}
A.~von Manteuffel, E.~Panzer, and R.~M. Schabinger, ``{Cusp and collinear
  anomalous dimensions in four-loop QCD from form factors},''
  \href{http://dx.doi.org/10.1103/PhysRevLett.124.162001}{{\em Phys. Rev.
  Lett.} {\bfseries 124} no.~16, (2020) 162001},
  \href{http://arxiv.org/abs/2002.04617}{{\ttfamily arXiv:2002.04617
  [hep-ph]}}.

\bibitem{Ablinger:2014lka}
J.~Ablinger, J.~Bl{\"u}mlein, A.~De~Freitas, A.~Hasselhuhn, A.~von Manteuffel,
  M.~Round, C.~Schneider, and F.~Wi\ss{}brock, ``{The Transition Matrix Element
  $A_{gq}(N)$ of the Variable Flavor Number Scheme at $O(\alpha_s^3)$},''
  \href{http://dx.doi.org/10.1016/j.nuclphysb.2014.02.007}{{\em Nucl. Phys.}
  {\bfseries B882} (2014) 263--288},
\href{http://arxiv.org/abs/1402.0359}{{\ttfamily arXiv:1402.0359 [hep-ph]}}.

\bibitem{Ablinger:2014vwa}
J.~Ablinger, A.~Behring, J.~Bl\"umlein, A.~De~Freitas, A.~Hasselhuhn, A.~von
  Manteuffel, M.~Round, C.~Schneider, and F.~Wi\ss{}brock, ``{The 3-Loop
  Non-Singlet Heavy Flavor Contributions and Anomalous Dimensions for the
  Structure Function $F_2(x,Q^2)$ and Transversity},''
  \href{http://dx.doi.org/10.1016/j.nuclphysb.2014.07.010}{{\em Nucl. Phys. B}
  {\bfseries 886} (2014) 733--823},
  \href{http://arxiv.org/abs/1406.4654}{{\ttfamily arXiv:1406.4654 [hep-ph]}}.

\bibitem{Ablinger:2014nga}
J.~Ablinger, A.~Behring, J.~Bl\"umlein, A.~De~Freitas, A.~von Manteuffel, and
  C.~Schneider, ``{The 3-loop pure singlet heavy flavor contributions to the
  structure function $F_2(x,Q^2)$ and the anomalous dimension},''
  \href{http://dx.doi.org/10.1016/j.nuclphysb.2014.10.008}{{\em Nucl. Phys. B}
  {\bfseries 890} (2014) 48--151},
  \href{http://arxiv.org/abs/1409.1135}{{\ttfamily arXiv:1409.1135 [hep-ph]}}.

\bibitem{Ablinger:2017tan}
J.~Ablinger, A.~Behring, J.~Bl\"umlein, A.~De~Freitas, A.~von Manteuffel, and
  C.~Schneider, ``{The three-loop splitting functions $P_{qg}^{(2)}$ and
  $P_{gg}^{(2, N_F)}$},''
  \href{http://dx.doi.org/10.1016/j.nuclphysb.2017.06.004}{{\em Nucl. Phys. B}
  {\bfseries 922} (2017) 1--40},
  \href{http://arxiv.org/abs/1705.01508}{{\ttfamily arXiv:1705.01508
  [hep-ph]}}.

\bibitem{Blumlein:2021enk}
J.~Bl\"umlein, P.~Marquard, C.~Schneider, and K.~Sch\"onwald, ``{The three-loop
  unpolarized and polarized non-singlet anomalous dimensions from off shell
  operator matrix elements},''
  \href{http://dx.doi.org/10.1016/j.nuclphysb.2021.115542}{{\em Nucl. Phys. B}
  {\bfseries 971} (2021) 115542},
  \href{http://arxiv.org/abs/2107.06267}{{\ttfamily arXiv:2107.06267
  [hep-ph]}}.

\bibitem{Billis:2019vxg}
G.~Billis, M.~A. Ebert, J.~K.~L. Michel, and F.~J. Tackmann, ``{A toolbox for
  $q_{T}$ and 0-jettiness subtractions at $\hbox {N}^3\hbox {LO}$},''
  \href{http://dx.doi.org/10.1140/epjp/s13360-021-01155-y}{{\em Eur. Phys. J.
  Plus} {\bfseries 136} no.~2, (2021) 214},
  \href{http://arxiv.org/abs/1909.00811}{{\ttfamily arXiv:1909.00811
  [hep-ph]}}.

\bibitem{Bruser:2018rad}
R.~Br\"user, Z.~L. Liu, and M.~Stahlhofen, ``{Three-Loop Quark Jet Function},''
  \href{http://dx.doi.org/10.1103/PhysRevLett.121.072003}{{\em Phys. Rev.
  Lett.} {\bfseries 121} no.~7, (2018) 072003},
  \href{http://arxiv.org/abs/1804.09722}{{\ttfamily arXiv:1804.09722
  [hep-ph]}}.

\bibitem{Banerjee:2018ozf}
P.~Banerjee, P.~K. Dhani, and V.~Ravindran, ``{Gluon jet function at three
  loops in QCD},'' \href{http://dx.doi.org/10.1103/PhysRevD.98.094016}{{\em
  Phys. Rev. D} {\bfseries 98} no.~9, (2018) 094016},
  \href{http://arxiv.org/abs/1805.02637}{{\ttfamily arXiv:1805.02637
  [hep-ph]}}.

\bibitem{Baranowski:2022khd}
D.~Baranowski, M.~Delto, K.~Melnikov, and C.-Y. Wang, ``{Same-hemisphere
  three-gluon-emission contribution to the zero-jettiness soft function at N3LO
  QCD},'' \href{http://dx.doi.org/10.1103/PhysRevD.106.014004}{{\em Phys. Rev.
  D} {\bfseries 106} no.~1, (2022) 014004},
  \href{http://arxiv.org/abs/2204.09459}{{\ttfamily arXiv:2204.09459
  [hep-ph]}}.

\bibitem{Baranowski:2021gxe}
D.~Baranowski, M.~Delto, K.~Melnikov, and C.-Y. Wang, ``{On phase-space
  integrals with Heaviside functions},''
  \href{http://dx.doi.org/10.1007/JHEP02(2022)081}{{\em JHEP} {\bfseries 02}
  (2022) 081}, \href{http://arxiv.org/abs/2111.13594}{{\ttfamily
  arXiv:2111.13594 [hep-ph]}}.

\bibitem{Chen:2022cvz}
W.~Chen, F.~Feng, Y.~Jia, and X.~Liu, ``{Double-Real-Virtual and
  Double-Virtual-Real Corrections to the Three-Loop Thrust Soft Function},''
  \href{http://arxiv.org/abs/2206.12323}{{\ttfamily arXiv:2206.12323
  [hep-ph]}}.

\bibitem{Ellis:2016jkw}
J.~Ellis, ``{TikZ-Feynman: Feynman diagrams with TikZ},''
  \href{http://dx.doi.org/10.1016/j.cpc.2016.08.019}{{\em Comput. Phys.
  Commun.} {\bfseries 210} (2017) 103--123},
  \href{http://arxiv.org/abs/1601.05437}{{\ttfamily arXiv:1601.05437
  [hep-ph]}}.

\bibitem{Cox:IVA}
D.~A. Cox, J.~Little, and D.~O'Shea,
  \href{http://dx.doi.org/10.1007/978-3-319-16721-3}{{\em {Ideals, Varieties,
  and Algorithms}}}.
\newblock Springer, Cham, 2018.

\bibitem{Singular}
W.~Decker, G.-M. Greuel, G.~Pfister, and H.~Sch{\"o}nemann,
  ``{\textsc{Singular} 4.2.1 --- A computer algebra system for polynomial
  computations},'' 2021.
\newblock \url{http://www.singular.uni-kl.de}.

\end{thebibliography}\endgroup
\end{document}